\documentclass{aastex63}

\shorttitle{Finding Galaxy Lenses}
\shortauthors{Knabel et al.}
\graphicspath{{./}{figures/}}
\usepackage{amsmath}

\begin{document}

\title{Galaxy And Mass Assembly: A Comparison between Galaxy-Galaxy Lens Searches in KiDS/GAMA.}

\author[0000-0001-5110-6241]{Shawn Knabel}
\affiliation{Department of Physics and Astronomy, 102 Natural Science Building, University of Louisville, Louisville KY 40292, USA}

\author[0000-0001-9537-5814]{Rebecca L. Steele}
\affiliation{Department of Physics and Astronomy, 102 Natural Science Building, University of Louisville, Louisville KY 40292, USA}

\author[0000-0002-4884-6756]{Benne W. Holwerda}
\affiliation{Department of Physics and Astronomy, 102 Natural Science Building, University of Louisville, Louisville KY 40292, USA}

\author[0000-0002-8584-1903]{Joanna S. Bridge}
\affiliation{Department of Physics and Astronomy, 102 Natural Science Building, University of Louisville, Louisville KY 40292, USA}

\author[0000-0001-9631-831X]{Alice Jacques}
\affiliation{Department of Physics and Astronomy, 102 Natural Science Building, University of Louisville, Louisville KY 40292, USA}

\author[0000-0002-6097-2747]{Andrew M. Hopkins}
\affiliation{Australian Astronomical Optics, Macquarie University, 105 Delhi Rd, North Ryde, NSW 2113, Australia}

\author[0000-0001-7821-7195]{Steven P. Bamford}	
\affiliation{School of Physics \& Astronomy, University of Nottingham, University Park, Nottingham, NG7 2RD, UK}

\author[0000-0002-1207-9137]{Michael J. I. Brown}
\affiliation{School of Physics \& Astronomy, Monash University, Clayton, VIC 3800, Australia}

\author[0000-0002-9796-1363]{Sarah Brough}	
\affiliation{School of Physics, University of New South Wales, NSW 2052, Australia}

\author[0000-0001-9395-4759]{Lee S. Kelvin}
\affiliation{Department of Astrophysical Sciences, Princeton University, 4 Ivy Lane, Princeton, NJ 08544, USA}

\author[0000-0002-3910-5809]{Maciej Bilicki}
\affiliation{Center for Theoretical Physics, Polish Academy of Sciences, Warsaw, Poland}

\author[0000-0003-0497-2651]{John Kielkopf}
\affiliation{Department of Physics and Astronomy, 102 Natural Science Building, University of Louisville, Louisville KY 40292, USA}

\begin{abstract}

Strong gravitational lenses are a rare and instructive type of astronomical object. Identification has long relied on serendipity, but different strategies -- such as mixed spectroscopy of multiple galaxies along the line of sight, machine learning algorithms, and citizen science -- have been employed to identify these objects as new imaging surveys become available.

We report on the comparison between spectroscopic, machine learning, and citizen science identification of galaxy-galaxy lens candidates from independently constructed lens catalogs in the common survey area of the equatorial fields of the Galaxy and Mass Assembly (GAMA) survey. In these, we have the opportunity to compare high-completeness spectroscopic identifications against high-fidelity imaging from the Kilo Degree Survey (KiDS) used for both machine learning and citizen science lens searches. 

We find that the three methods -- spectroscopy, machine learning, and citizen science -- identify 47, 47, and 13 candidates respectively in the 180 square degrees surveyed. These identifications barely overlap, with only two identified by both citizen science and machine learning. We have traced this discrepancy to inherent differences in the selection functions of each of the three methods, either within their parent samples (i.e. citizen science focuses on low-redshift) or inherent to the method (i.e. machine learning is limited by its training sample and prefers well-separated features, while spectroscopy requires sufficient flux from lensed features to lie within the fiber). These differences manifest as separate samples in estimated Einstein radius, lens stellar mass,  and lens redshift. The combined sample implies a lens candidate sky-density $\sim0.59$ deg$^{-2}$ and can inform the construction of a training set spanning a wider mass-redshift space. A combined approach and refinement of automated searches would result in a more complete sample of galaxy-galaxy lens candidates for future surveys. 
\end{abstract}



\section{Introduction}

Elliptical galaxies' structure, kinematics, and formation histories are a compelling test of the Cold Dark Matter ($\Lambda$CDM) paradigm as they are the end-product of galaxy formation \citep{De-Lucia06}. Furthermore, those that act as strong gravitational lenses appear in every respect to be just like other elliptical galaxies, so results of their study can therefore be generalized to all spheroidal galaxies in the observed mass range \citep{slacs9}. These lensing systems provide a highly accurate measurement of the {\em total} mass inside the Einstein radius, and therefore the dark matter content, in these elliptical galaxies that compares well with current galaxy evolution models and assumptions \citep[][and reference therein]{slacs12}. To date, gravitational lenses have proven General Relativity to be correct with high accuracy over galaxy-wide scales \citep{Collett18} and may provide an excellent test case for other theories of gravity, such as the Emergent Gravity recently proposed by \cite{Verlinde17}, through a combination of lensing and kinematic measurements \citep[see][]{Tortora18a}. 

Gravitational lensing has been a powerful technique to measure the masses of the {\em most massive} elliptical galaxies \citep{slacs4, slacs5}, as well as helping to understand their Fundamental Plane \citep{slacs7}, the stellar population's mass-to-light ratio, and thus their initial mass function \citep[IMF;][]{slacs9,Hopkins18}.
The observational drive is now to measure their mass content throughout the spheroidal galaxy mass function \citep{slacs1}, explore the Fundamental Plane in different environments \citep{slacs8}, constrain the stellar mass-to-light ratio in nearby ellipticals \citep{slacs2,Collier18a,Collier18b}, discover dark matter substructure in known strong galaxy lens cases \citep{Vegetti12, Cyr-Racine19}, and independently measure $H_0$ through time-delay cosmography 
\citep[e.g. H0LiCOW,][]{Suyu17, Chen19b}. These observational studies need larger samples of lenses. Only with a significant expansion of the lensing sample can the effects of evolution and stellar mass be decoupled \citep[e.g.,][]{slacs2}.

Thus far, lensing arcs have predominantly been identified in {\em massive} ($>10^{11}M_\odot$) lens systems thanks to selection biases: SDSS spectroscopic targets are volume-weighted to greater mass (intrinsically bright galaxies are included over a greater volume), and visual identification favors well-separated arc and lens \citep[][]{slacs12}. With the GAMA spectroscopy-selected sample \citep[][]{Holwerda15}, a greater range in lens masses is now available. However, the drive for much larger samples has led to increased searches using machine learning to constrain $\Lambda$CDM in detail \citep{Petrillo17,Petrillo18,Petrillo19,Speagle19,Huang20a,Huang20b,Jacobs19,Li20c}.

The success and completeness of the different identification techniques are difficult to test against one another. This has motivated our study here that benefits from three independent identifications -- spectroscopic, machine learning and citizen science -- on the same target fields. Our aim is to compare all three techniques to map out an optimal path for future searches. In the following calculations we adopt a flat $\Lambda$CDM cosmological model with  $h_0 = 0.738$, where $h_0 = \frac{H_0}{100}$ $km s^{-1} Mpc^{-1}$ as indicated in \cite{Riess11} and $\Omega_{m,0} = 0.262$.

\section{Identifying Strong Galaxy-Galaxy Lenses}

The principal selection technique for galaxy-galaxy lenses has been the identification of double spectral profiles in a single aperture based spectrum \citep[blended spectra, see][Steele et al. {\em in prep.}]{Bolton04,Holwerda15}. Searches for blended spectra in the Sloan Digital Sky Survey (SDSS) have been highly successful in identifying strong-lens candidate galaxies by identifying spectra containing both a low-redshift passive galaxy and emission lines from a much higher-redshift lensed source. In order to confirm such sources as true lensing systems, one requires significantly higher spatial resolution imaging than SDSS can provide. 

85 cases of strong lenses have already been confirmed with Hubble Space Telescope (HST) imaging through the SLACS (Sloan Lens ACS) survey \citep{slacs2,slacs3,slacs4,slacs5,slacs6,slacs7,slacs8,slacs9}. The SLACS survey was efficient in finding these rare strong lensing objects: approximately 50\% of the blended spectra (with a dispersion estimate of the Einstein radius) targeted were confirmed with HST observations, and its success has expanded into the BOSS survey and higher redshifts \citep[BELLS, $0.4<z<0.7$,][]{Brownstein12, Shu16}. 

However, due to the depth and completeness limitations of SDSS spectroscopy -- main survey depth is $m_r < 17.7$ \citep{Eisenstein01}\footnote{The spectroscopic luminous red galaxy (LRG) sample is limited to $m_r < 19.5$ thanks to the 4000\AA{} break.} -- typically only massive foreground galaxy lenses can be identified ($10^{11}-10^{12} M_\odot$, SDSS-Luminous Red Galaxy (LRG) sample, \cite{Eisenstein01}) at intermediate redshifts ($z=0.05-0.5$). According to \cite{Hilbert08} the lensing cross section drops rapidly below $z\sim0.5$, so naturally there are fewer lenses at lower redshifts. However, \cite{Sonnenfeld15}, referencing \cite{Arneson12} and \cite{Gavazzi14}, points to the Einstein radius as the main quantity determining the detection probability as opposed to the lensing cross section. Lens candidate identification through any spectroscopic method requires sufficient flux from the background galaxy in order to obtain a second spectral match. For lensing systems whose Einstein radius exceeds the radius of the instrument's aperture ($1.5^{\prime\prime}$-radius for SDSS), the probability of detection goes down significantly and rapidly \citep[][and reference therein]{Sonnenfeld15}.

\pagebreak

\subsection{GAMA Spectroscopic Identification}

The Galaxy and Mass Assembly survey \citep[GAMA;][]{Driver09,Driver11} is a multi-wavelength survey built around a deep and highly complete redshift survey of five fields with the Anglo-Australian Telescope. GAMA has three major advantages over SDSS in the identification of blended spectra: 
(1) the spectroscopic limiting depth is 2 magnitudes deeper ($m_r < 19.8$ mag.), 
(2) the completeness is close to 98\% \citep{Liske15}, and 
(3) the {\sc autoz} redshift algorithm easily identifies spectra with signal from two different redshifts \citep{Baldry14}. 

In the GAMA survey, \cite{Holwerda15} identified 104 strong lensing candidates from their blended spectra, all of which showed a passive galaxy (PG) with an emission line galaxy (ELG) at higher redshift. \cite{Chan16} found 10 of a subset of 14 of these spectroscopically identified lens candidates to be probable lenses using deep Subaru imaging. This GAMA spectroscopy sample is dominated by lower-mass spheroidal galaxies and higher-redshift galaxies (Steele et al. {\em in prep.}). These GAMA strong lens candidates from \cite[][Steele et al. {\em in prep.}]{Holwerda15} extend the stellar mass range and provide a medium-redshift observation in between the SLACS, S4TM, and BELLS samples. 

As mentioned above, the spectroscopic approach is intrinsically limited by the aperture of the spectroscopy: if the lensed features fall outside it, the signal of the lensed (source) galaxy will be weak and unlikely to be detected \citep[][and reference therein]{Sonnenfeld15}. As a result of this, GAMA spectroscopy's $1^{\prime\prime}$-radius aperture structurally misses lower redshift and higher mass lens candidates. This point is discussed further in Sections \ref{section_mass_redshift} and \ref{section_maxmass} in relation to other identified candidates. GAMA has improved the spectroscopic identification of lower mass strong lenses, but a much larger sample of lower redshift ($z<0.1$) lenses is needed to constrain mass-to-light ratios in ellipticals.

\subsection{Beyond Spectroscopic Identification of Lenses}

\subsubsection{Machine Learning Identification of Lenses}

Machine learning is gaining popularity as a method for identifying galaxy-galaxy lens candidates, e.g. in Subaru Hyper-Supreme Cam  \citep{Speagle19}, DECAM \citep{Huang20b}, and Dark Energy Survey data \citep{Jacobs19}. 
\cite{Petrillo17, Petrillo18, Petrillo19} introduced and developed a machine learning technique to visually identify strong lens candidates by training the convolutional neural networks to recognize the characteristic arcs that appear next to a lensing elliptical galaxy using simulated images as their training set. These ``mock lenses" are created by simulating lens features around the images of real galaxies that are selected using color-magnitude criteria modified from SDSS-LRG selection \citep{Eisenstein01} and with the Einstein radius parameter drawn from a logarithmic distribution in the range of 1.0-5.0 arcseconds, intentionally resembling SLACS lenses. Their selection process included a follow-up visual inspection of each candidate by seven members of the team, which provided a score for each candidate between 0 (reflecting low confidence) and 70 (reflecting high confidence). This new identification method resulted in the $\sim$1300 candidates of the galaxy-galaxy Lenses in the Kilo-Degree Survey (LinKS) sample \citep[KiDS;][]{de-Jong13,de-Jong15,de-Jong17,Kuijken19}, which overlaps with 100\% of the equatorial fields of the GAMA survey (fields G09, G12, and G15). These identifications by \cite{Petrillo18} have shown that there are many more strong lenses to be found in the same survey area using identification methods other than spectroscopic. This approach has succeeded in finding more candidate strong lenses similar to the simulated massive elliptical galaxies on which the neural network was trained, ie. large ellipticals (LRGs) with characteristics similar to those of SLACS lens candidates.

\subsubsection{Citizen Science Identification of Lenses}

GalaxyZoo \citep{Lintott08,Marshall16} has classified KiDS postage stamps of galaxies on the same area that corresponds to the GAMA/KiDS survey overlap \citep[Kelvin et al. {\em in prep},][]{Holwerda19} using a question tree design shown below in Figure \ref{fig:q_tree}. It leads participants through a series of questions that (given the object is a galaxy) will arrive at a question that prompts them to identify any number of the following ``odd features" in the image: ``None", ``Ring", ``Lens or arc", ``Dust lane", ``Irregular", ``Other", and ``Overlapping." The GalaxyZoo team imposed a redshift restriction of $z < 0.15$ in the pre-selection of images to be classified by participants. With the voting completed on the GAMA/KiDS equatorial fields, we can now analyze the results of this citizen science approach.

With three independent techniques to identify strong gravitational lenses in the same three fields, we have an opportunity to test how well different identification techniques can be calibrated against one another, discover implicit selection effects in each, and approximate the on-sky density of strong galaxy-galaxy lens candidates. 
\begin{figure*}
    \centering
     \includegraphics[width=0.8\textwidth]{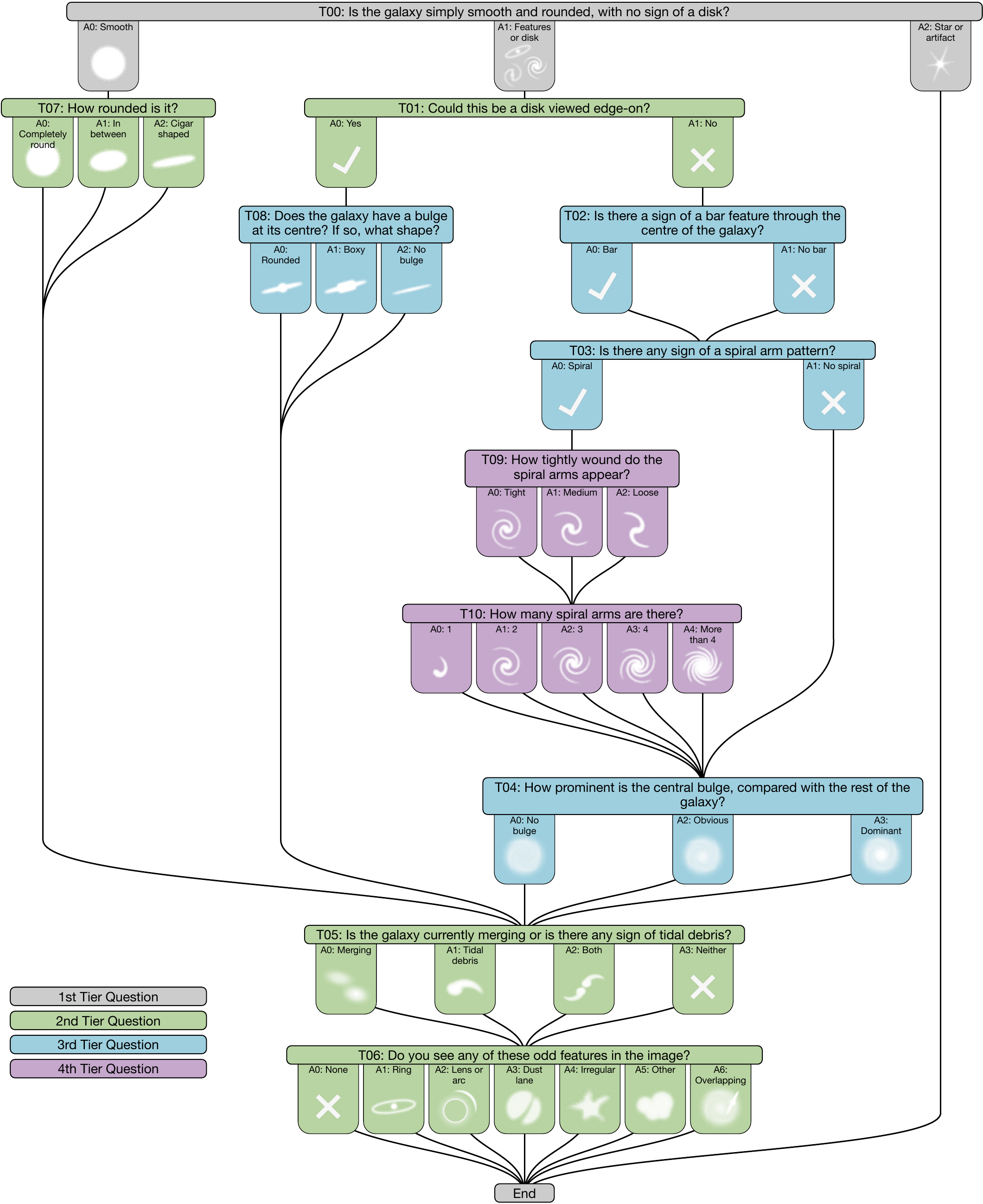}
    \caption{The question tree employed by GalaxyZoo citizen science, which volunteer participants use to classify images from the Kilo-Degree Survey (KiDS). 2nd Tier Question T06 prompts participants to identify ``odd features," of which an option is ``Lens or arc." Participants can choose more than one option in T06. We used results from this question to determine the final cut for our GalaxyZoo sample selection.}
    \label{fig:q_tree}
\end{figure*}
\clearpage

\section{Data and Observations}

\begin{figure}[h!]
    \centering
    \includegraphics[width=\linewidth]{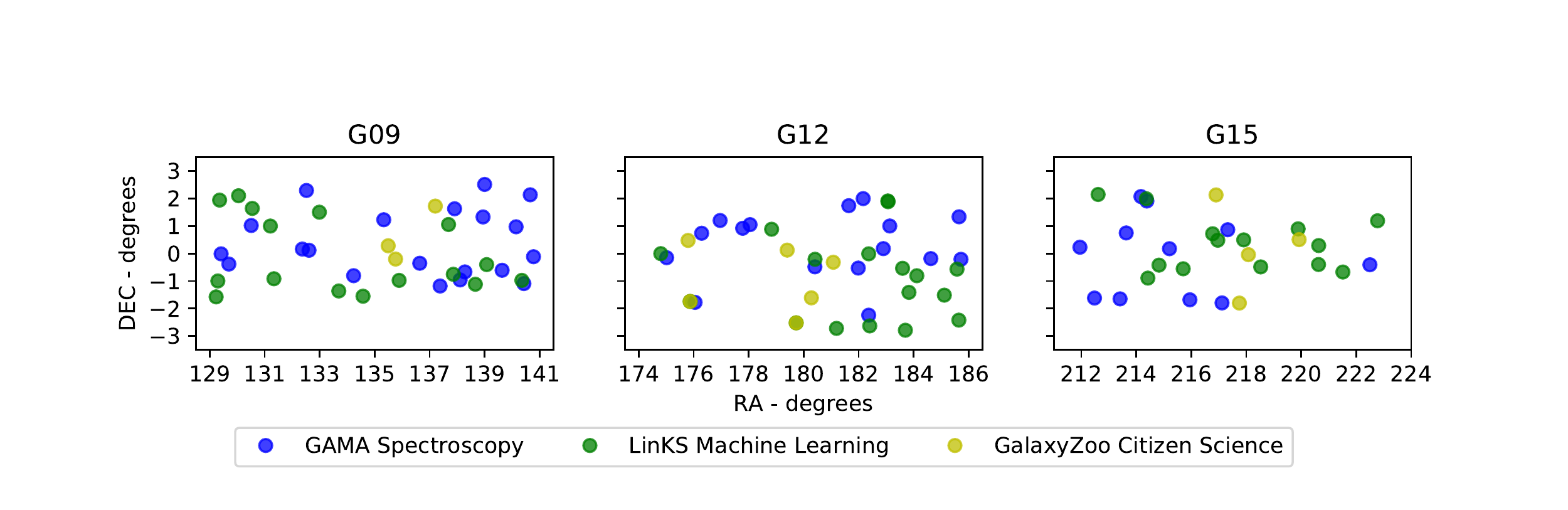}
    \caption{The three GAMA/KiDS fields (G09, G12 and G15) within which data from all three lensing identification techniques are available. GAMA spectroscopic, LinKS machine learning, and GalaxyZoo citizen science galaxy-galaxy lens candidates are identified in all three fields. The candidates shown here are those that passed final selection criteria.}
    \label{fig:gama_regions}
\end{figure}

In order to obtain a valid comparison of the three techniques, the catalogs of candidate lenses obtained by GAMA blended spectra \citep{Holwerda15}, LinKS machine learning \citep{Petrillo18}, and GalaxyZoo citizen science (Kelvin et al. {\em in prep.}) were cut to represent only those candidates identified within the equatorial GAMA fields G09, G12, and G15, which is the area of overlap between GAMA and KiDS. We show the presence of candidates identified by each method in each of those fields in Figure \ref{fig:gama_regions}. This cut resulted in usable catalogs consisting of 85 spectroscopically-identified lens candidates, 421 candidates identified by LinKS machine learning, and a misleading total of 12934 GalaxyZoo candidates with ``Lens or arc" scores of 0 or higher. Further cuts (see below in Sections \ref{spec_catalog} and \ref{gz_catalog}) to the catalogs were then made to account for false positives in each of the samples.  

Stellar mass estimates and redshifts for each candidate were taken from the GAMA 
{\sc LAMBDAR} photometric catalog \citep{Wright16} converted to stellar masses using the prescription from \cite{Taylor11} from GAMA DR2/3 \citep{Liske15,Baldry18}\footnote{\url{http://www.gama-survey.org/dr3/schema/dmu.php?id=9}}. We note that the stellar masses from the {\sc MagPhys} \citep{Wright18}\footnote{\url{http://www.gama-survey.org/dr3/schema/table.php?id=82}} catalog are a factor 2 smaller than the {\sc LAMBDAR} photometric estimates \citep{Taylor20}. 
Both the {\sc MagPhys} and {\sc LAMBDAR} photometry stellar masses are based on the same {\sc LAMBDAR} photometry. The documentation notes that both are not corrected for aperture using the S\'ersic fits from \cite{Kelvin12} because {\sc LAMBDAR} uses matched apertures. The {\sc MagPhys} is a full SED treatment from ultraviolet to sub-mm, and {\sc LAMBDAR} stellar mass estimates use u--Y photometry. Both use Chabrier IMFs and the \cite{Bruzual03} stellar template models. The {\sc MagPhys} uses a mix of dust models, while the  {\sc LAMBDAR} mass catalog uses a screen with \cite{Calzetti99} extinction law. Scatter between stellar mass estimates from photometry using M/L ratio prescriptions is typically 0.3 dex \citep[see, e.g.][for the discrepancy between  {\sc MagPhys} and WISE photometry mass estimates]{Cluver14,Kettlety18}.
Given that both are based on the same photometry, without aperture corrections, and that {\sc MagPhys} was developed with star-forming galaxies in mind and the {\sc LAMBDAR} mass catalog calibrated for ellipticals \citep[see also][]{Taylor20}, we opt for the {\sc LAMBDAR} mass catalog as the most appropriate for our sources. Candidates were matched between independent catalogs by GAMA ID, except for the LinKS sample, which was matched based on RA and DEC.

\subsection{GAMA Blended Spectra and LinKS Machine Learning Catalogs} \label{spec_catalog}

\begin{figure}[h!]
    \centering
        \includegraphics[width=\linewidth]{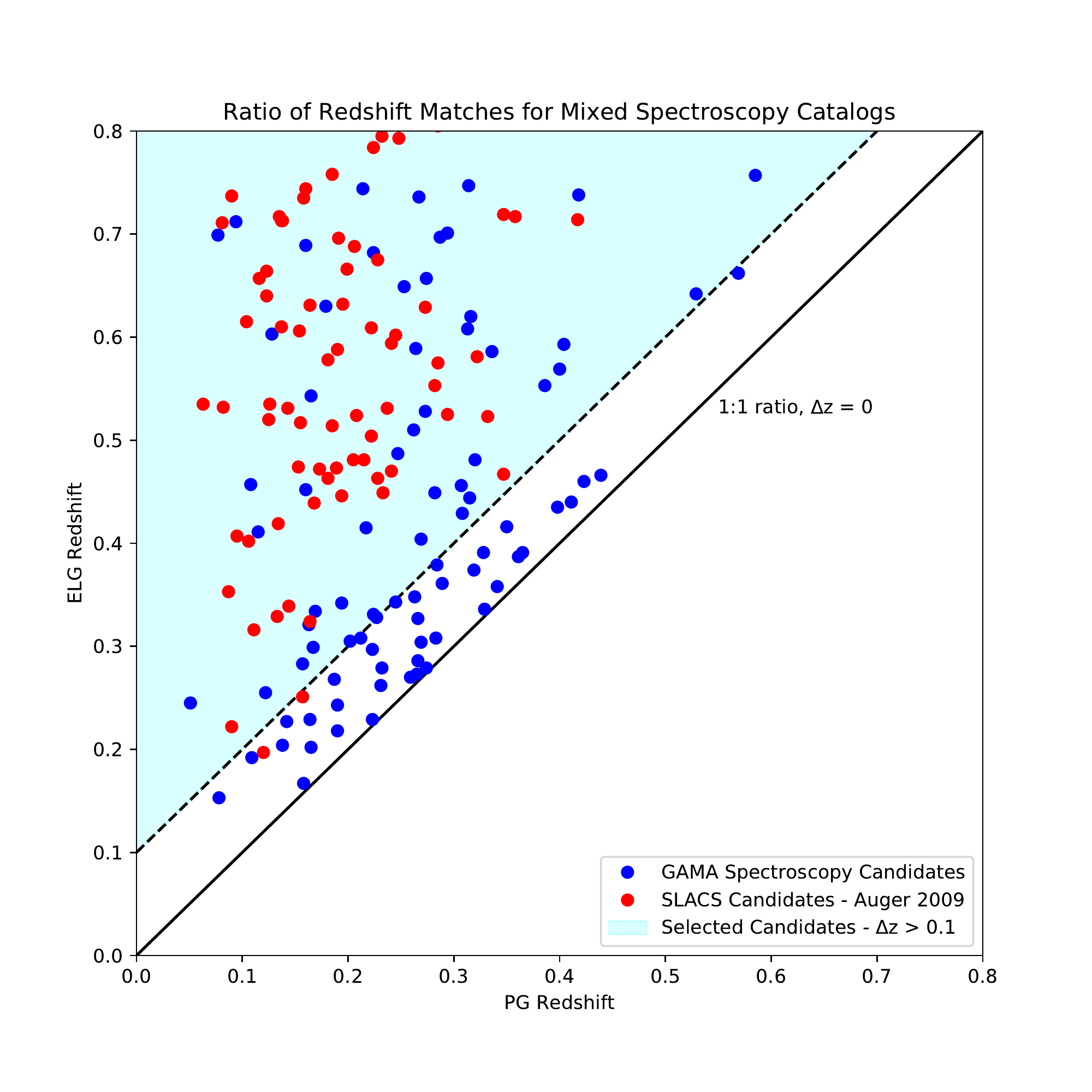}
    \caption{Selection of GAMA spectroscopy candidates based on redshift difference between lens and source. Blue markers denote GAMA spectroscopy candidates from \cite{Holwerda15}. In all cases, the passive galaxy match is at lower redshift than the emission line galaxy match. The shaded area represents redshift differences greater than 0.1, which indicate likely lens candidates. We remove all GAMA spectroscopy candidates outside this shaded region, which are more likely interacting or close paired galaxies. Red markers show redshift difference for SLACS candidates taken from \cite{Auger09}.}
    \label{fig:spec_selection}
\end{figure}

The lens candidate sample provided by \cite{Holwerda15} is based on spectral match, and therefore does not include a subjective follow-up visual inspection. However, in order to attain a more pure sample for consideration here, we selected candidates with a minimum difference of $\Delta z > 0.1$ between the redshift of the passive galaxy (PG) spectral match and an emission line galaxy (ELG) match at higher redshift. This selection is shown in Figure \ref{fig:spec_selection} in comparison with Grade-A SLACS lenses from \cite{Auger09}. The result of this selection is 47 candidates identified by GAMA spectroscopy, 31 of which have stellar mass estimates from the GAMA {\sc LAMBDAR} catalog.

The LinKS machine learning catalog of 421 candidates obtained from \cite{Petrillo18} included all those objects whose visual inspection score was greater than 0. In order to compare only the highest quality candidates, a score threshold of $> 17$ was taken from \cite{Petrillo17}. This reduced the LinKS machine learning catalog to 47, of which 46 have stellar mass estimates in the {\sc LAMBDAR} catalog. All following mentions of the GAMA spectroscopy and LinKS machine learning samples refer to these reduced selections of candidates unless otherwise specified.

\subsection{GalaxyZoo Citizen Science Catalog} \label{gz_catalog}

GalaxyZoo presented the most variables to consider when analyzing its reliability as a method for identifying strong lens candidates. The percentage of votes for lensing features in each candidate is taken to be a subjective score given in a similar manner to the LinKS machine learning catalog. However, since the program was not designed specifically to identify only lenses, each candidate's score must also be considered relative to scores of other choices within the same question level. For each candidate in question, other ``odd features" could potentially pull votes away from the ``Lens or arc" classification. This aspect is specific to GalaxyZoo citizen science among the three methods considered here: GalaxyZoo simultaneously considers a range of other classification results that add false positives and noise to the data when attempting to focus on one type of object, e.g. strong lens candidates. All other classifications must be accounted for in the cleaning of the GalaxyZoo sample.

\begin{figure*}
    \centering
    \includegraphics[width=0.4\columnwidth]{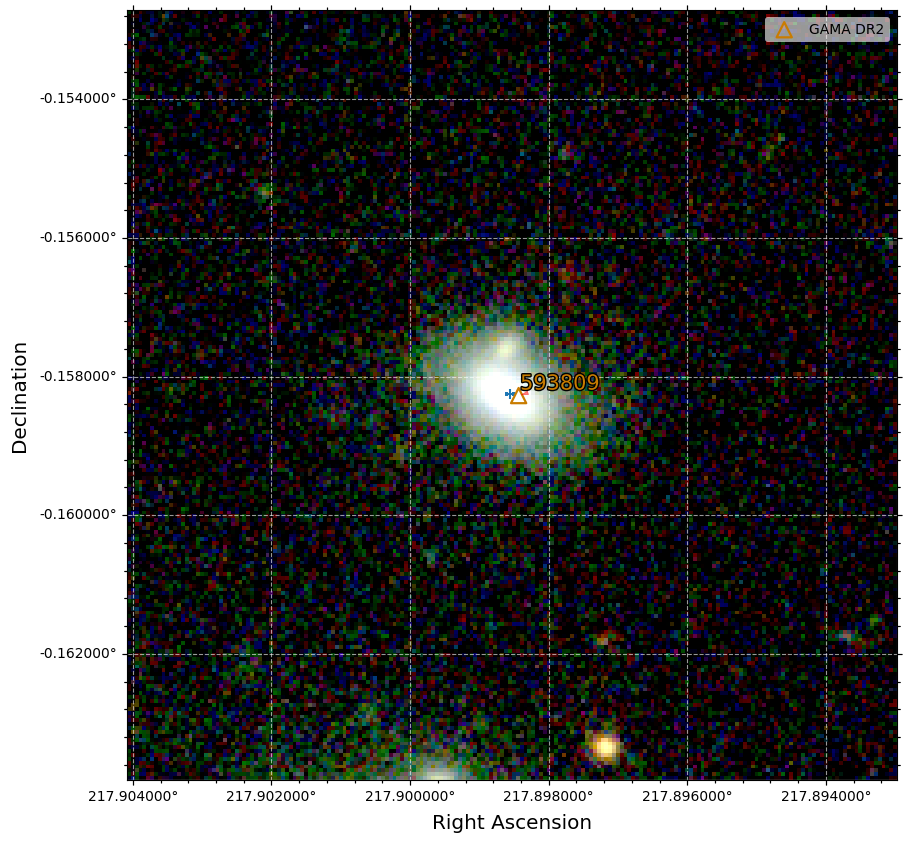}
    \includegraphics[width=0.4\columnwidth]{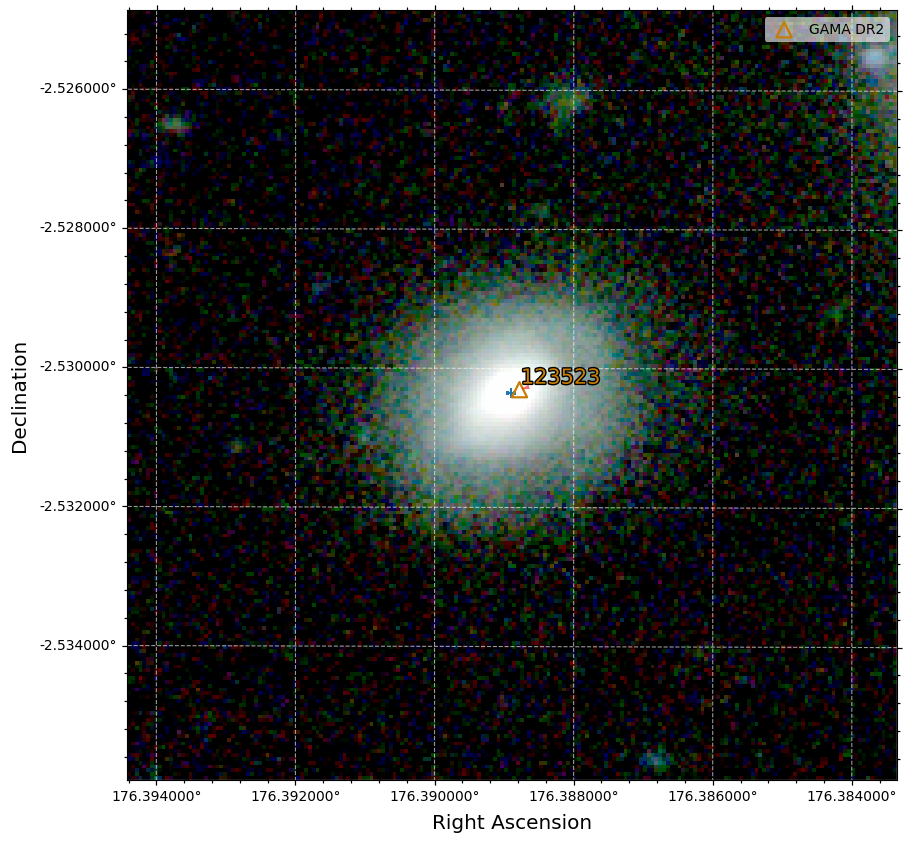}
    \includegraphics[width=0.4\columnwidth]{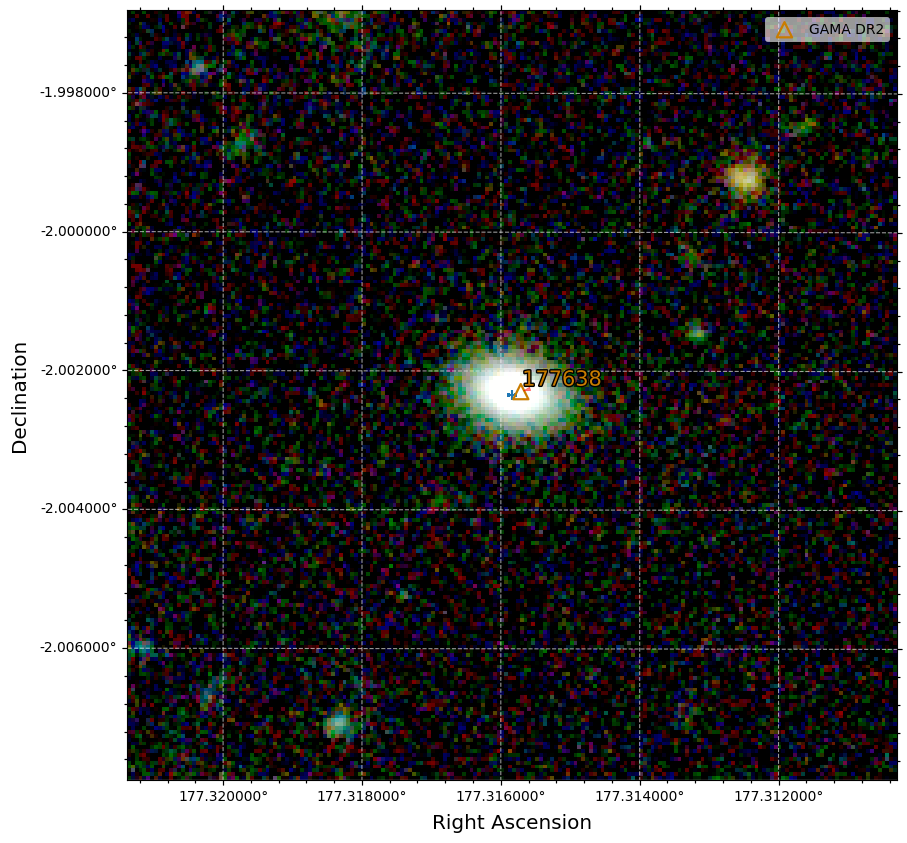}
    \includegraphics[width=0.4\columnwidth]{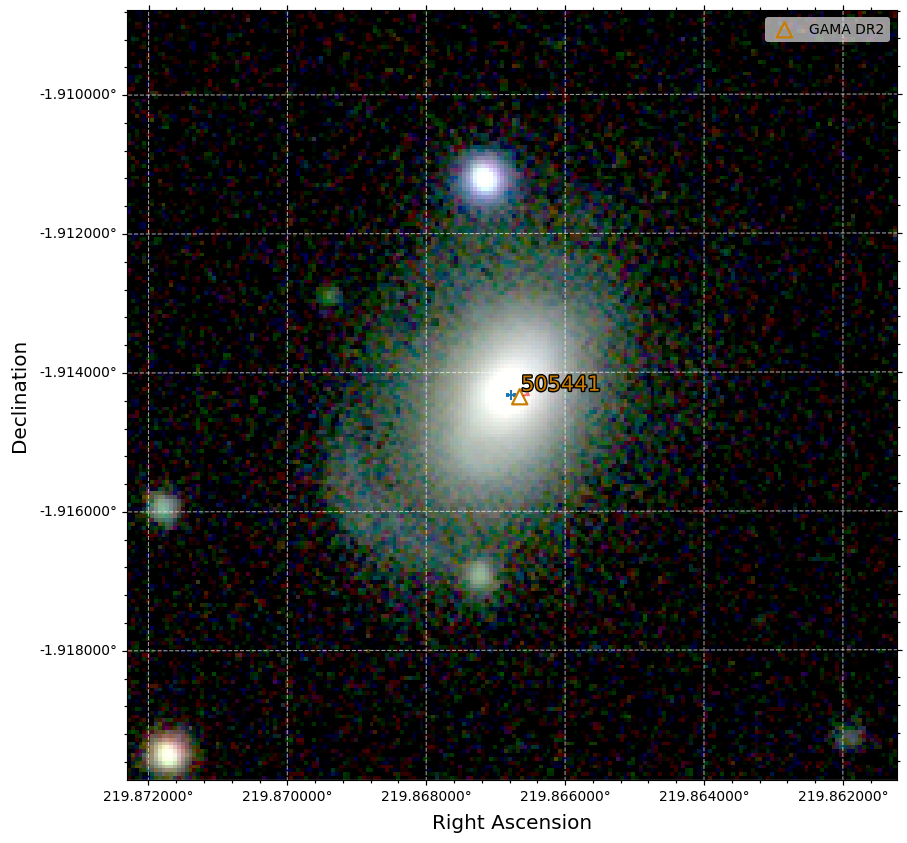}
    \includegraphics[width=0.4\columnwidth]{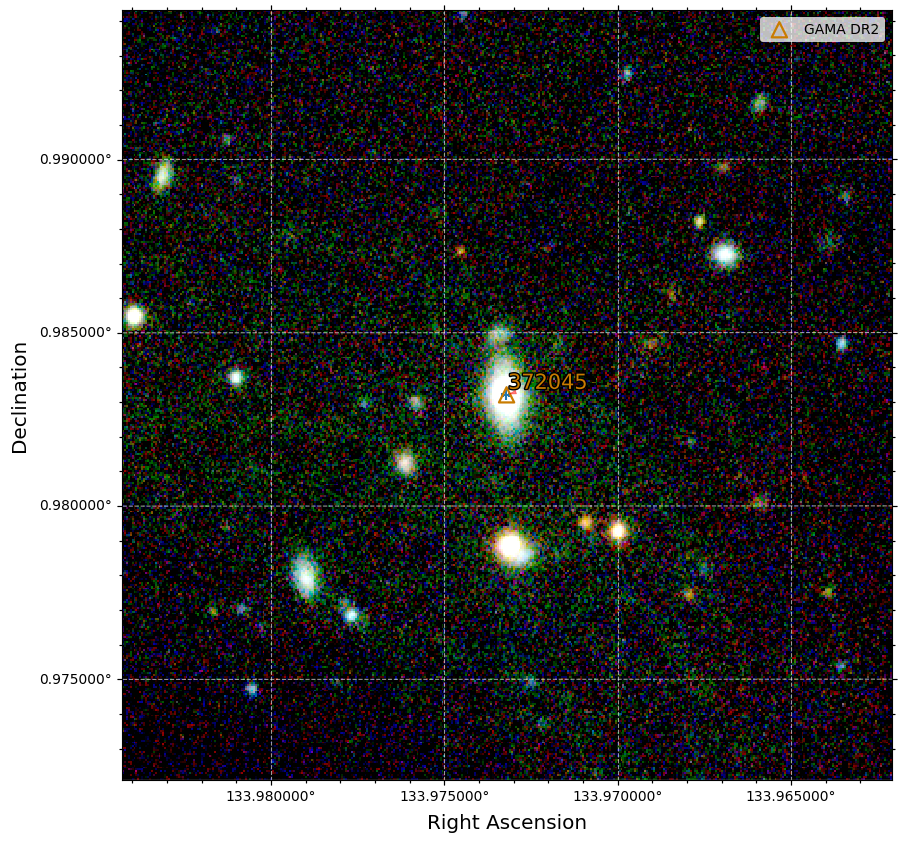}
    \includegraphics[width=0.4\columnwidth]{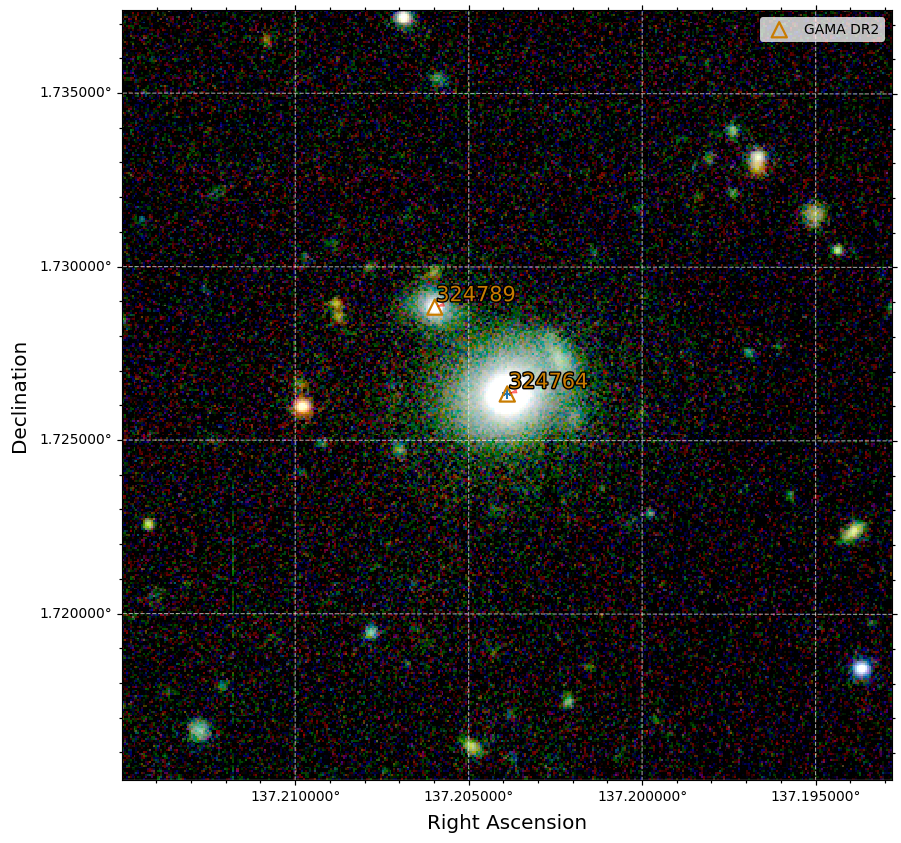}
    \caption{Examples of lens candidates identified by GalaxyZoo citizen science, taken from KiDS (\textit{g-, r-,} and \textit{i}-bands). G593809 (top left) represents a ``Lens or arc" score of $\sim10\%$ of votes. G123523 (top right) represents $\sim20\%$. G177638 (middle left) represents $\sim25\%$. G505441 (middle right) represents $\sim30\%$. G372045 (bottom left) represents $\sim35\%$. G324764 (bottom right) represents $\sim40\%$. These are representative of the images that would be viewed by GalaxyZoo participants, as well as informing our final cutoff for the GalaxyZoo selection considered here.}
    \label{fig:zoo_examples}
\end{figure*}

\begin{figure*}
    \centering
    \includegraphics[width=0.40\columnwidth]{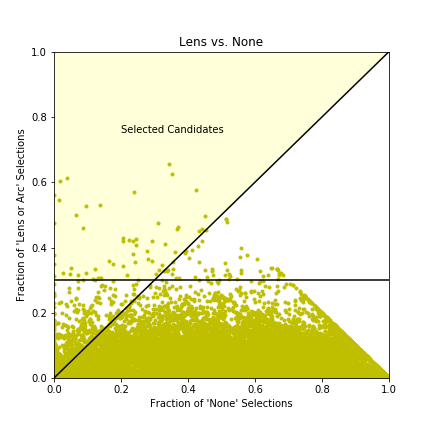}
    \includegraphics[width=0.40\columnwidth]{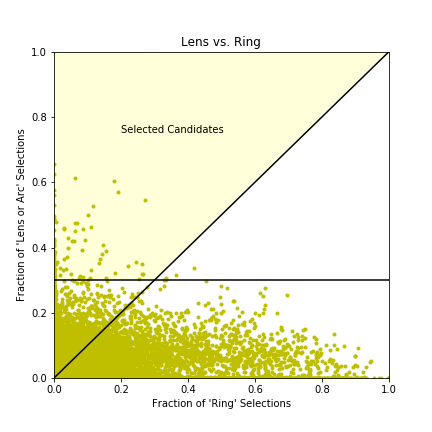}
    \includegraphics[width=0.40\columnwidth]{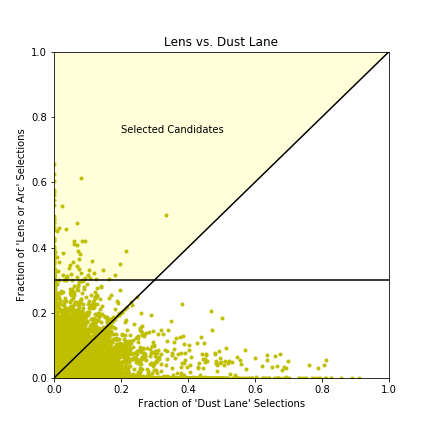}
    \includegraphics[width=0.40\columnwidth]{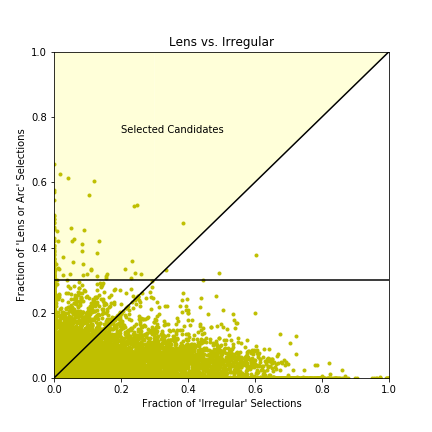}
    \includegraphics[width=0.40\columnwidth]{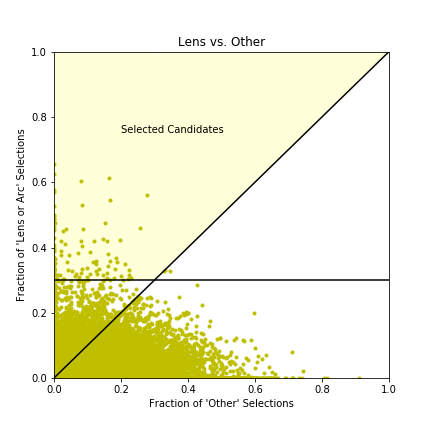}
    \includegraphics[width=0.40\columnwidth]{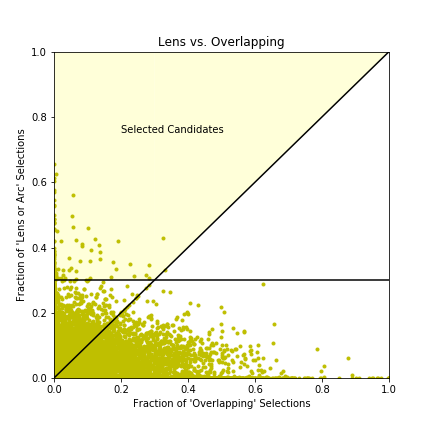}
    \caption{Ratios of the fractions of ``Lens or arc" selections on the vertical axis with respect to the fractions of selections of the other six ``odd features" options from GalaxyZoo Question Tree question T06, shown above in Figure \ref{fig:q_tree}, on the horizontal axis.  Solid lines indicate boundaries used in the final cut for the GalazyZoo selection. Only points that fall within the shaded region of all six plots are selected for our final GalaxyZoo sample.}
    \label{fig:zoo_features_vs_lens}
\end{figure*}

Selection votes could be distributed unpredictably across the seven options, so we took a multi-faceted selection informed by visual comparison of candidates of a range of ``Lens or arc" scores, examples of which are shown in Figure \ref{fig:zoo_examples}, as well as comparing them with LinKS machine learning candidates, in addition to analyzing ratios of ``Lens or arc" scores to the scores given to each of the other six ``odd feature" options, which we demonstrate in Figure \ref{fig:zoo_features_vs_lens}. Our final selection considered only those candidates with the following characteristics:
\begin{itemize}
    \item[1.] ``Lens or Arc" score was greater than all other ``odd feature" options' scores.
    \item[2.] ``Lens or Arc" score was greater than 30 percent.
    \item[3.] Thin-lens Einstein radius estimate (Section \ref{section_theta_e_estimates} and \ref{section_theta_e_results} for details) is larger than the point spread function (PSF) for KiDS imaging (0.65 arcsec).
\end{itemize}{}

This final cut reduced the GalaxyZoo citizen science catalog to 13 candidates (listed in Table \ref{GZ_table} in the Appendix) of comparable reliability to the other two techniques, according to our own visual inspection of various scoring lens candidates. These 13 candidates all have reliable stellar mass estimates from the GAMA {\sc LAMBDAR} mass catalog. All following mentions of the GalaxyZoo citizen science sample refer to this reduced selection of candidates unless otherwise specified.

 \subsection{Einstein Radius Estimation}\label{section_theta_e_estimates}

Limited by the exclusion of velocity dispersion measurements of the GAMA/KiDS fields, we estimate the Einstein radius for each candidate based on empirical fits of total stellar mass $M_*$ to lensing parameters $M_E$ (the total mass enclosed within the Einstein radius) and $\sigma_{SIS}$ (velocity dispersion).

\subsubsection{Thin-Lens Estimate}

We approximate the lens galaxy as a thin-lens system with the lens and source positioned along the same line of sight, where the Einstein radius ($\theta_E$) is given by 

\begin{equation} \label{theta_e_pm_eqn}
\theta_E = \left(\frac{M_E}{10^{8.09}M_\odot}\right)^{1/2}\left(\frac{D_{LS}}{D_{L}D_{S}} Mpc\right)^{1/2} \textnormal{arcsec}
\end{equation}
where $M_E$ is the total mass enclosed by the Einstein radius, $D_{LS}$ is the distance from lens to source, $D_{L}$ is the distance from observer to lens, and $D_{S}$ is the distance from observer to source. All distances are angular diameter distances calculated from redshift. We emphasize that $M_E$ does not denote the total mass of the galaxy but only the mass enclosed.
\cite{Auger10} modeled and analyzed 73 SLACS lenses and presented a linear relation between the log of the total (lensing) mass enclosed within half the effective radius, a close match to the typical Einstein radius, and the log of stellar mass. Using following relation
\begin{equation} \label{enclosed_mass_eqn}
M_E = 0.0011 M_{\odot} \left(\frac{M_*}{M_{\odot}}\right)^{1.25}
\end{equation}
and utilizing stellar mass measurements from GAMA {\sc LAMBDAR} stellar mass estimates \citep{Taylor11}, we estimate the total mass enclosed for use in Equation (\ref{theta_e_pm_eqn}).
Only the GAMA spectroscopy candidates include source redshift measurements; for these candidates, $D_{LS}$ and $D_S$ are calculated from this redshift. For LinKS machine learning and GalaxyZoo citizen science candidates, the lens galaxy is assumed to be positioned approximately halfway between the observer and source, which is the distance at which the most dramatic strong-lensing features should be observable. For comparison, we conduct a second Einstein radius estimate that accounts for mass distribution.

\subsubsection{Singular Isothermal Sphere (Velocity Dispersion) Estimate}

Approximated as a singular isothermal sphere (SIS), the Einstein radius takes a new form:
\begin{equation} \label{theta_e_sis_eqn}
\theta_E = \left(\frac{\sigma_{SIS}}{186 \ km/s}\right)^2 \frac{D_{LS}}{D_S} \textnormal{arcsec}
\end{equation}
According to \cite{Petrillo17} the $\sigma_{SIS}$ parameter can be substituted with stellar velocity dispersion $\sigma_*$ as a first approximation. In order to obtain fiducial velocity dispersions, we adopt a combination of empirical $M_*\sigma$ (total stellar mass to line of sight central stellar velocity dispersion) relations reported in \cite{Zahid16}. For low-redshift galaxies, these relations are fits to measured data from $\sim370,000$ objects with $\log(M_*/M_{\odot})>9$ and $0<z<0.2$ from SDSS DR-12 \citep{Alam15}. The intermediate redshift relation is a fit to a sample of 4585 galaxies from the Smithsonian Hectoscopic Lens Survey (SHELS, \cite{Geller06, Geller14} and \cite{Geller16}). These relations take the form
\begin{align} \label{sigma_lowz_eqn}
    \sigma_{z_{low}}&=10^{2.073}  \left(\frac{M*}{M_b}\right)^{\alpha} \textnormal{km/s} & (z &< 0.2 \textnormal{, SDSS}) \\ 
    \sigma_{z_{int}}&=10^{2.071} \left(\frac{M*}{M_b}\right)^{0.281}\textnormal{km/s} &   (0.2 &\leq z < 0.65 \textnormal{, SHELS}) \label{sigma_intz_eqn}
\end{align}
\begin{align*}
    \textnormal{where} \qquad \alpha&=0.403 \qquad \textnormal{for}\qquad M_*\leq M_b \\
    \alpha&=0.293 \qquad \textnormal{for} \qquad M_*> M_b \\
    \textnormal{and} \qquad M_b &= 10^{10.26} M_{\odot} \textnormal{.}
\end{align*}

These equations are applied to each candidate with reliable GAMA {\sc LAMBDAR} mass estimates, and the resulting central stellar velocity dispersion estimate is taken to be $\sigma_*$ as an approximation to the parameter $\sigma_{SIS}$ for use in Equation (\ref{theta_e_sis_eqn}).
As with the thin-lens estimate, $D_{LS}$ and $D_S$ are calculated from measured spectroscopic redshifts for GAMA spectroscopy candidates. For LinKS machine learning and GalaxyZoo citizen science candidates, the ratio of $\frac{D_S}{D_L}$ is assumed to be equal to two.

\section{Results}

\begin{figure*}
    \centering
    \includegraphics[width=\textwidth]{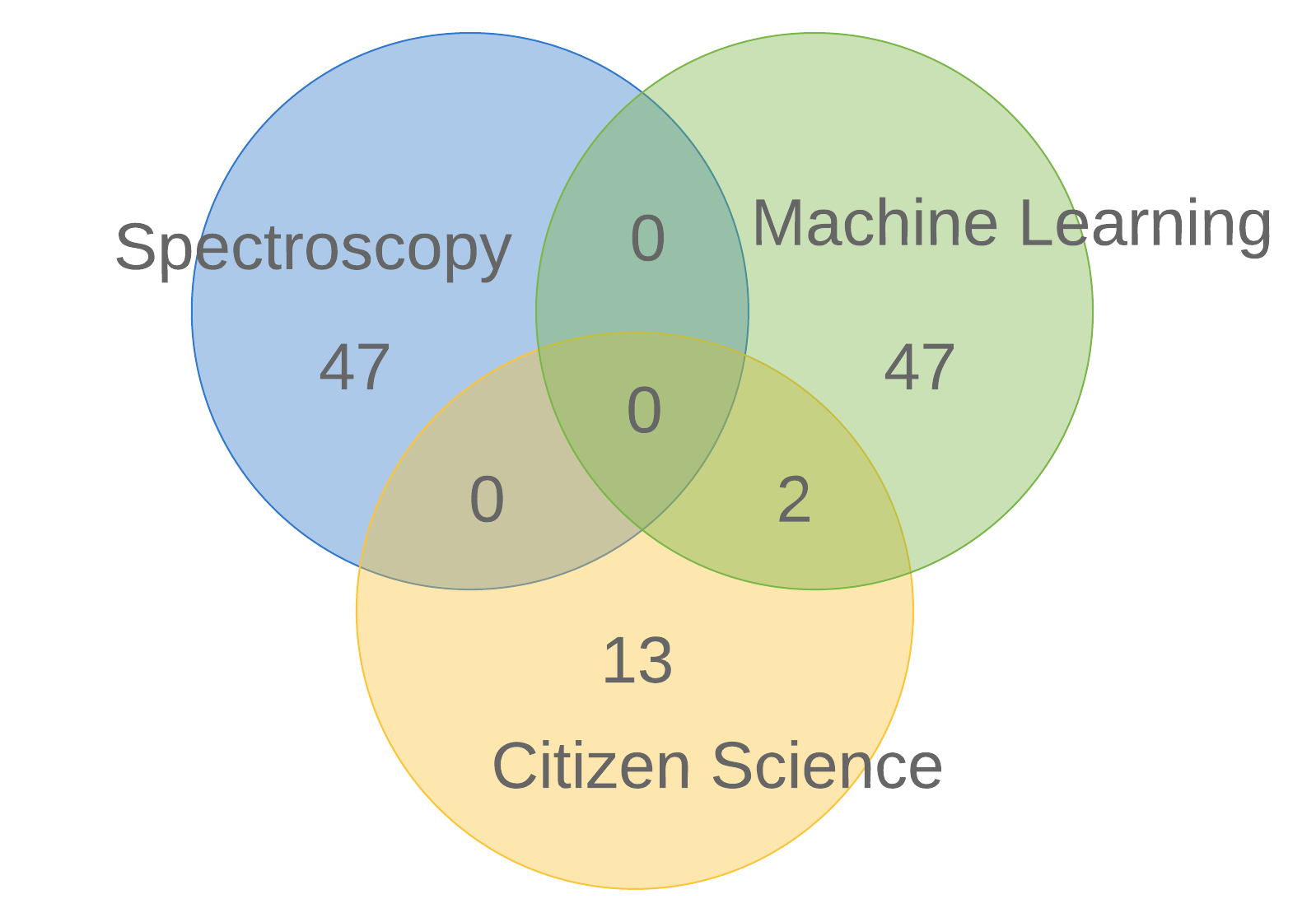}
    \caption{Venn diagram showing the number of lens candidates identified by each of the three methods following our final sample selections. Overlapping regions indicate the number of lens candidates identified by both (or all three) candidates, while numbers in each methods' region indicate the total number of candidates identified by that method. The two overlaps occurred between 47 LinKS machine learning candidates and the 13 GalaxyZoo citizen science candidates.}
    \label{fig:venn_hardcut}
\end{figure*}

We found remarkably little overlap between catalogs of candidates obtained by the three methods. Our final cuts of the three catalogs included no candidate common to all three methods and only two candidates common to two methods, both of which were between LinKS machine learning and GalaxyZoo citizen science, as shown in Figure \ref{fig:venn_hardcut}. 

\subsection{Results of Einstein Radius Estimates} \label{section_theta_e_results}

\begin{figure}
    \centering
     \includegraphics[width=0.45\textwidth]{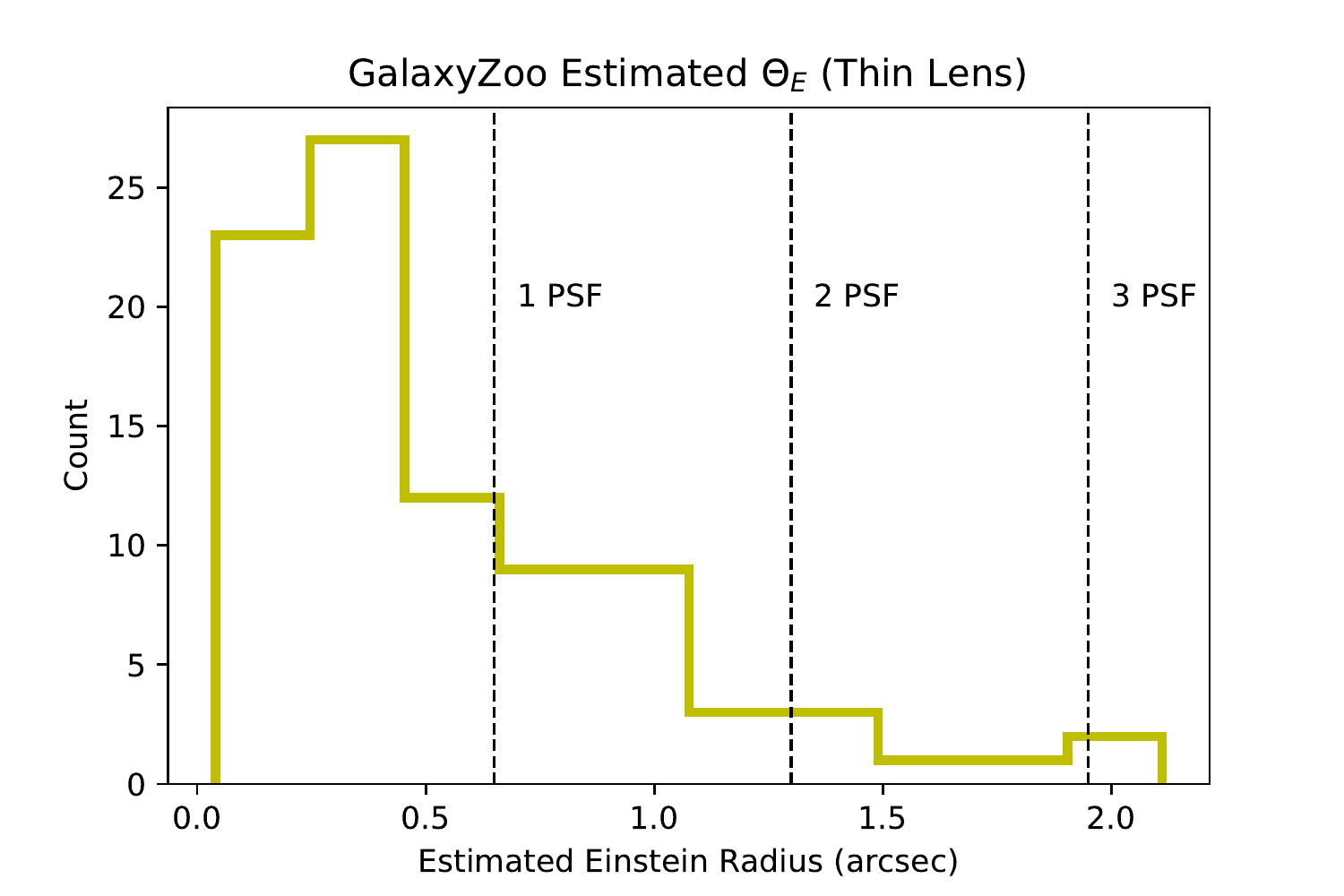}
     \includegraphics[width=0.45\textwidth]{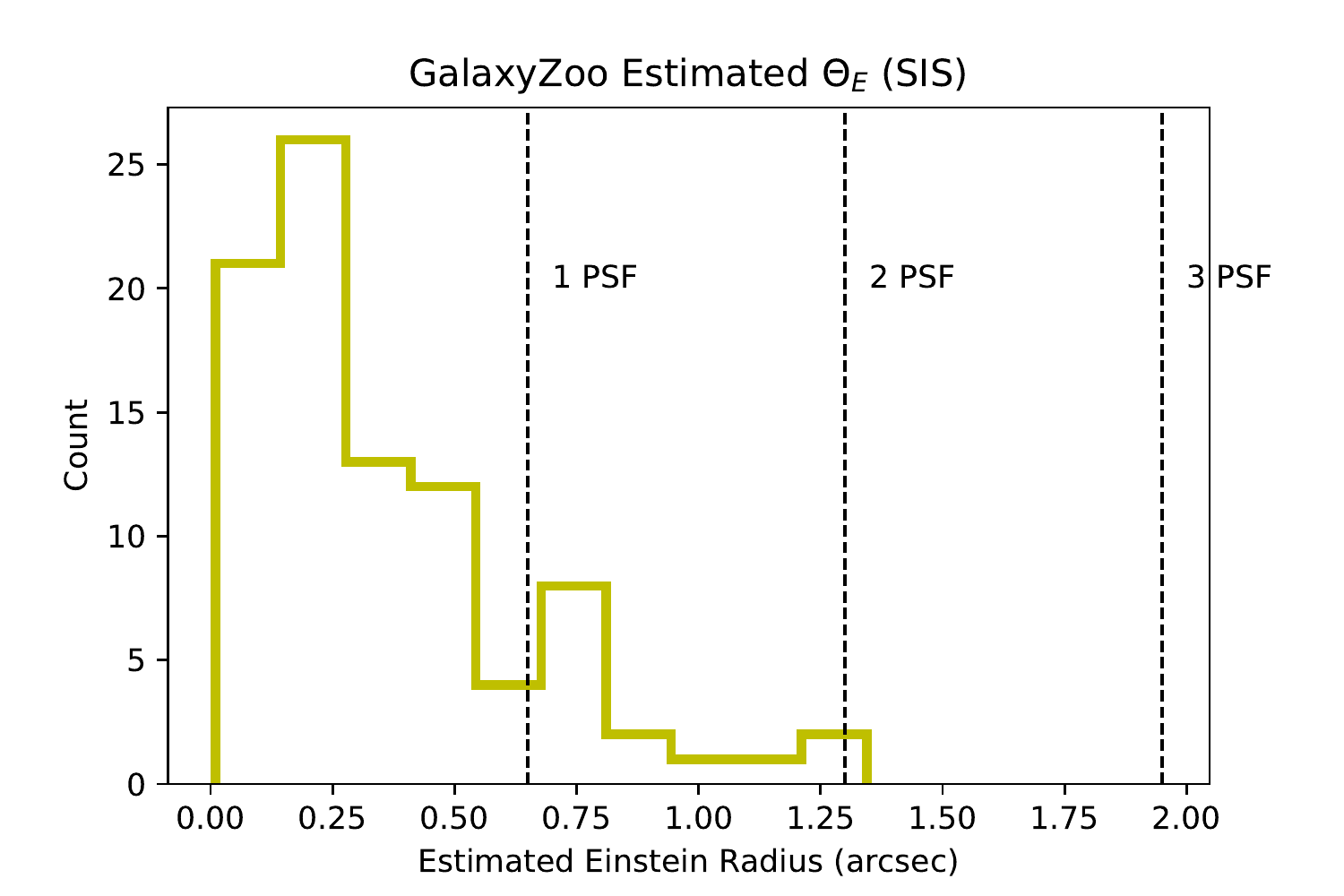}
     \includegraphics[width=0.45\textwidth]{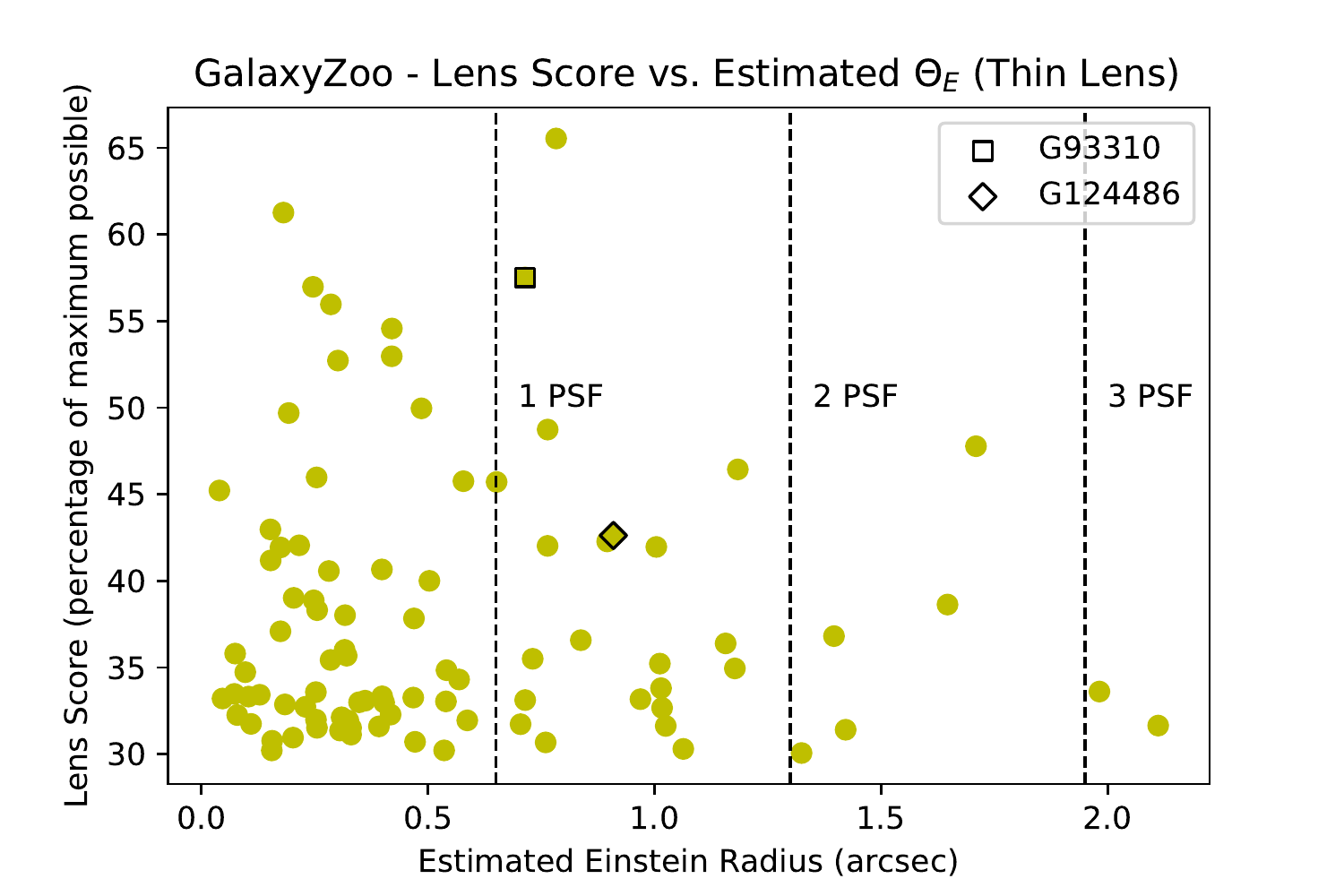}
     \includegraphics[width=0.45\textwidth]{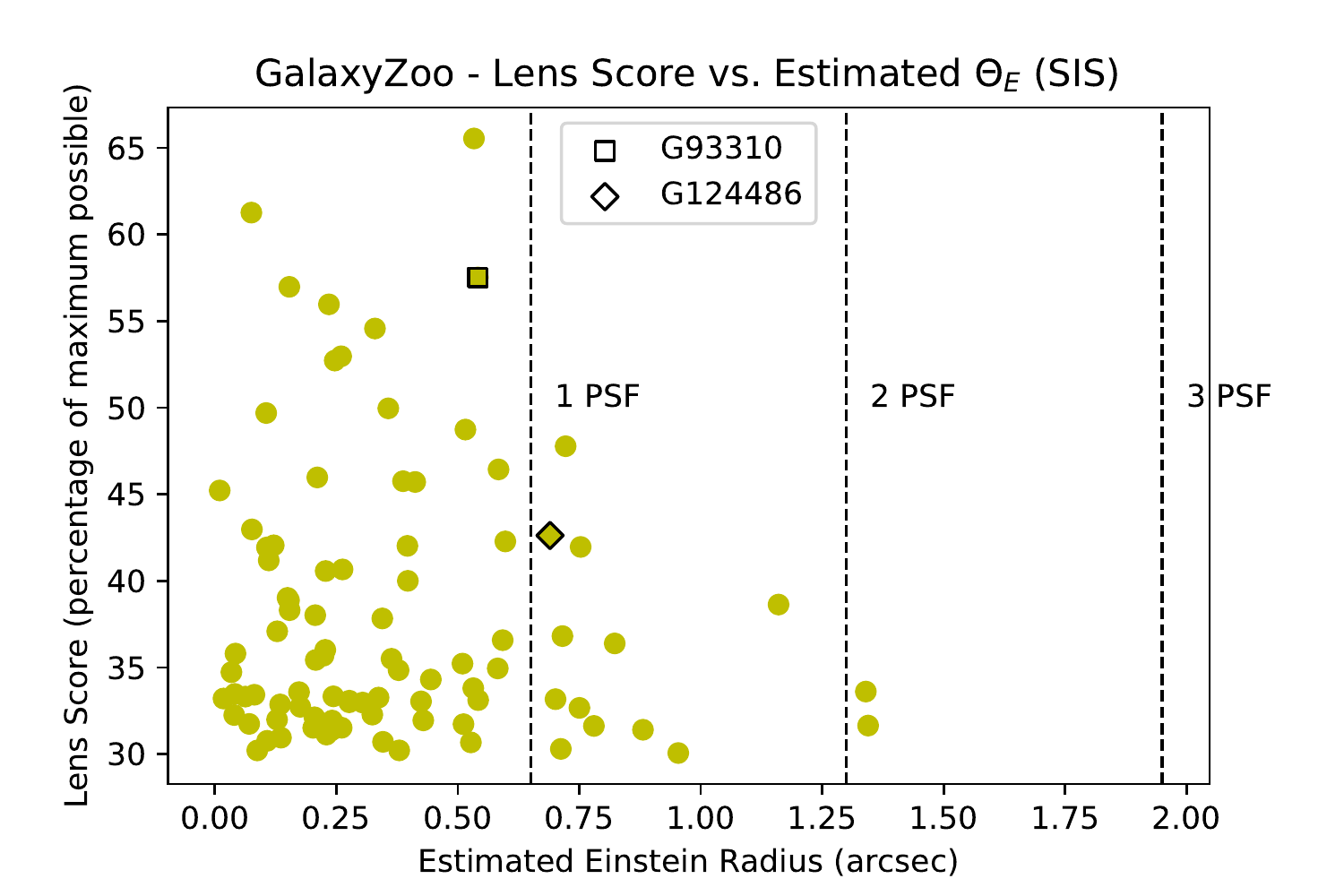}
    \caption{These figures show all GalaxyZoo citizen science candidates with ``Lens or arc" scores higher than 30\%. (Top) Histogram showing the distribution of Einstein radius estimates. (Bottom) Vertical axis shows ``Lens or arc" score as percentage of votes. Candidates with estimated Einstein radii lower than the KiDS PSF are considered poor candidates and are removed from the final sample. The high scoring GalaxyZoo objects in this removed subsample are good candidates for tidal features. Note overlapping candidates (LinKS - GalaxyZoo) G93310 and G124486 emphasized by the black square and diamond.}
    \label{fig:gz_theta_e}
\end{figure}

\begin{figure}
    \centering
     \includegraphics[width=0.45\textwidth]{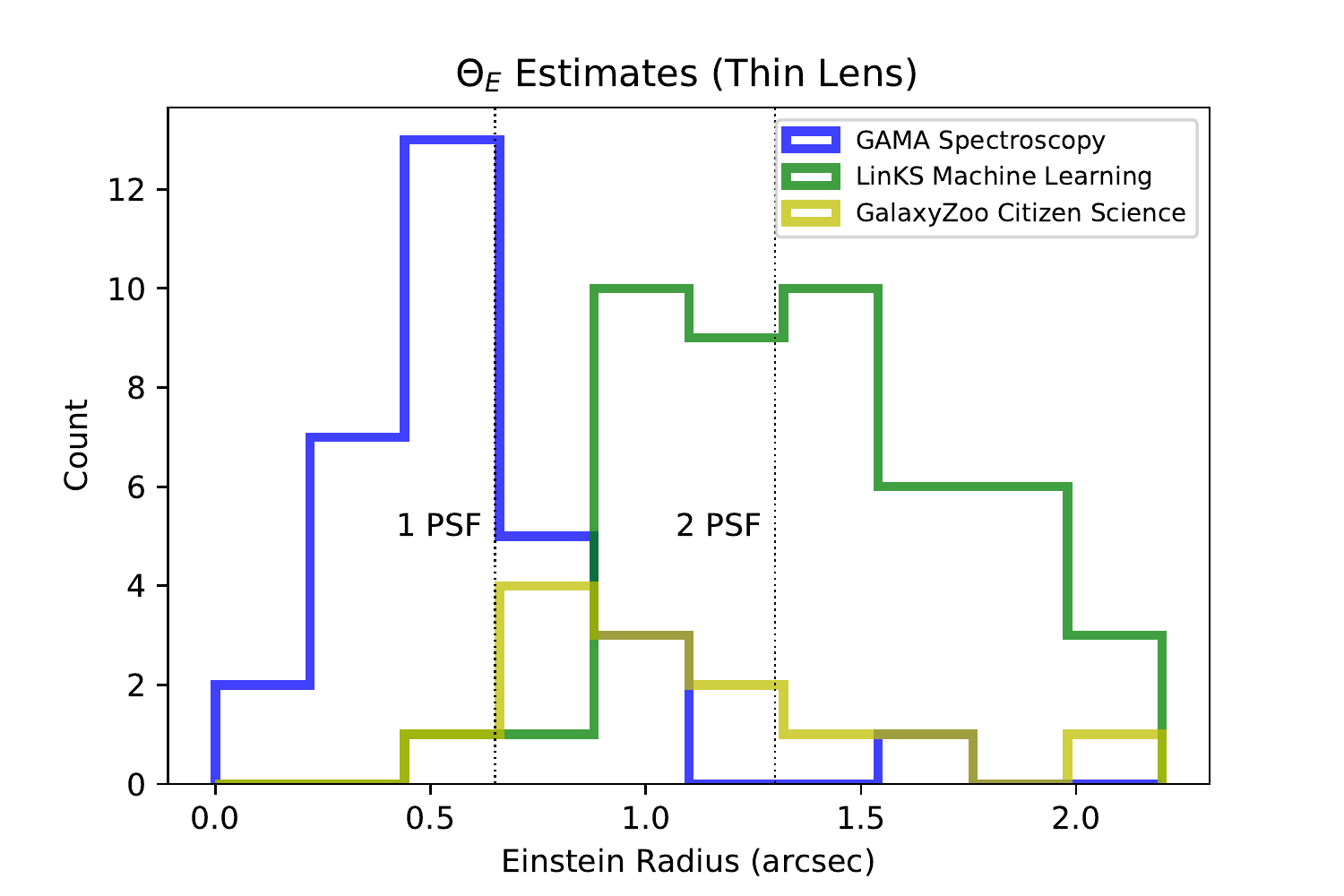}
     \includegraphics[width=0.45\textwidth]{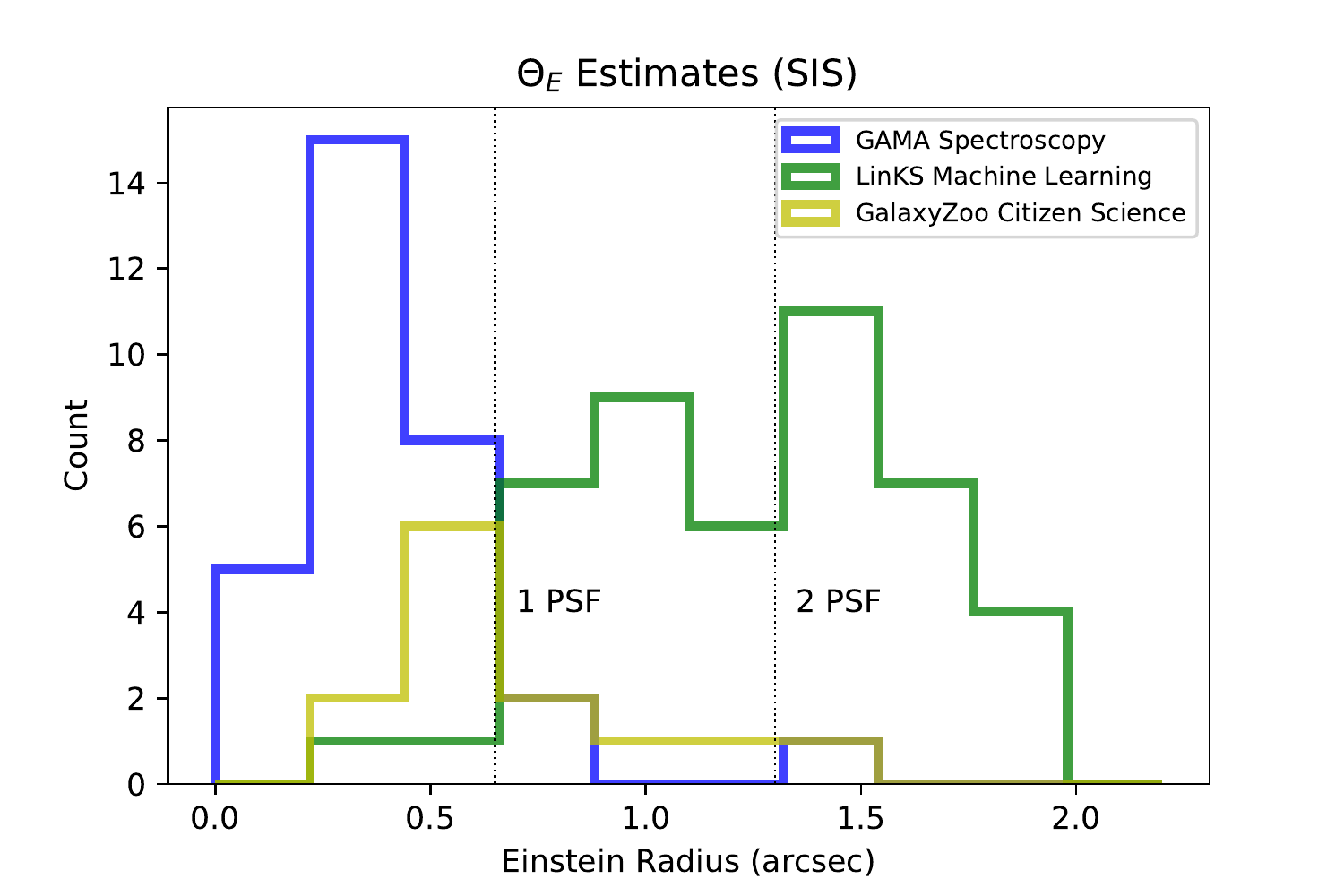}
     \includegraphics[width=0.45\textwidth]{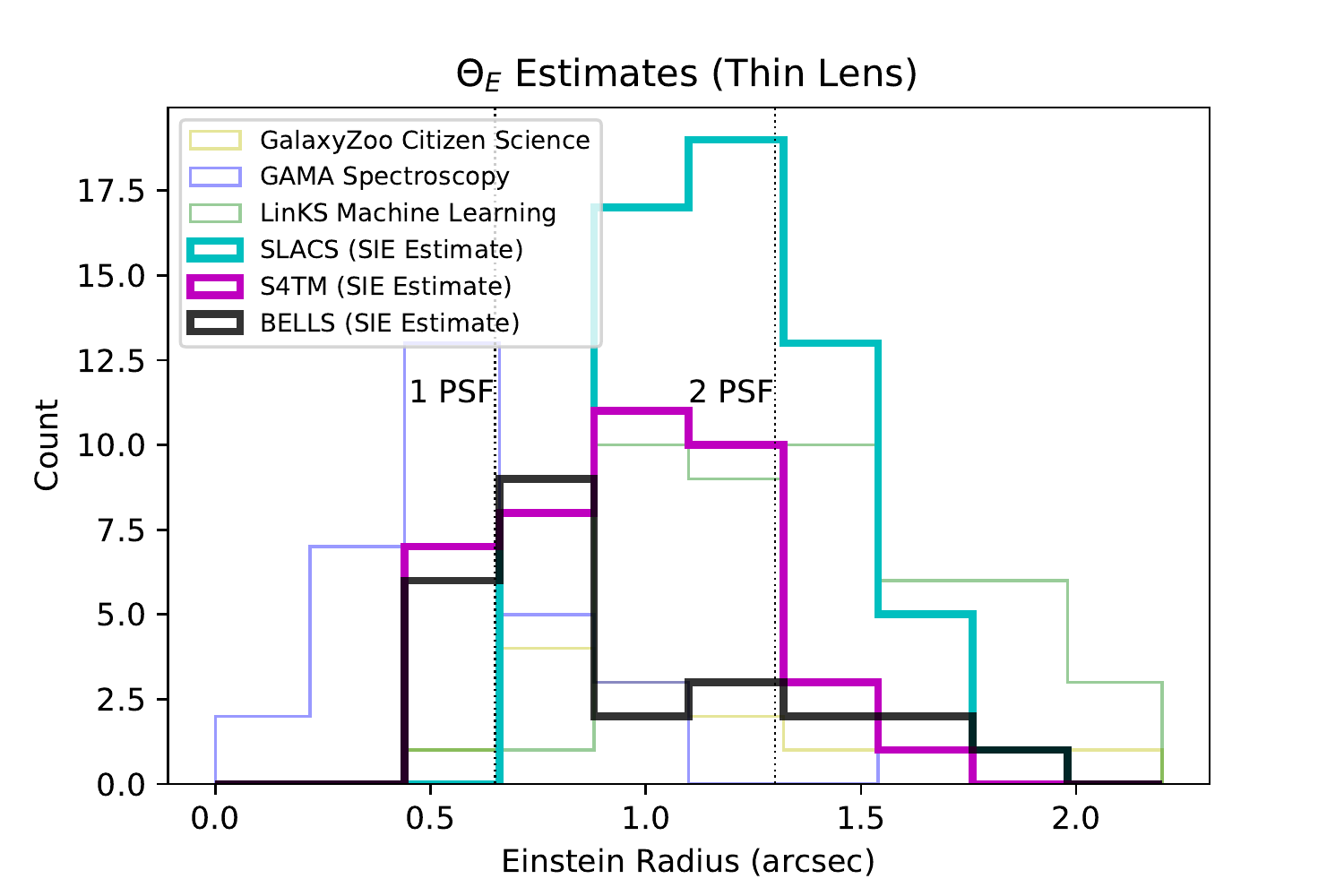}
     \includegraphics[width=0.45\textwidth]{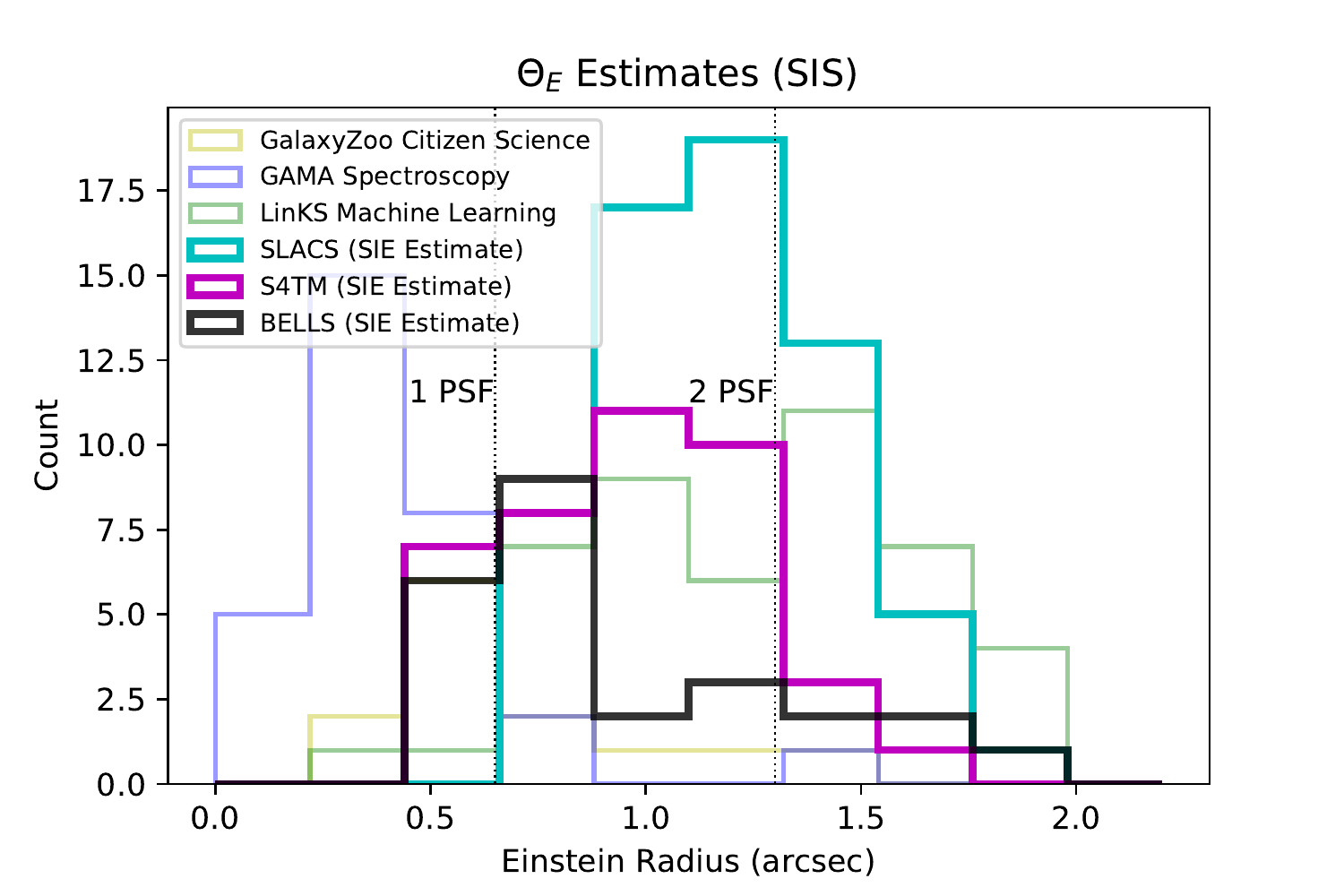}
    \caption{(Top) Thin-lens and SIS Einstein radius estimates for candidates of the three methods examined in this study, and (Bottom) Einstein radius estimates of candidates identified by previous spectroscopic lens surveys for comparison. Section \ref{section_theta_e_estimates} details calculations for the three GAMA/KiDS methods, while SIE estimates for SLACS, S4TM, and BELLS candidates are taken from \cite{Bolton08}, \cite{Shu17}, and  \cite{Brownstein12}. For all GAMA/KiDS candidates, thin-lens Einstein radius estimates are slightly larger than SIS estimates.}
    \label{fig:theta_e}
\end{figure}

\begin{figure}[h!]
    \centering
    \includegraphics[width=0.45\textwidth]{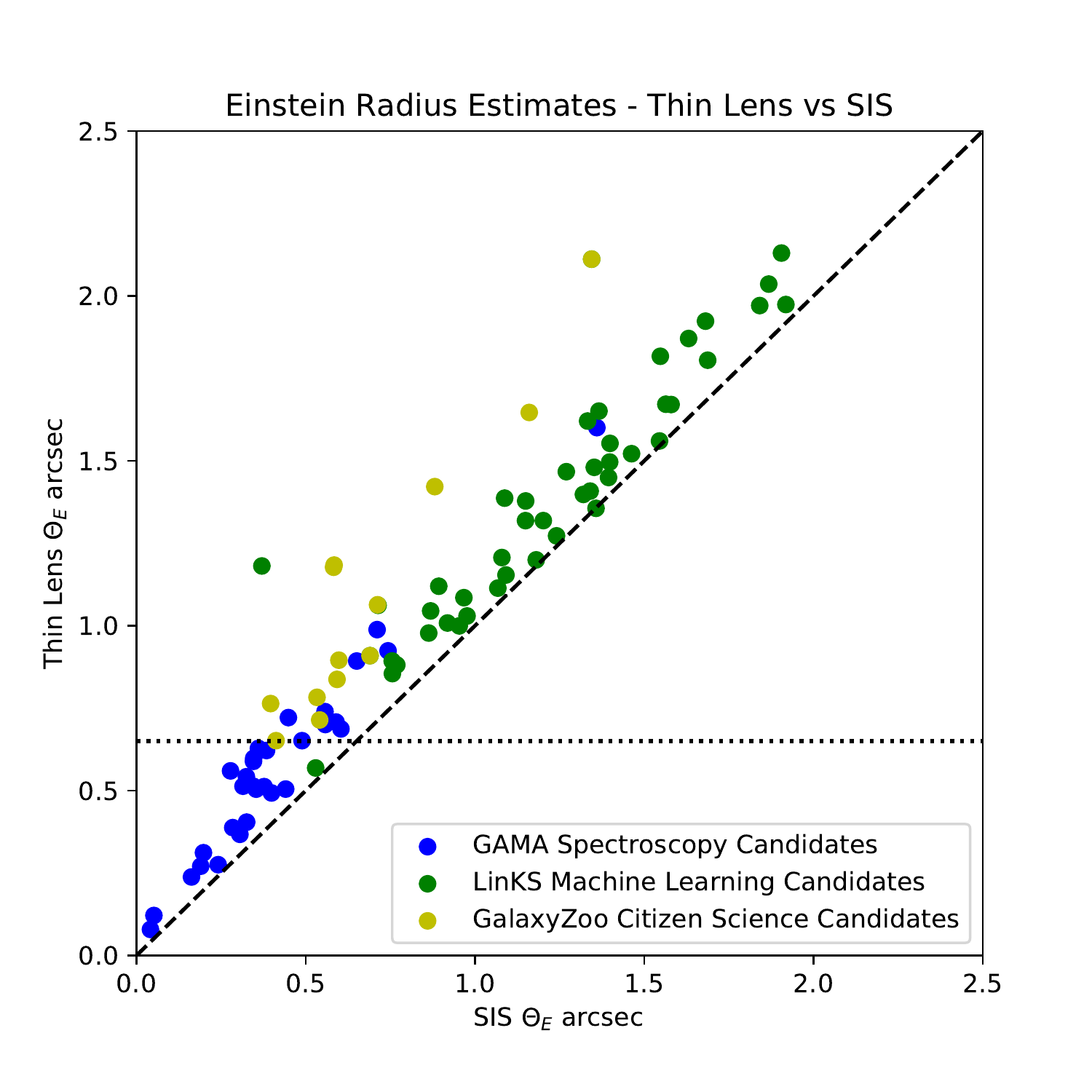}
    \includegraphics[width=0.45\textwidth]{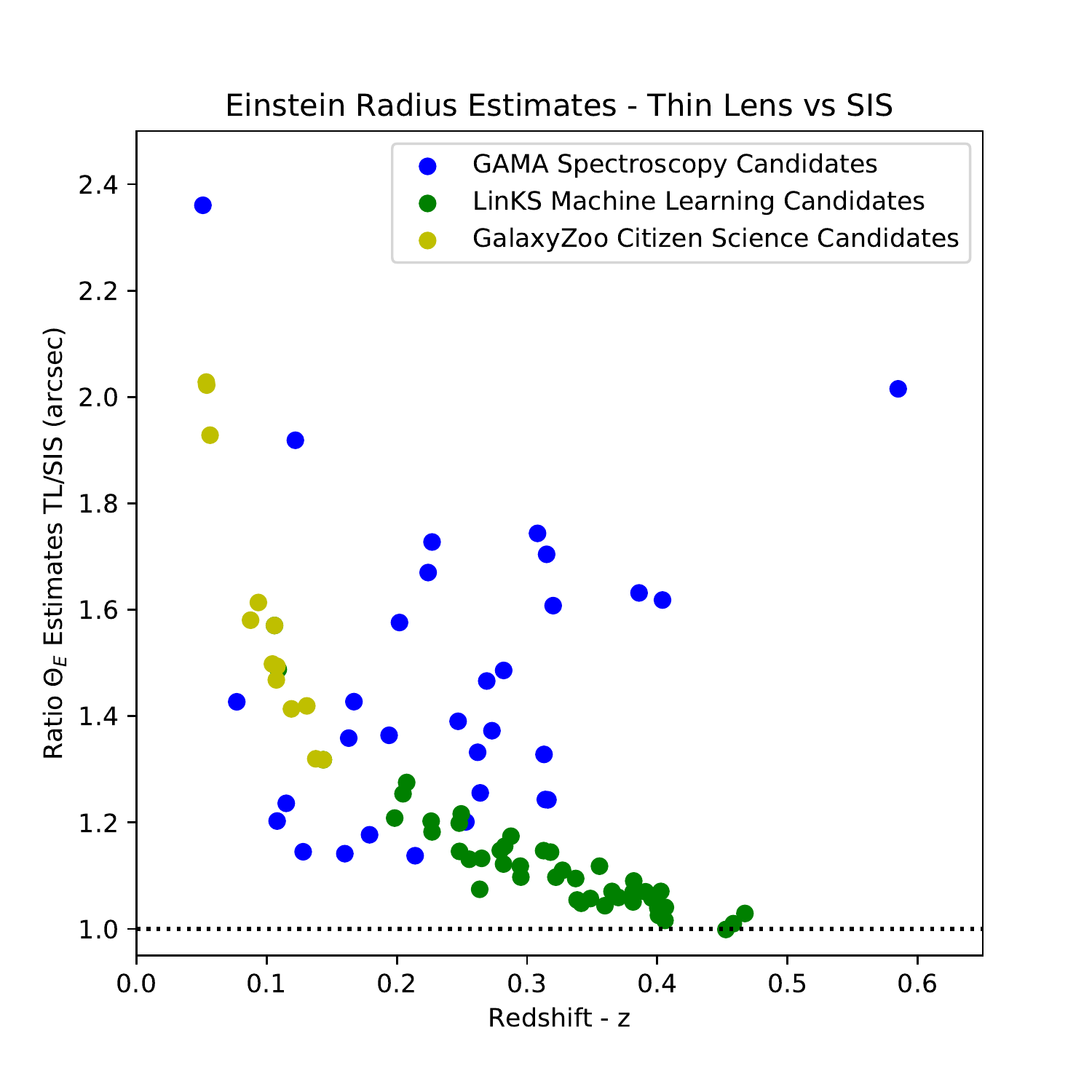}
    \caption{(Left) Thin-lens estimates of the Einstein radius are larger than SIS velocity dispersion estimates for the same candidates. We take each of these to represent a range of probable values. (Right) Because thin-lens estimates are proportional to the inverse square of the lens distance, given a fixed ratio of source to lens distance (as we assume 2:1 for LinKS machine learning and GalaxyZoo citizen science), the thin-lens estimate decreases with redshift. As a result, the differences between estimates of the same candidates are pronounced at low redshift. This trend is further discussed in Section \ref{section_theta_e_differences.}}
    \label{fig:theta_e_difference}
\end{figure}

The Einstein radius is a fundamental feature of a lensing system that determines to a large degree the probability of detection by any lens finding method \citep[][and reference therein]{Sonnenfeld15}. Note that fiducial source redshifts are utilized in the estimation of Einstein radii for LinKS machine learning and GalaxyZoo citizen science candidates, so comparisons to GAMA spectroscopy candidates and previously studied systems whose source redshifts are known should be considered with discretion. Shown in Figure \ref{fig:gz_theta_e}, the third selection criterion for GalaxyZoo candidates, based on the capabilities of the instrument, removes several high-scoring candidates from consideration. These removed objects are worth follow-up as candidates for galaxies showing tidal features. The results of the estimates applied to all three catalogs, displayed in the upper two plots of Figure \ref{fig:theta_e}, show that in general the thin-lens estimate is larger than the SIS estimate for the same candidates. This difference is further demonstrated in Figure \ref{fig:theta_e_difference} and discussed in greater detail in Section \ref{section_fits_estimates}. Mean and median values for each sample for both models are given in Table \ref{theta_e_table}.

\begin{table}[h!]
\caption{Mean and median Einstein radius estimates for each candidate sample.}
\label{theta_e_table}
\begin{center}
\begin{tabular}{l l l l l}
\hline
& Thin-Lens Model & & SIS Model & \\
 & Mean $\theta_E$ ($\prime\prime$) & Median $\theta_E$ ($\prime\prime$) & Mean $\theta_E$ ($\prime\prime$) & Median $\theta_E$ ($\prime\prime$)\\
\hline
GAMA Spectroscopy &  0.57 & 0.54 & 0.41 & 0.35 \\
LinKS Machine Learning &  1.39 & 1.38 & 1.21 & 1.25 \\
GalaxyZoo Citizen Science &  1.09 & 0.91 & 0.59 & 0.69 \\

\hline
\end{tabular}
\end{center}
\label{t:hla}
\end{table}%

For both models, the majority of GAMA spectroscopy candidates have estimated $\theta_E \lesssim 1^{\prime\prime}$, the GAMA spectroscopy aperture radius. The majority of candidates from the image-based machine learning and citizen science methods have characteristic estimated $\theta_E \gtrsim 1^{\prime\prime}$, mostly above $0.65^{\prime\prime}$, the PSF where images of the lens galaxy and arc features can be distinctly separable. We note that some candidates from both GalaxyZoo citizen science and LinKS machine learning have estimated Einstein radii close to or below the PSF, particularly for SIS estimates at low redshift. We have chosen to retain all LinKS candidates as well as those GalaxyZoo candidates with $\theta_E > 0.65^{\prime\prime}$ in the thin-lens estimate. Because our basic models estimate the Einstein radius with the lens and source positioned along the line of sight, lensing features for a given Einstein radius may in reality extend beyond the estimated Einstein radius due to asymmetry across the line of sight axis, allowing for the possibility of detection of lens candidates with Einstein radii smaller than the PSF by image-based techniques.

Reference $\theta_E$ estimates based on SIE models of SLACS \citep{Bolton08}, S4TM \citep{Shu17}, and BELLS \citep{Brownstein12, Shu16} grade-A lens candidates are shown in the lower plot of Figure \ref{fig:theta_e}. The LinKS machine learning sample identified candidates with comparable estimated $\theta_E$ to the SLACS sample (which was the guide for its training sample), and S4TM and BELLS can be seen to approach the smaller estimated Einstein radii of candidates identified in GAMA spectroscopy. GAMA spectroscopy candidates extend to lower estimated Einstein radii than other spectroscopic selections because GAMA's high completeness reduces mass bias.

\begin{figure*}
    \centering
     \includegraphics[width=0.45\textwidth]{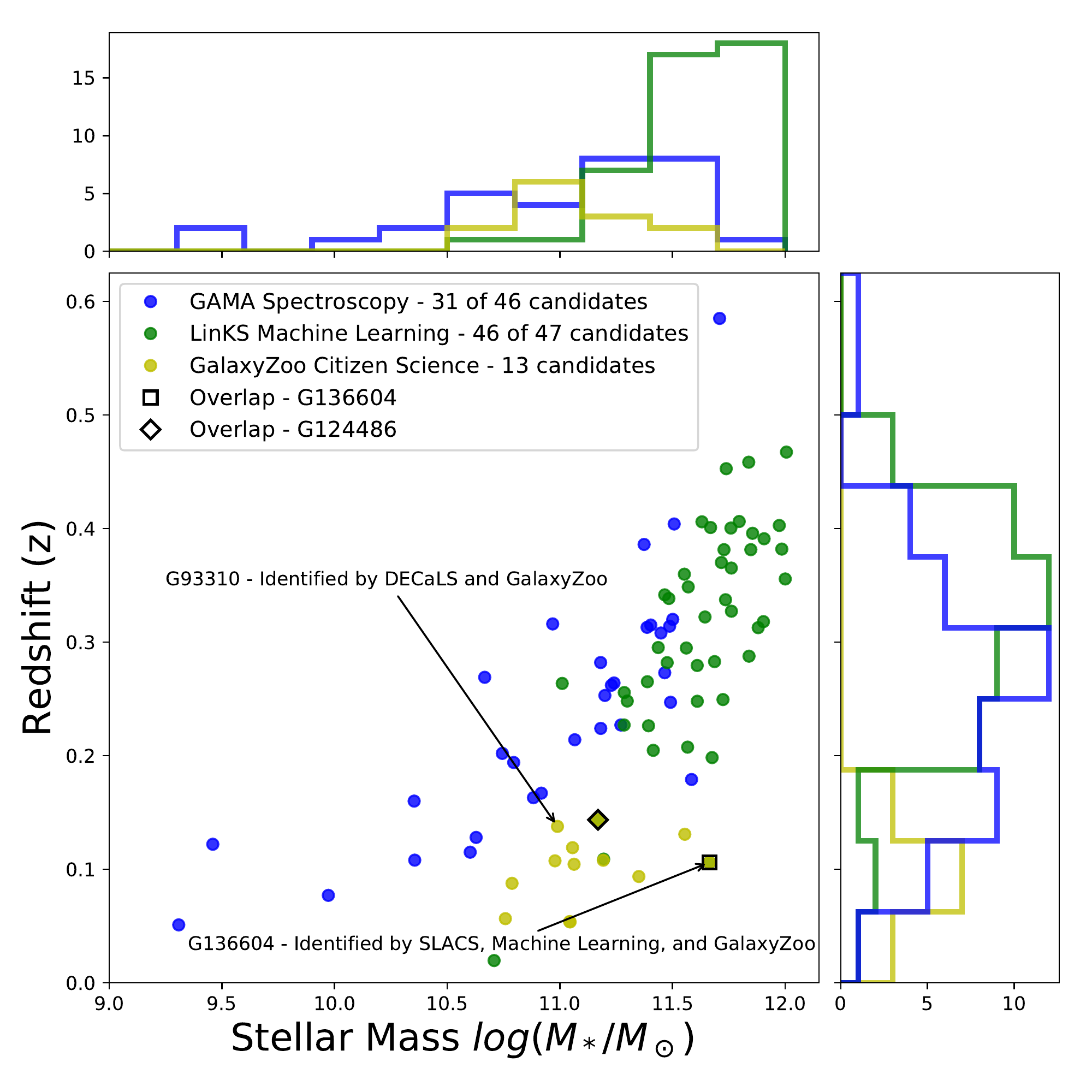}
     \includegraphics[width=0.45\textwidth]{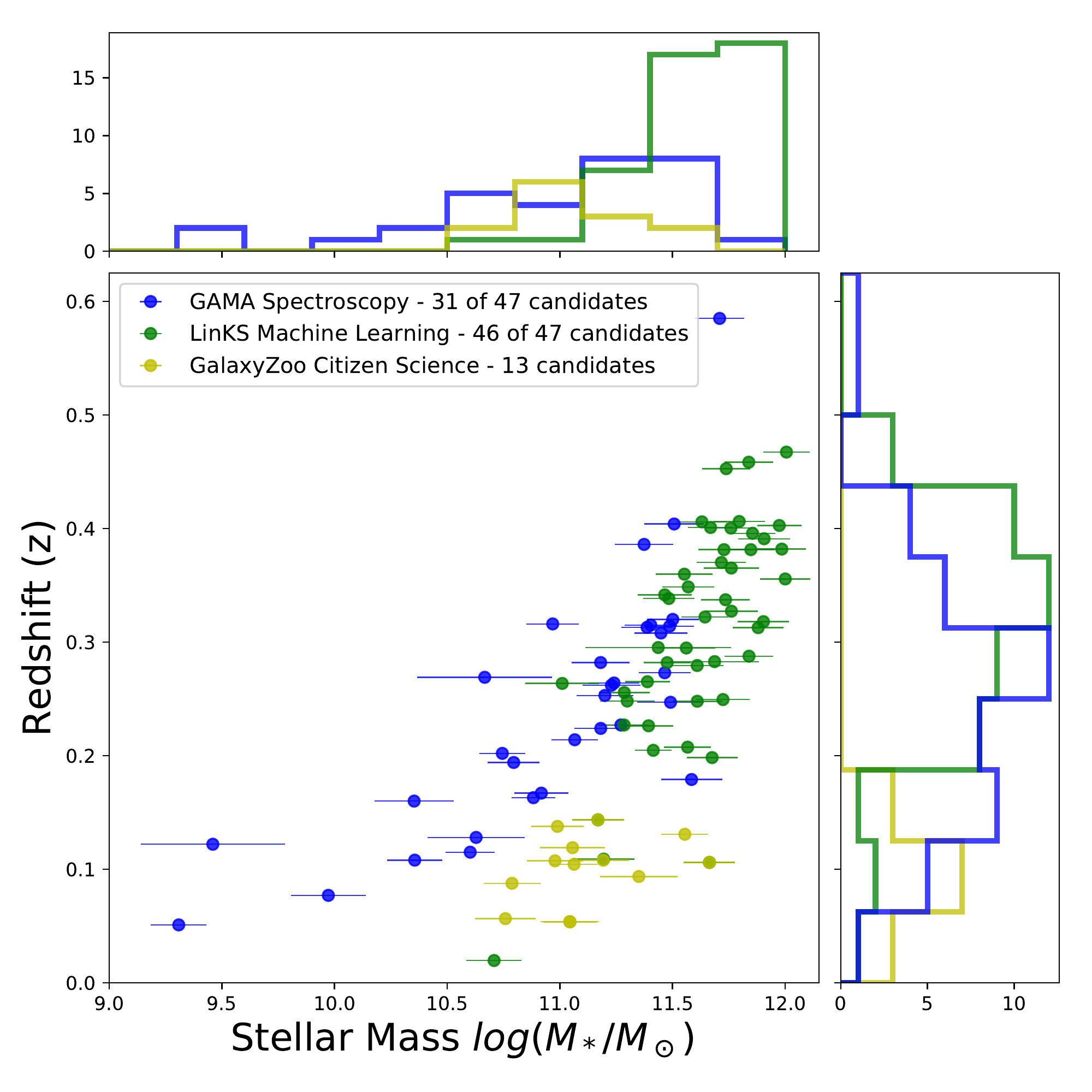}     
    \caption{Distribution of lens candidates identified by each of the three methods in terms of each candidate's stellar mass and redshift. Histograms above and to the right of the scatter plot indicate the number of candidates identified by each method within specific ranges of stellar mass and redshift respectively. Each method's catalog occupies a unique range of both characteristics. (Left) The two candidates that overlap between GalaxyZoo citizen science and LinKS machine learning samples are marked with square and diamond outlines, and we point out candidates G93310 and G136604 that were previously identified in DECaLS \citep{Huang20b} and SLACS \citep{Bolton08} respectively. (Right) Error bars represent uncertainty in stellar mass estimates. Note that 16 GAMA spectroscopy candidates and one LinKS candidate lack reliable stellar mass estimates from the GAMA {\sc LAMBDAR} catalog and are therefore not shown in the scatter plots.}
    \label{fig:big_plot}
\end{figure*}

\begin{figure}[h!]
    \centering
     \includegraphics[width=0.5\textwidth]{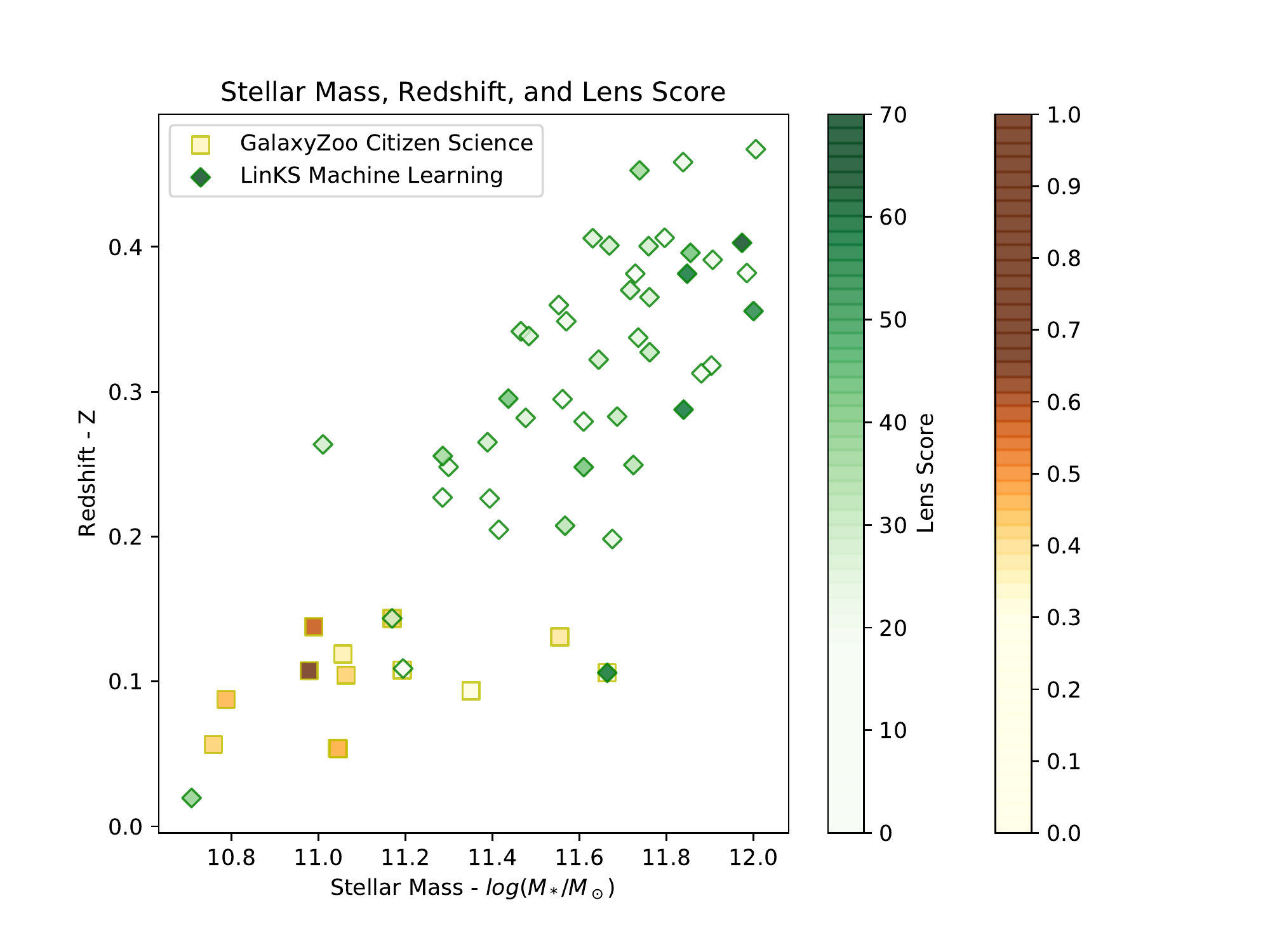}
     \includegraphics[width=0.45\textwidth]{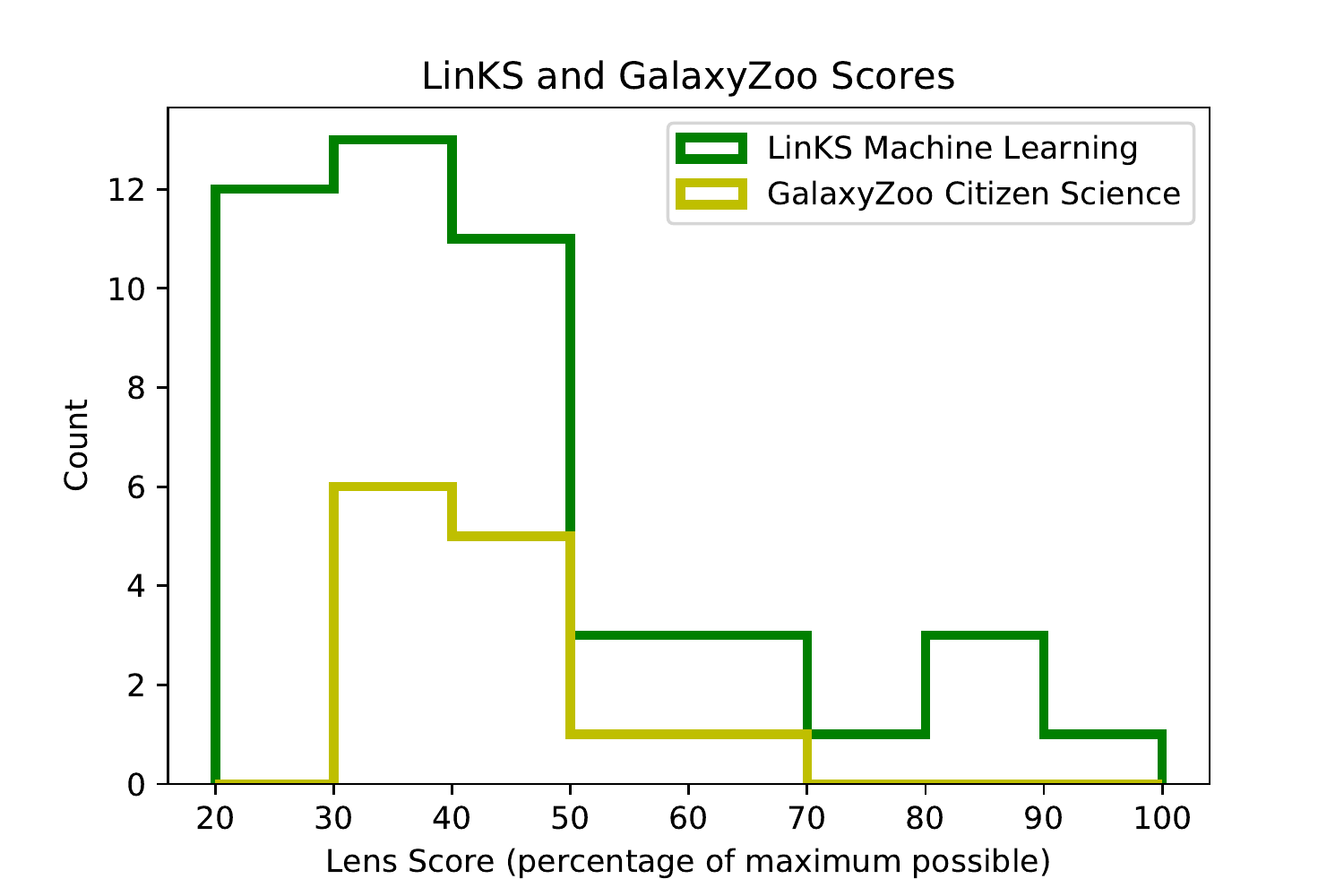}
    \caption{(Left) Green diamond markers show the stellar mass and redshift of strong lens candidates identified by LinKS machine learning from \protect\cite{Petrillo18}. LinKS lens score reflected by the green color bar refers to the authors' visual inspection and scoring of candidates on a scale of 0-70, with higher numbers reflecting higher confidence. Mid-scoring objects occur throughout the mass and redshift range, with the highest scoring above ${\rm log}(M_*/M_\odot) \sim 11.5$.
    Yellow-to-brown square markers indicate the stellar mass and redshift of our final selection of GalaxyZoo candidates (Kelvin et al {\em in prep}). GalaxyZoo lens score shown by the yellow-to-brown color bar refers to the fractional score out of 1 indicating the fraction of votes for ``Lens or arc" for each candidate, with higher numbers reflecting higher confidence. Highest-scoring objects occur middle of the stellar mass range of the GalaxyZoo candidates (${\rm log}(M_*/M_\odot) \sim 11$). (Right) Histogram shows the distribution of lens scores for LinKS machine learning and GalaxyZoo citizen science candidates as a percentage of the maximum possible score (70 and 1.0 respectively).}
    \label{fig:mac_massvsredshift}
\end{figure}

\subsection{Stellar Mass and Redshift Space}
\label{section_mass_redshift}

Analyzing the properties of each catalog in terms of the candidates' stellar mass and redshift, which we illustrate in Figure \ref{fig:big_plot}, we discovered that each method identified galaxies within a distinct region of the parameter space.

The candidates identified through GAMA spectroscopy tended to be found between $z\sim0.1 - 0.4$ (mean 0.250, median 0.253) and with stellar masses between ${\rm log}(M_*/M_\odot) \sim 10.5 - 11.5$ (mean 10.98, median 11.18), with a few between $\rm log(M_*/M_\odot) \sim 9.5-10.5$. For each redshift in the range, there is a characteristic mass we term the \textit{fiber radius mass} for which the Einstein radius fits in the spectroscopic aperture of the GAMA survey (angular radius = $1^{\prime\prime}$). This is lower than the SDSS fiber radius mass for the same candidate due to narrower fiber aperture (SDSS fiber aperture is $1.5^{\prime\prime}$). See Section \ref{section_maxmass}.

Those candidates identified through LinKS machine learning spanned a slightly higher redshift range (mean 0.316, median 0.322) and tended to be more massive, with the majority of candidates above ${\rm log}(M_*/M_\odot)\sim11.5$ (mean 11.61, median 11.67). As Figure \ref{fig:mac_massvsredshift} shows, mid-range scoring LinKS candidates can be found throughout the redshift-mass range, with the highest scoring in the high-mass end above ${\rm log}(M_*/M_\odot) \sim 11.5$. This is because the training set for the \cite{Petrillo18} LinKS machine learning method is based on intermediate redshift massive galaxies (LRGs) that would have been identified in SDSS spectroscopy, i.e. the SLACS identified sample.

GalaxyZoo identified candidates definitively below the catalog's $z<0.15$ cutoff (mean 0.104, median 0.107) distributed in the mass range ${\rm log}(M_*/M_\odot) \sim 10.5 - 11.5$ (mean 11.13, median 11.06). The highest-scoring GalaxyZoo candidates occur in the middle of the stellar mass range of the sample (${\rm log}(M_*/M_\odot) \sim 11$).

\subsection{Overlapping Candidates}

\begin{figure}
    \centering
    \includegraphics[height=0.45\columnwidth]{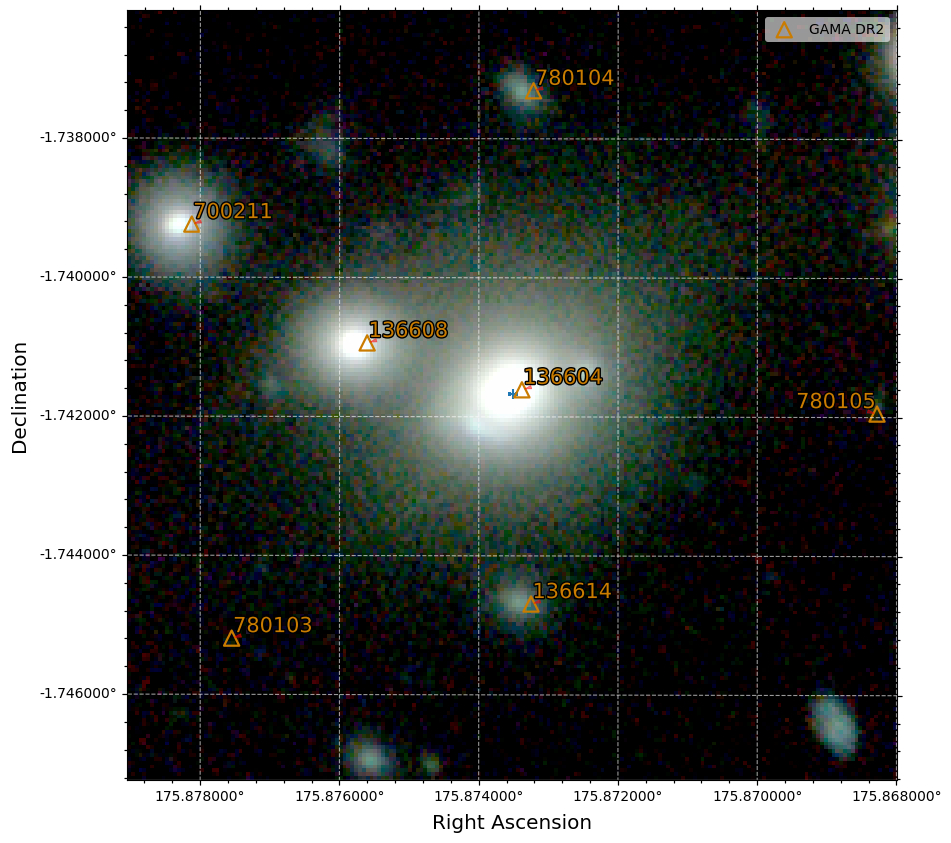}
    \includegraphics[height=0.45\columnwidth]{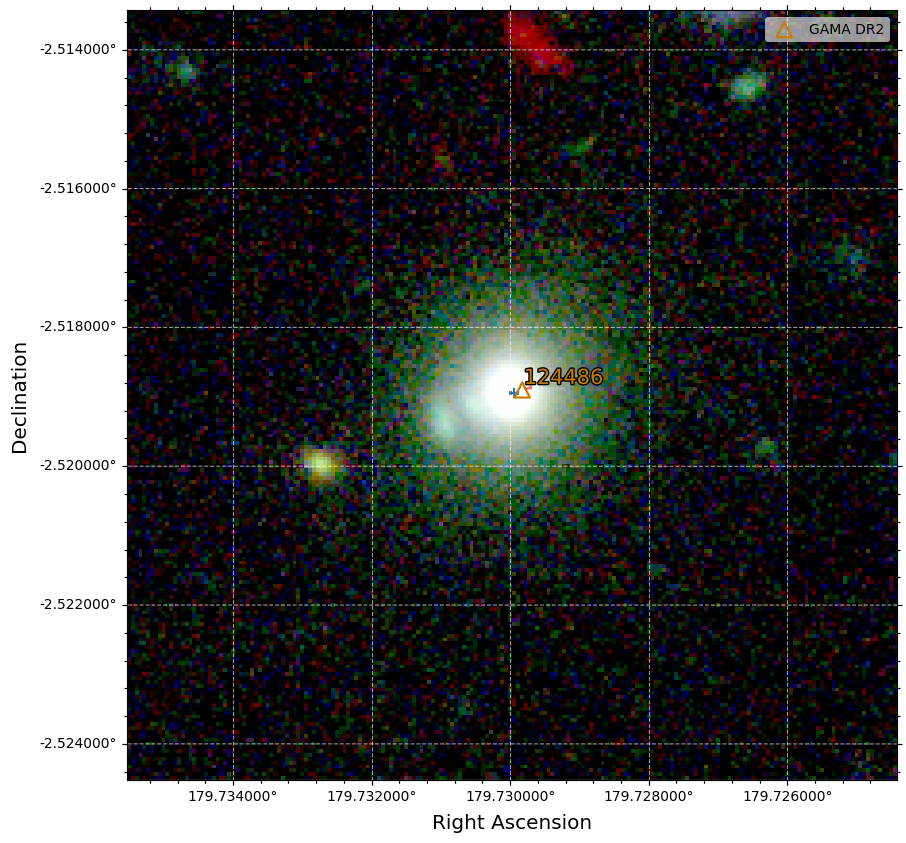}
    \caption{G136604 and G124486, the two candidates identified by both LinKS machine learning and GalaxyZoo citizen science. 
    }
    \label{fig:mac_zoo_overlap}
\end{figure}

The two candidates (G136604 and G124486) common to both LinKS machine learning and GalaxyZoo citizen science fall within the overlap of the parameter space occupied by the two methods' samples in terms of stellar mass and redshift, as shown in Figure \ref{fig:big_plot}. Table \ref{overlap_table} compares these two candidates, and their KiDS cutouts are shown in Figure \ref{fig:mac_zoo_overlap}. Their high scores make these two of the most promising candidates; however, it is worth noting the disagreement between scores given to G136604 by each method. This candidate scored very highly in the LinKS catalog but barely passed GalaxyZoo score selection criteria. G124486 scored around 40\% of the maximum possible score for both methods.

\begin{table}
\caption{Overlap of LinKS Machine Learning and GalaxyZoo citizen science Samples.}
\label{overlap_table}
\begin{center}
\begin{tabular}{l l l l l l l l l}
\hline
GAMAID & RA & DEC & $(M_*/M_\odot)$ & z & $\theta_{E,TL}$ (arcsec) & $\theta_{E,SIS}$ (arcsec) & ML Score & GZ Score\\
\hline
G136604 & 175.87 & -1.74 & $2.87 \times 10^{11}$ & 0.106 & 2.112 & 1.345 & 58 & 31.65\% \\
G124486 & 179.73 & -2.52 & $8.54 \times 10^{10}$ & 0.144 & 0.909 & 0.690 & 28 & 42.62\%\\

\hline
\end{tabular}
\end{center}
\label{t:hla}
\end{table}%

\section{Selection Effects}\label{section_selection}

\begin{figure}
    \centering
    \includegraphics[width=0.45\columnwidth]{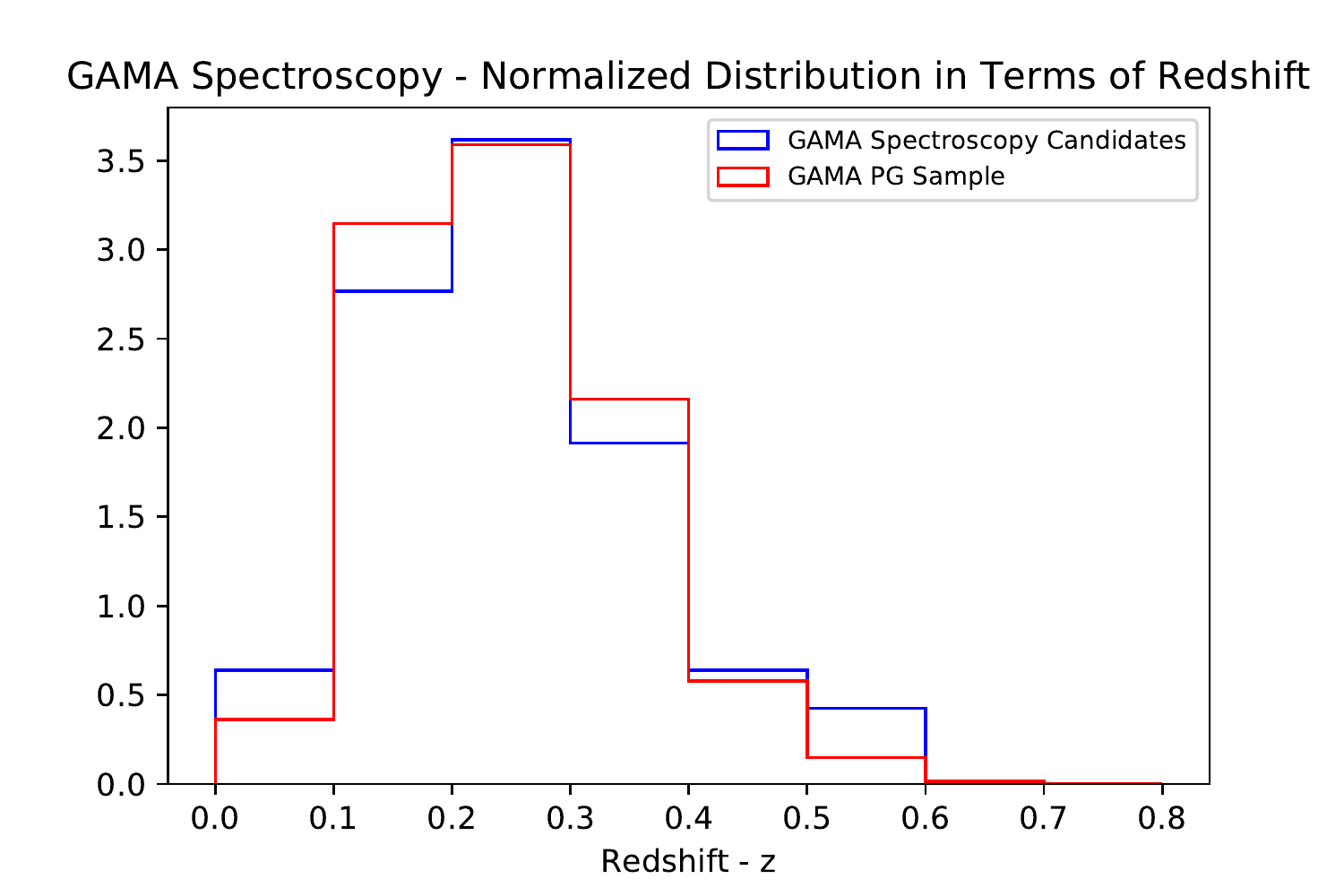}
    \includegraphics[width=0.45\columnwidth]{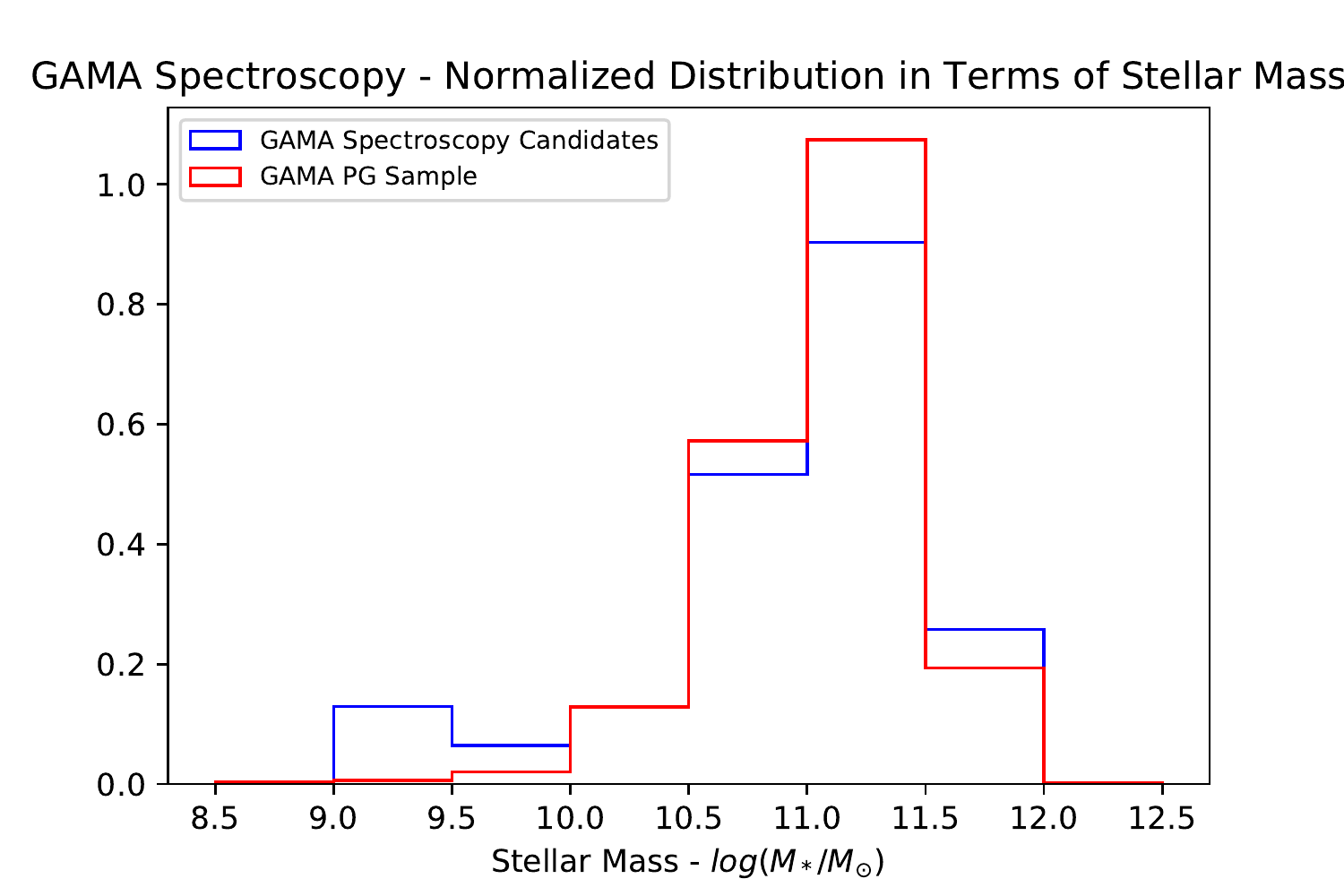}
    \includegraphics[width=0.45\columnwidth]{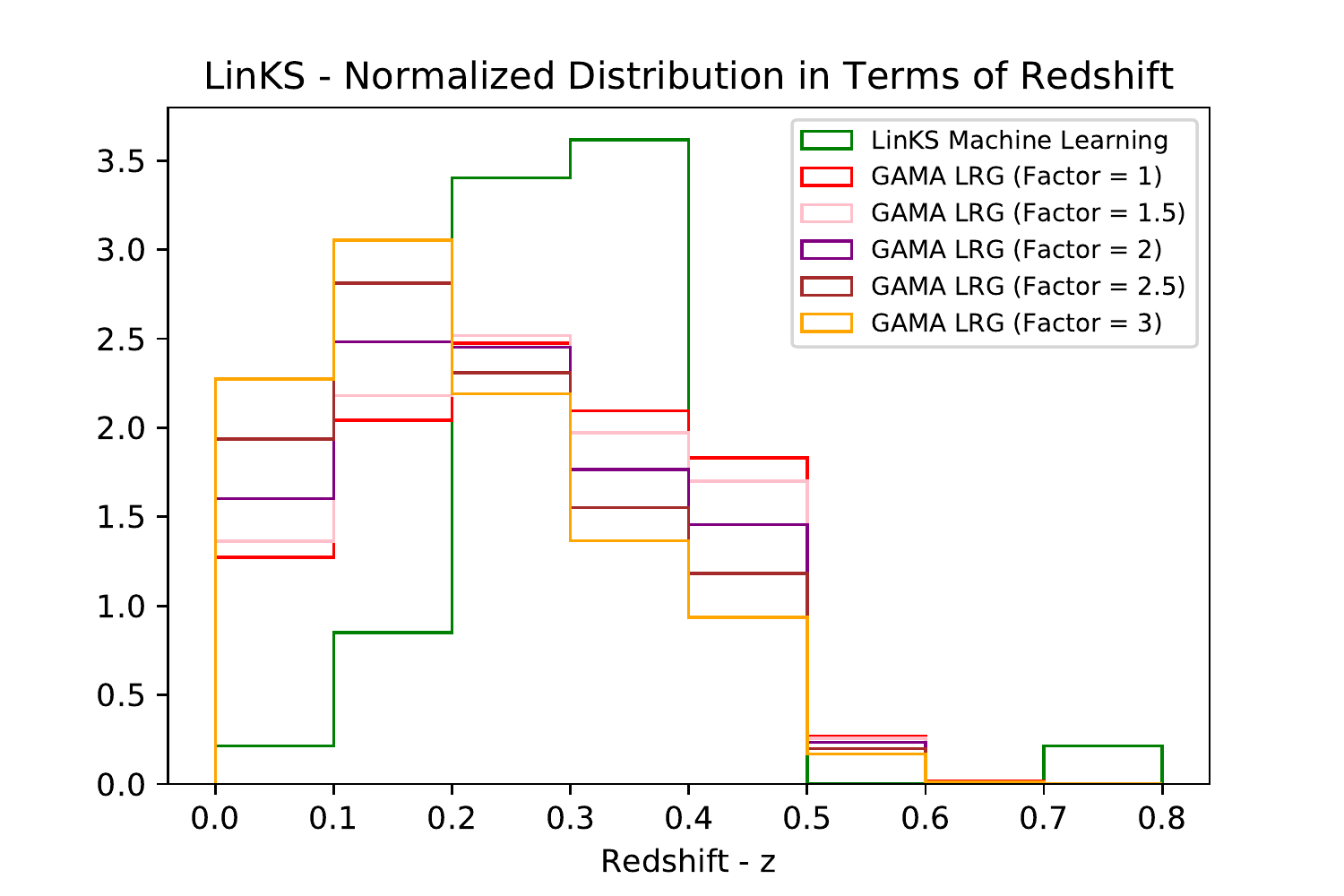}
    \includegraphics[width=0.45\columnwidth]{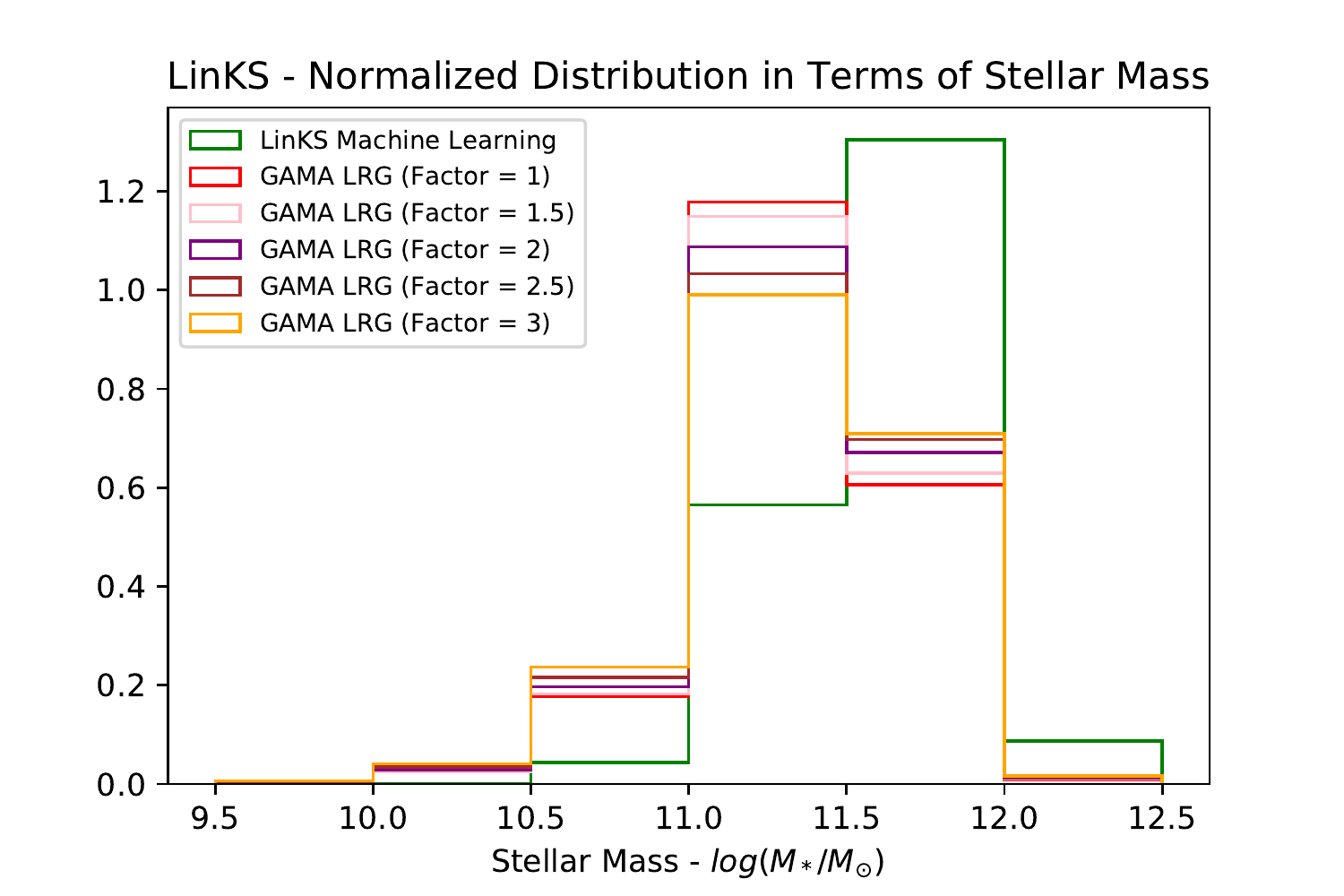}
    \includegraphics[width=0.45\columnwidth]{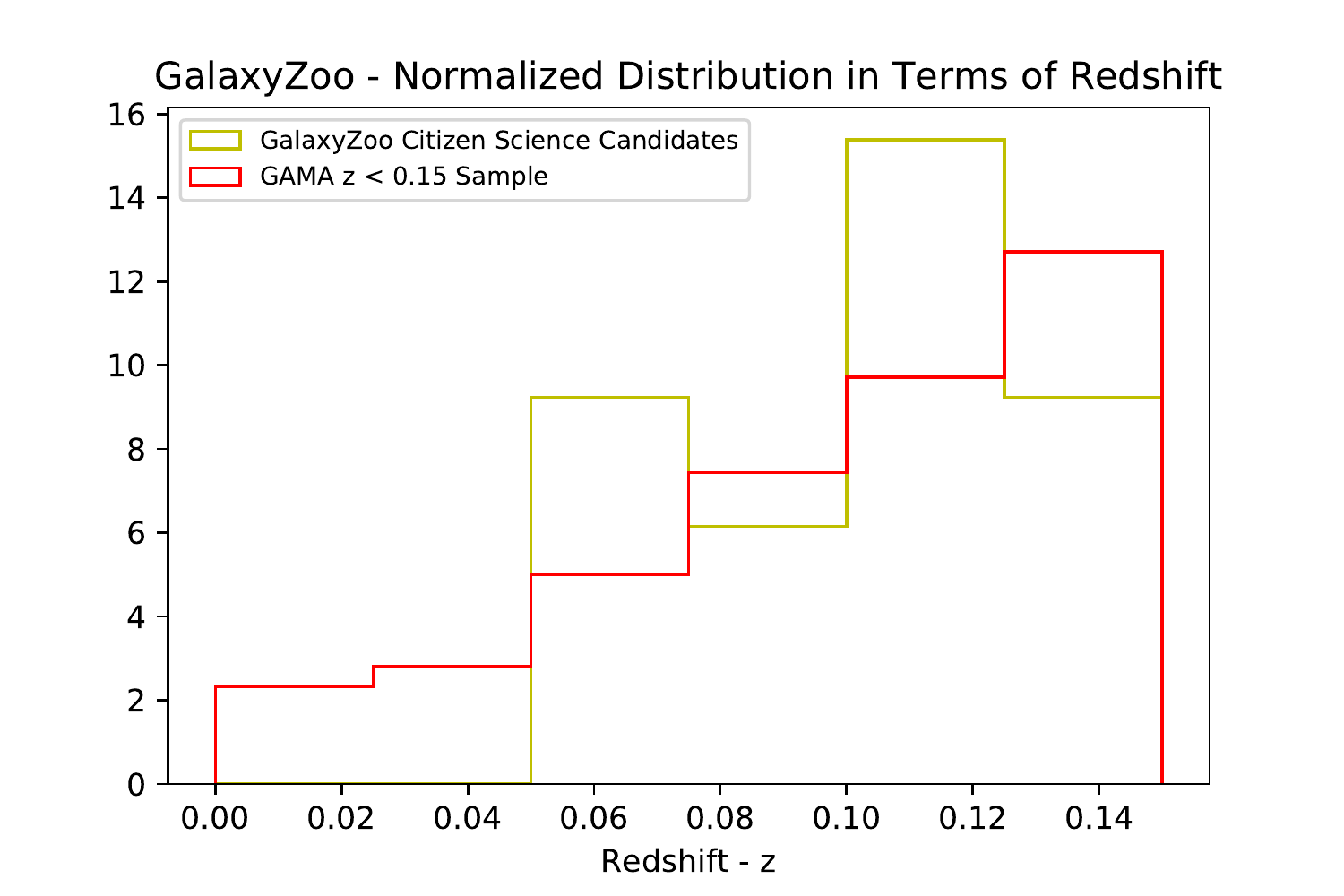}
    \includegraphics[width=0.45\columnwidth]{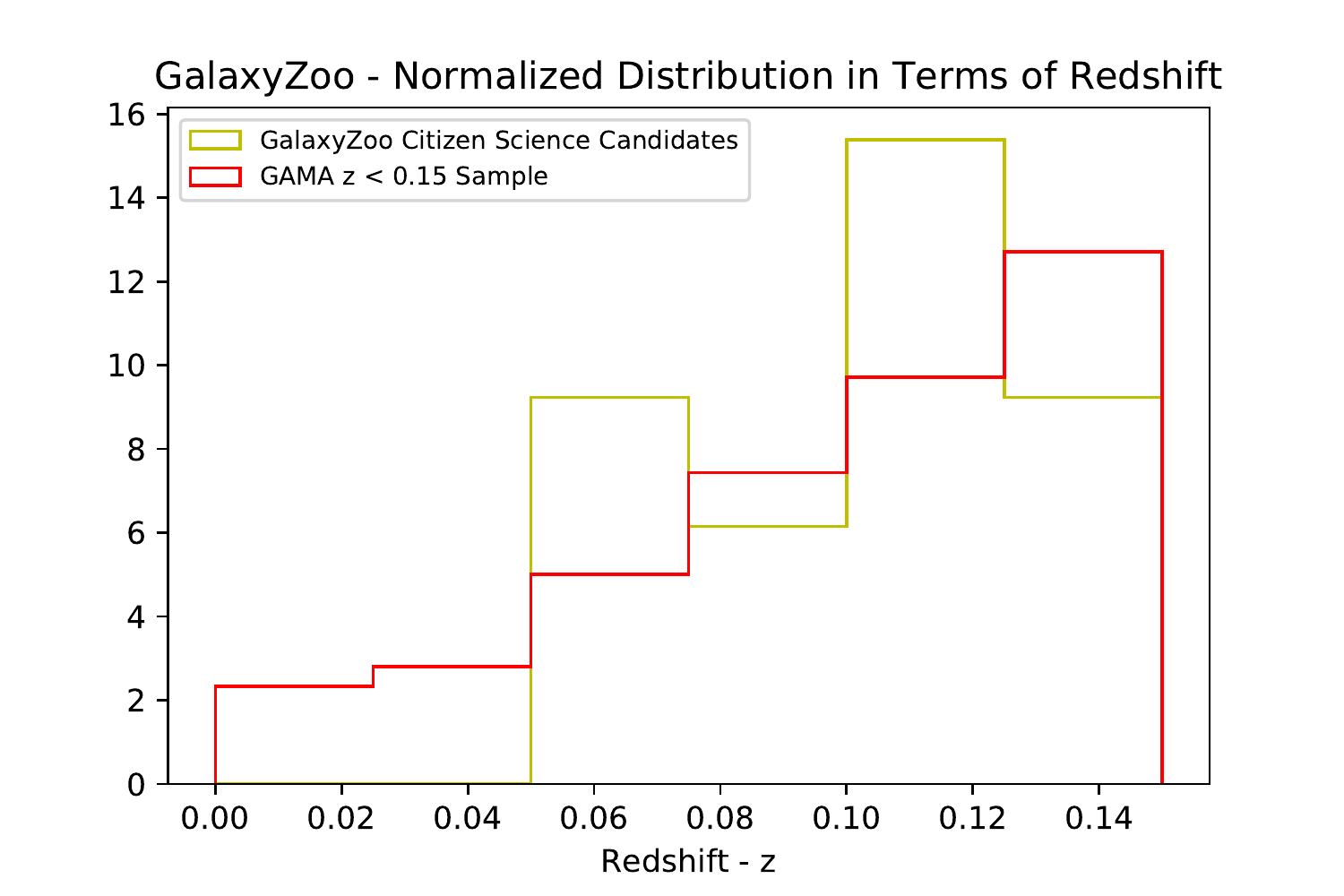}
    \caption{Normalized histograms showing the distributions of the three candidate samples in comparison with their parent samples in terms of redshift and stellar mass. The LinKS machine learning parent samples include a range of values for an empirical scaling factor that significantly affects the distribution of the parent samples and is shown in the corresponding K-S metrics of Table \ref{table_mac_ks}.}
    \label{fig:ks_histograms}
\end{figure}

GAMA spectroscopy candidates are selected from a parent subsample of spectroscopic targets in the equatorial regions of GAMA DR3 with spectral template matches of greater than 90\% confidence to one of three passive galaxy templates in the {\sc autoz} algorithm. Stellar masses and redshifts are taken from the GAMA {\sc Lambdar} catalog, and the resulting sample of 38278 passive galaxies in GAMA fields G09, G12, and G15 is taken to be the ``parent sample" for comparison with the GAMA spectroscopy lens candidate sample. Kolmogrov-Smirnoff tests between the lens candidate sample and the passive galaxy sample reveal K-S metrics and $p$-values of 0.0799 and 0.926 in terms of redshift and 0.374 and $4.01\times10^{-6}$ in terms of stellar mass. Note that stellar mass analysis is conducted for only the 31 GAMA spectroscopy candidates with reliable stellar mass estimates. All 47 redshifts are used. This analysis is suggestive (but not significantly, with high p-value) of parity between the parent sample and candidate sample in terms of redshift. The high metric and low p-value for stellar mass indicate a difference between the samples, showing that GAMA spectroscopy does present bias in the mass range of identifiable candidates. Figure \ref{fig:ks_histograms} offers a visual description of the distributions of each parameter.

The LinKS sample \citep{Petrillo18} of machine learning candidates, like its training sample, was selected using color-magnitude cuts modified from the SDSS-LRG (Large Red Galaxy) \citep{Eisenstein01, Petrillo17, Petrillo19} low-z selection criteria, with magnitudes taken from S-Extractor {\sc MagAuto}. Additionally, only those objects with effective radii greater than the FWHM of the PSF (times an empirical factor) were selected, in order to remove stellar contaminants. The parent sample for candidates selected within the GAMA equatorial regions therefore includes all galaxies within those regions that satisfy those criteria. Color-magnitude selection includes the following:
$$r < 20$$
$$c_{perp} < 0.2$$
$$r < 14 + \frac{c_{par}}{0.3}$$
where
$$c_{par} = 0.7(g-r) + 1.2[(r-i)-0.18)]$$ 
$$c_{perp} = (r - i) - \frac{g-r}{4.0} - 0.18$$ 
We take AB magnitudes from GAMA {\sc LAMBDAR} SDSS $g-$, $r-$, and $i-$ catalogs. Stellar mass and redshift are also taken from GAMA {\sc LAMBDAR}. Effective radii are taken from single-component Sersic fits to the 2D surface brightness distribution in the SDSS r-band from GAMA SersicPhotometry catalog. We select only those objects with effective radii greater than 0.65 arcseconds, the FWHM of KiDS PSF, times an emprical factor. Because the empirical factor utilized in \cite{Petrillo17, Petrillo18, Petrillo19} is not specified, we adopted five values from 1 to 3 to determine the factor's effect on the characteristics of the resulting parent sample selections, the results of which are shown in Table \ref{table_mac_ks}. The K-S metric for redshift appears to increase monotonically with the increase in this empirical factor, which is also evident in the middle left plot of Figure \ref{fig:ks_histograms}. Increasing the value of this empirical factor shifts the distribution of the resulting parent sample to lower redshift. The effect is milder for the stellar mass distribution. The results of these tests indicate to high significance that the LinKS machine learning sample is not representative of its parent sample for all adopted values of the empirical factor, as it is optimized for identifying high-mass candidates at intermediate redshift.

\begin{table}
    \centering
    \caption{K-S tests between LiNKS candidate sample and parent LRG samples utilizing five different empirical scaling factors.} \label{table_mac_ks}
    \begin{tabular}{l l l l l l}  
    \hline
        Factor & Objects & Redshift &  & Stellar Mass & \\
        & & K-S Metric & p-value & K-S Metric & p-value \\ \hline
        1 & 20009 & 0.2409 & 0.00864 & 0.47 & $1.93\times10^{-9}$ \\ 
        1.5 & 18646 & 0.2634 & 0.00299 & 0.46 & $4.82\times10^{-9}$ \\ 
        2 & 15739 & 0.3196 & 0.000139 & 0.44 & $2.70\times10^{-8}$ \\ 
        2.5 & 12761 & 0.3871 & 1.61E-06 & 0.423 & $1.10\times10^{-7}$ \\ 
        3 & 10421 & 0.4429 & 2.12E-08 & 0.412 & $2.48\times10^{-7}$ \\ \hline
    \end{tabular}
\end{table} 

GalaxyZoo selects only objects whose redshift $z < 0.15$, so the parent sample for the GalaxyZoo lens candidate sample contains all 40903 galaxies within the GAMA equatorial fields below that redshift upper limit. Redshifts and stellar masses are taken again from the GAMA {\sc LAMBDAR} catalog, and K-S tests of redshift and stellar mass yield (K-S metric, $p$-value) of (0.175, 0.823) and (0.781, $2.55\times10^{-7}$) respectively. This indicates a preference for high-mass candidates within its narrow redshift range, which is also indicated in Figure \ref{fig:ks_histograms}.

Lensing features become more easily observed in the high mass range, allowing for easier detectability of objects at higher mass and leading to bias in the candidate population toward these objects by image-based surveys like LinKS and GalaxyZoo, as well as by the inclusion of LRG selection utilized in some mixed spectroscopy efforts. GAMA spectroscopy, which has an upper mass constraint for likelihood of identification at a given redshift if the Einstein radius exceeds the radius of the aperture (See Section \ref{section_maxmass}), is an exception. GAMA spectroscopy and GalaxyZoo citizen science show little redshift bias from their respective parent sample due to the former's spectroscopic completeness and the latter's narrow redshift range. LinKS machine learning shows a combined redshift and mass difference between the candidate sample and the color-magnitude selected parent sample, suggesting that some bias is instituted after the parent sample selection.

\section{Discussion}

\subsection{$\theta_E$ Estimates and Empirical Fits} \label{section_fits_estimates}

\subsubsection{Systematic Differences Between Thin Lens and SIS $\theta_E$ Estimates} \label{section_theta_e_differences}

\begin{figure}
    \centering
     \includegraphics[width=0.7\columnwidth]{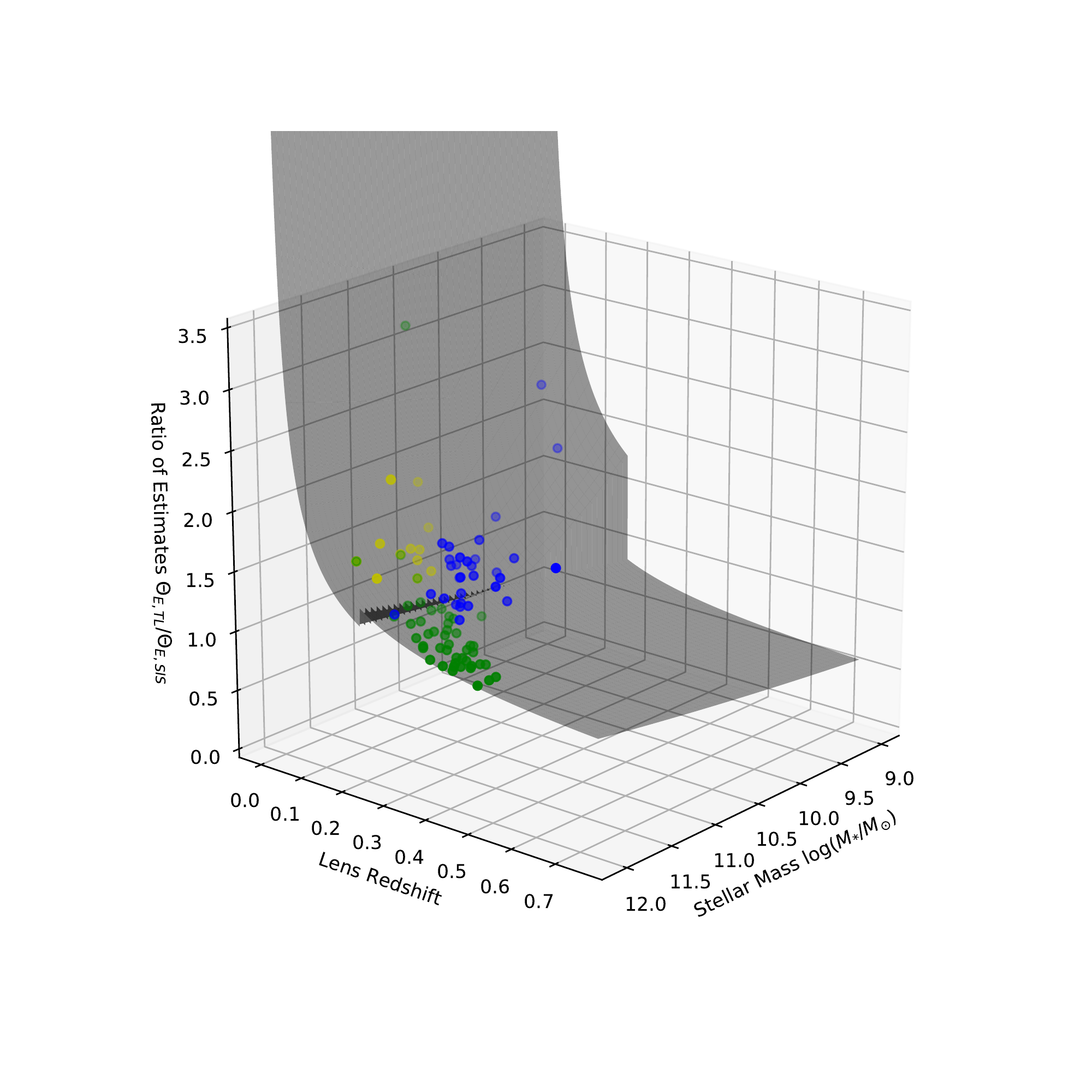}
     \includegraphics[width=0.7\columnwidth]{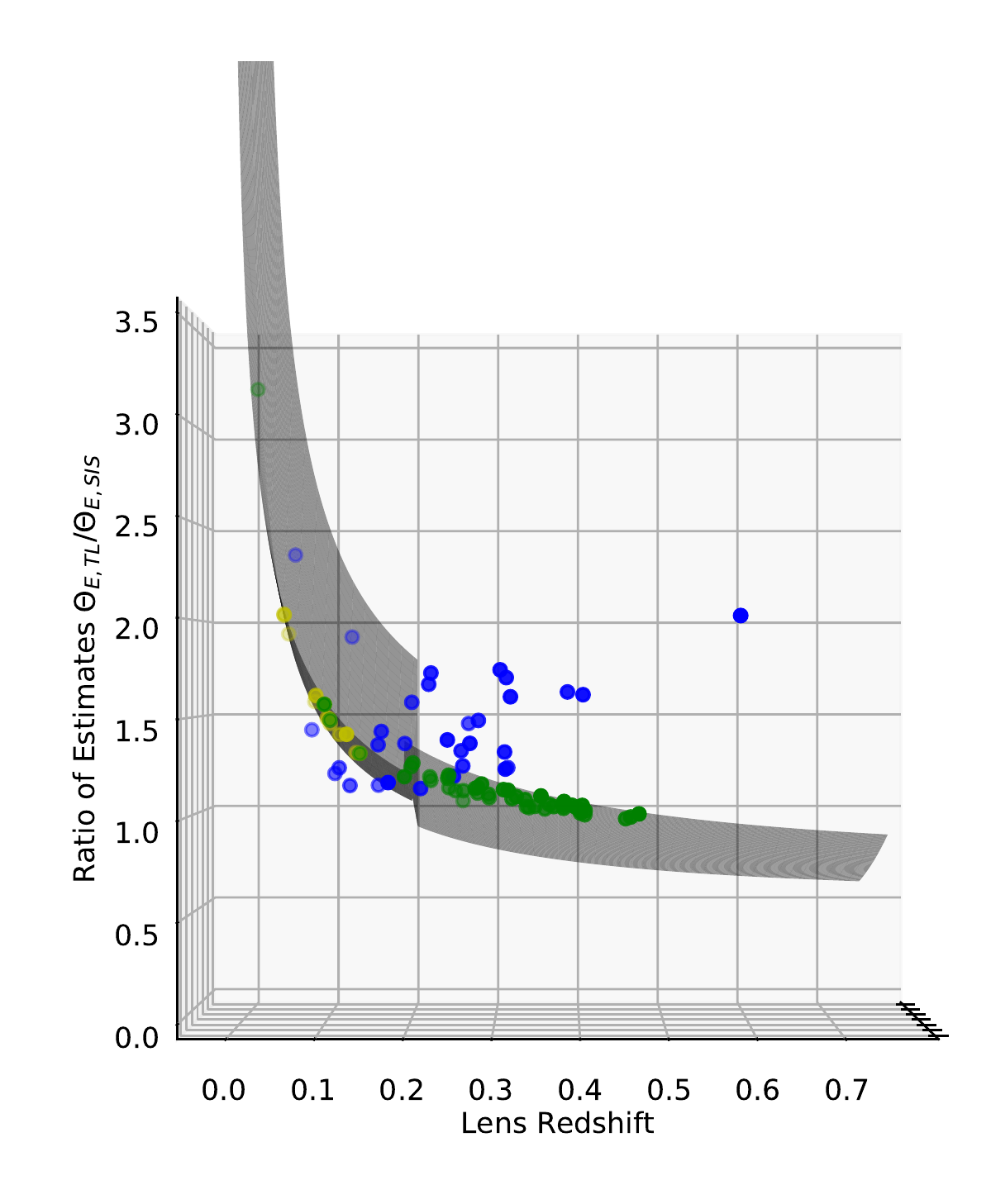}
     \includegraphics[width=0.7\columnwidth]{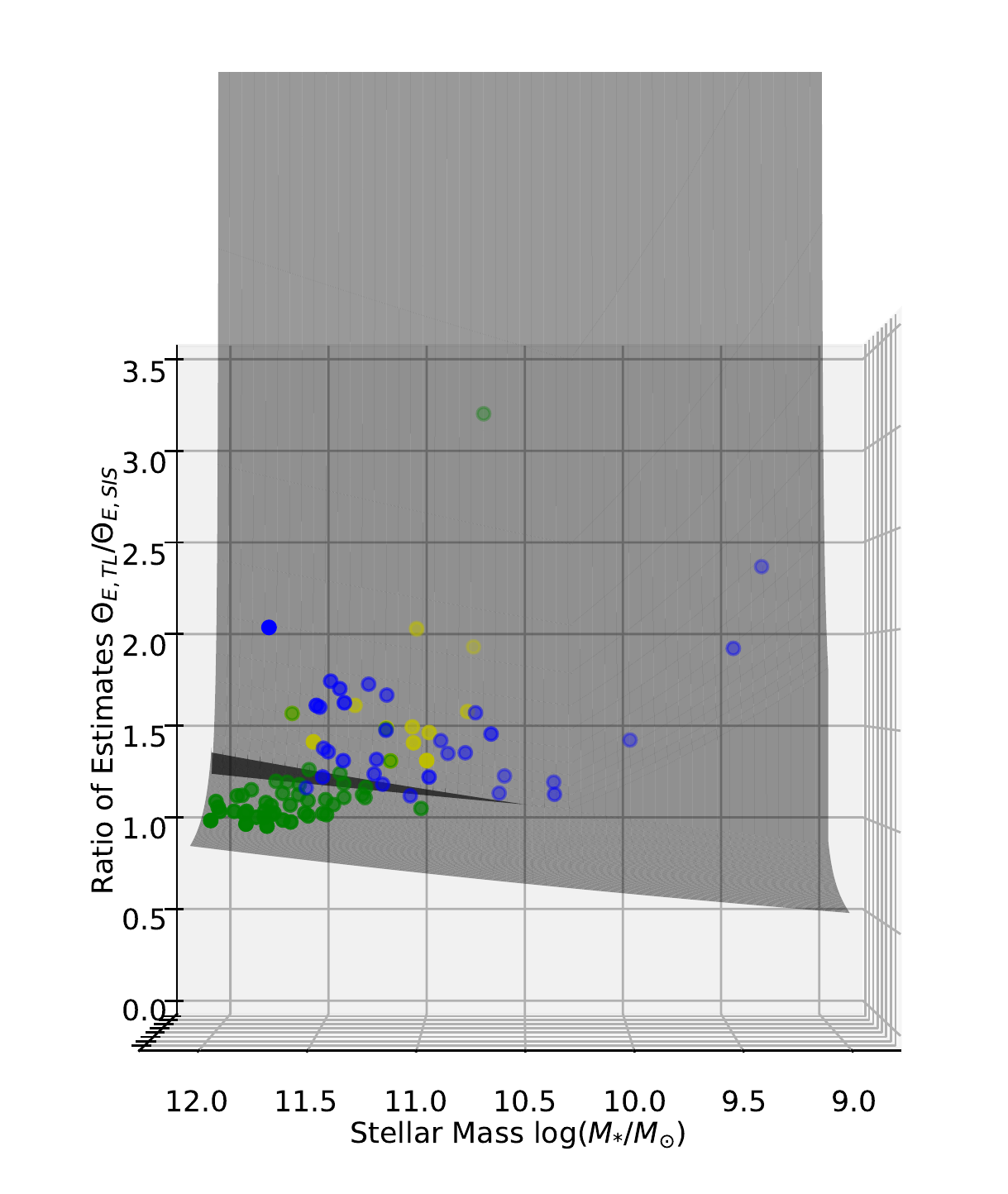}
    \caption{The ratio between thin lens and SIS estimates of Einstein radius as a function of two parameters, lens redshift and lens stellar mass. The black surface indicates the relationship between these parameters when assuming the simplifying 2:1 ratio of source distance to lens distance. All LinKS machine learning and GalaxyZoo citizen science candidates are intersected by this surface. The surface is defined by three distinct piecewise regions as a result of the different velocity dispersion estimates adopted from \cite{Zahid16} and described in Section \ref{section_theta_e_estimates}. }
    \label{fig:3d_theta_e_ratio}
\end{figure}

We further explore the trends introduced in Section \ref{section_theta_e_results} and Figure \ref{fig:theta_e_difference} regarding the difference between thin lens and SIS Einstein radius estimates. In Figure \ref{fig:3d_theta_e_ratio}, we plot the ratio of thin lens to SIS estimates for each candidate as a function of lens redshift and $\log({M_*/M_{\odot}})$. In order to understand the relationship between these parameters for LinKS and GalaxyZoo candidates, we set a ratio of Equations (\ref{theta_e_pm_eqn}) and (\ref{theta_e_sis_eqn}).

\begin{align}
    \frac{\theta_{E,TL}}{\theta_{E,SIS}} & = \left(\frac{4 G M_E D_{LS}}{c^2 D_S D_L}\right)^{1/2}\left(\frac{D_S c^2}{4 \pi \sigma_{SIS}^2 D_{LS}}\right) \label{ratio_eqn_1}
    \\
     & = \left(\frac{G^{1/2} c }{2 \pi}\right) \left(\frac{M_E^{1/2}}{\sigma_{SIS}^2}\right) \left(\frac{D_S}{D_L D_{LS}}\right)^{1/2} \label{ratio_eqn_2}
    \\
     & = \beta \left(\frac{M_*}{M_{\odot}}\right)^{\Gamma} \left(\frac{D_S}{D_{LS}}\right)^{1/2} \left(\frac{1}{D_L}\right)^{1/2}
     \label{ratio_eqn_3}
\end{align}

where Equation (\ref{ratio_eqn_3}) substitutes Equations (\ref{enclosed_mass_eqn}, \ref{sigma_lowz_eqn} and \ref{sigma_intz_eqn}) for the values of $M_E$ and $\sigma_{SIS}$ in Equation (\ref{ratio_eqn_2}), and

\begin{align*}
    \beta = \left(1.90\times10^6 \: \textnormal{Mpc}\right)^{1/2}, \quad \Gamma&=-0.181 \qquad  \textnormal{for} \qquad z < 0.2 \:, \qquad M_* \leq M_b \\
    \beta = \left(57.4 \: \textnormal{Mpc}\right)^{1/2}, \quad  \Gamma&=0.039 \qquad \quad \textnormal{for} \qquad  z < 0.2 \:, \qquad M_* > M_b \\
    \beta = \left(19.1 \: \textnormal{Mpc}\right)^{1/2}, \quad  \Gamma&=0.063 \qquad \quad \textnormal{for} \qquad 0.2 \ z <0.65 
\end{align*}

We assume the source distance to be twice the lens distance, so the middle parentheses term of Equation (\ref{ratio_eqn_3}) becomes $\sqrt{2}$, and the resulting function describes a surface in this three-dimensional parameter space that intersects each point representing the LinKS and GalaxyZoo candidates, as shown in Figure \ref{fig:3d_theta_e_ratio}. This function shows a clear dependence of the value of this ratio on the inverse square root of the lens distance. This dependence holds for each of the three distinct conditional regions of redshift and stellar mass described in Equations (\ref{sigma_lowz_eqn}) and (\ref{sigma_intz_eqn}). This shows a pronounced difference between the estimates for these candidates at lower redshift. There is a strong correlation between mass and redshift for these candidates, where stellar mass increases with redshift. All LinKS and GalaxyZoo candidates lie in the region above $M_* = M_b=10^{10.26}$, where this ratio value decreases with decreasing stellar mass. Therefore, the observed trend of a greater difference in $\theta_E$ estimates for candidates at lower redshift does not appear to be due to a mass dependence, which is in fact a competing dependence.

The surface shown in Figure \ref{fig:3d_theta_e_ratio} reveals an interesting trend at stellar masses below $M_* = M_b=10^{10.26}$ and redshifts below $z = 0.2$, which shows a steep rise with decreasing stellar mass due to the fit introduced in Equation (\ref{sigma_lowz_eqn}) for lower mass and lower redshift. Only GAMA spectroscopy candidates, for which an assumed distance ratio was not applied, reside in this region. The few data points that appear to follow that trend cannot be considered to be direct evidence for it, since they were calculated from measured source redshifts. Of course there is the possibility of other redshift related effects in the estimation of Einstein radius and in how accurately these approximations determine the true Einstein radius of a system, but the redshift trend observed in the data as first shown in Figure \ref{fig:theta_e_difference} appears to be a result of the chosen assumptions for GalaxyZoo and LinKS, which isolate the lens distance as a unique factor to one estimate (thin lens) and not the other (SIS). Importantly, this analysis is meant only to explore the limitations of the manner in which these estimates have been specifically applied to the candidates in this work so as to better understand how to properly consider the results of the application. The adopted fits from \cite{Auger10} and \cite{Zahid16} described in Section \ref{section_theta_e_estimates} play a significant role in the algebraic relations discussed here.

\subsubsection{Limitations of Adopted Empirical Fits}

We also consider more deeply the limitations of the application of these empirical fits to the estimation of Einstein radii for the candidates under consideration. We note that the adopted relation between $M_E$ and $M_*$ (Equation (\ref{enclosed_mass_eqn}) in Section \ref{section_theta_e_estimates}) is calculated (a) from SLACS lenses with higher stellar masses than some of our candidates examined here, (b) within half the object's effective radius and not necessarily the Einstein radius, and (c) assuming a power-law profile for mass distribution. Through rigorous calculations as detailed in \cite{Tortora10}, \cite{Petrillo17} show $\sigma_{SIS}$ and thereby the calculated Einstein radius estimate to be higher than its stellar counterpart. This suggests that the Einstein radii calculated with the SIS model detailed above in Section \ref{section_theta_e_estimates} may be underestimates due to the approximation of $\sigma_*$ to $\sigma_{SIS}$ in Equation (\ref{theta_e_sis_eqn}). Conversely, the stellar velocity dispersion as presented in \cite{Petrillo17} is measured within the SDSS and BOSS fibers (with $3^{\prime\prime}$ and $2^{\prime\prime}$ apertures respectively), while the velocity dispersion in \cite{Zahid16} has been corrected to a fiducial physical aperture of 3 kpc, which is progressively smaller in angular size at higher redshift. As a result, the $\sigma_*$ of \cite{Zahid16} may be higher than the $\sigma_{*}$ used in \cite{Petrillo17}, which could account for the difference. 

\begin{figure}
    \centering
    \includegraphics{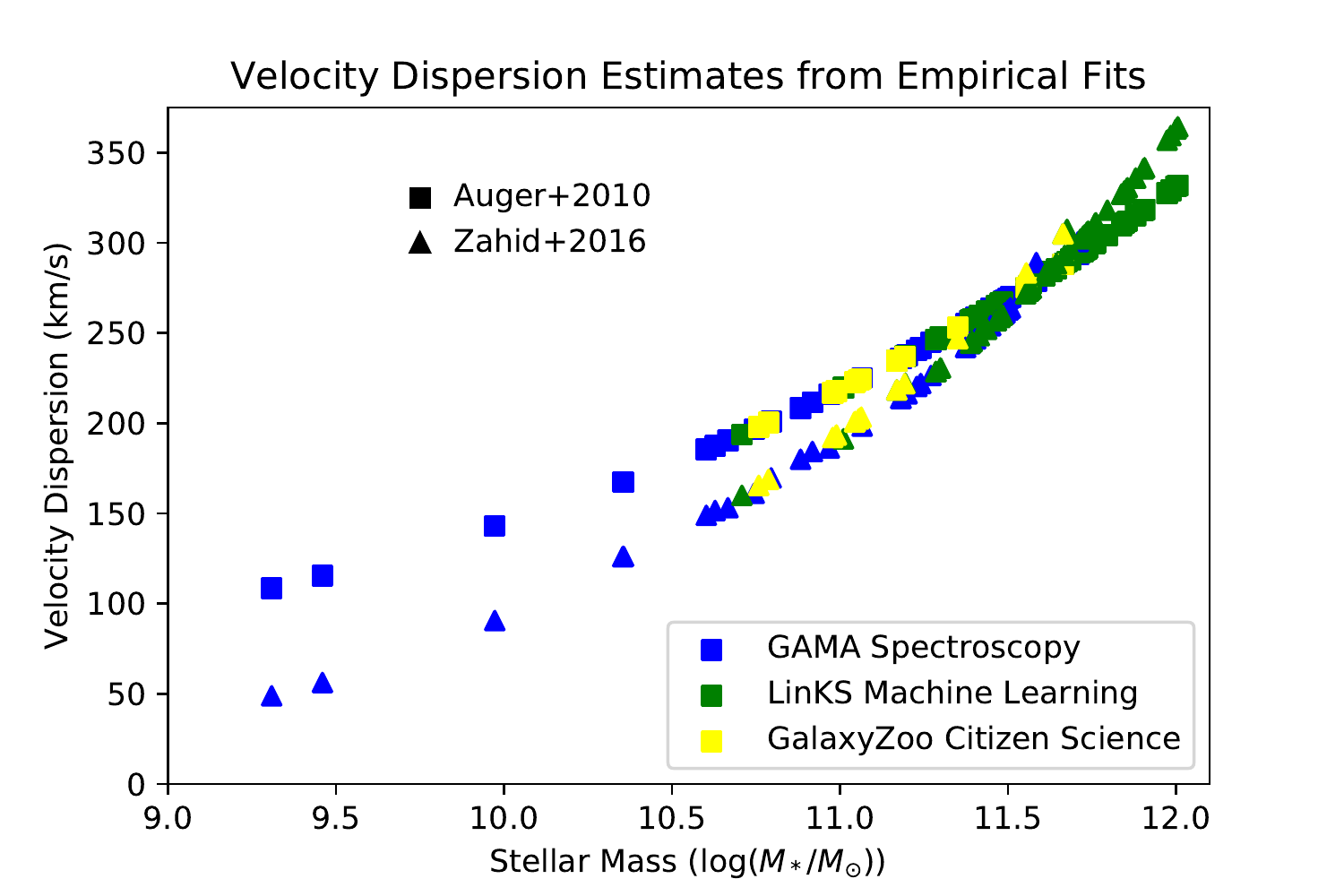}
    \caption{Velocity dispersions estimated from stellar mass of candidates derived from empirical fits taken from \cite{Auger10} (squares) and from \cite{Zahid16} (triangles). The differences that manifest in the lowest and highest ends of the stellar mass range result in greater differences in the estimated Einstein radii for those candidates.}
    \label{fig:auger_vs_zahid}
\end{figure}

As an additional sanity check, we calculate the velocity dispersion by utilizing another fit from the SLACS lens analysis of \cite{Auger10}, which relates the velocity dispersion within half the effective radius to the total stellar mass of the galaxy. The relation can be written as
\begin{equation} \label{sigma_auger_relation}
\sigma_* = 10^{2.34}\left(\frac{M_*}{10^{11}M_{\odot}}\right)^{0.18} \textnormal{km/s}
\end{equation}
the result of which is shown in comparison to velocity dispersions derived from the \cite{Zahid16} relation in Figure \ref{fig:auger_vs_zahid}. Relative to the velocity dispersions derived from the \cite{Zahid16} relation, the \cite{Auger10} velocity dispersions are higher for candidates at lower stellar mass and lower for candidates at higher stellar mass. These differences result in greater differences in Einstein radius for candidates in those ranges. The difference between these two estimates may be due to the differences in parent samples. The \cite{Zahid16} relation is based on a range of masses as well as accounting for different ranges of redshift ($\log(M_*/M_{\odot})>9$ from SDSS at $z < 0.2$, and $\log(M_*/M_{\odot})>9.5$ from SHELS at $z > 0.2$) while the \cite{Auger10} relation is fit to a much smaller sample of SLACS lenses with a mean $\log(M_*/M_{\odot})$ of 11.33, minimum 10.43, and maximum 11.79. This suggests that either prescription may be more appropriate for different selections depending on the mass ranges involved.

\begin{table}
\caption{SLACS Lenses in the GAMA {\sc LAMBDAR} Catalog.
Values for Einstein radius estimates are computed using the procedure as outlined in Section \ref{section_theta_e_estimates}, utilizing the true source redshift as well as adopting the 2:1 ratio assumed for those candidates in this study whose source redshifts are unknown. Here, we present the result of utilizing the true source redshift before the assumed as (true/assumed).}
\label{slacs_table}
\begin{center}
\begin{tabular}{l l l l l l l l l}
\hline
GAMA ID & SDSS ID & $\log(M_*/M_\odot)$ & $z_{lens}$ & $z_{source}$ & $\theta$ ('') & & & \\
& & & & & True (Bolton+) & TL (Auger+) & SIS (Zahid+) & SIS (Auger+) \\
\hline
G136604 & J1143−0144 & 11.66 & 0.106 & 0.324 & 1.68 & 2.106/2.112 & 1.421/1.345 & 1.141/1.197 
\\
G216398 & J0912+0029 & 11.87 & 0.164 & 0.402 & 1.63 & 2.394/2.344 & 1.724/1.763 & 1.538/1.415
\\
\hline

\end{tabular}
\end{center}
\label{t:hla}
\end{table}%

\subsubsection{SLACS Lenses in GAMA {\sc LAMBDAR} Catalog}

Two confirmed and well-studied SLACS grade-A lenses are present in the GAMA {\sc LAMBDAR} catalog, which allows us to compare the three Einstein radius estimates to the measured Einstein radius of these two examples from the detailed modeling and analysis by \cite{Bolton08}. We estimate the Einstein radii for these two lenses utilizing both the true source redshifts and the assumed 2:1 ratio of source distance to lens distance that we apply to GAMA/KiDS candidates with unknown source redshifts. These serve to inform both the GAMA spectroscopy estimates that include source redshifts as well as the LinKS and GalaxyZoo estimates that require the fiducial assumption. For G136604, which is one of the two overlaps between the LinKS machine learning and GalaxyZoo citizen science candidate catalogs, the true Einstein radius \citep{Bolton08} lies between the SIS estimate from the \cite{Zahid16} relation and the thin-lens estimate, while the SIS estimate from \cite{Auger10} underestimates the true value by around half an arcsecond. The SIS estimate utilizing the true source redshift and the  \cite{Zahid16} relation most closely approximates the true Einstein radius. In the case of G216398, it appears that the \cite{Zahid16} fit slightly overestimates the true Einstein radius, which does lie between the lowest SIS estimate \citep{Auger10} and the thin lens estimate. This may be a result of the fact that the \cite{Zahid16} velocity dispersion fit is less appropriate than the \cite{Auger10} fit for the highest-mass galaxies, though again the \cite{Zahid16} estimate is in fact the closest of the three models we have applied to the true value for both examples. 

Based on the questions and insights discussed throughout this Section \ref{section_fits_estimates}, we consider each estimate to inform the range of probable Einstein radius values. For the two SLACS examples, the SIS estimate approximates the measured value of the Einstein radius closely. However, given the low number statistics and the uncertainties introduced by adopting fiducial values for $\sigma_{SIS}$, $M_E$ and $z_{source}$, we consider this far from conclusive. We elect to use the thin lens estimate to pare down the GalaxyZoo selection as discussed in Section \ref{section_selection} because it is more inclusive and allows for occasional better KiDS PSF as well as GalaxyZoo selection improvement using color. 
We elect to use the $M_*\sigma$ conversion from \cite{Zahid16} instead of \cite{Auger10} because the former is fit to a larger sample that covers the entire range of masses occupied by our candidates objects. Given accurate source redshifts, each candidate could be modeled in a more complex manner to check these estimates more thoroughly. This followup validation, though worthwhile, is outside the scope of this paper. Deeper analysis and modeling of the other candidates presented in this paper will shed more light on the limitations of each of these models and assumptions. These first estimates of Einstein radii suffice to select the samples and plan for follow-up observations in a future study.

\subsection{LinKS Machine Learning Training Set}

\begin{figure}
    \centering
    \includegraphics[width=0.45\textwidth]{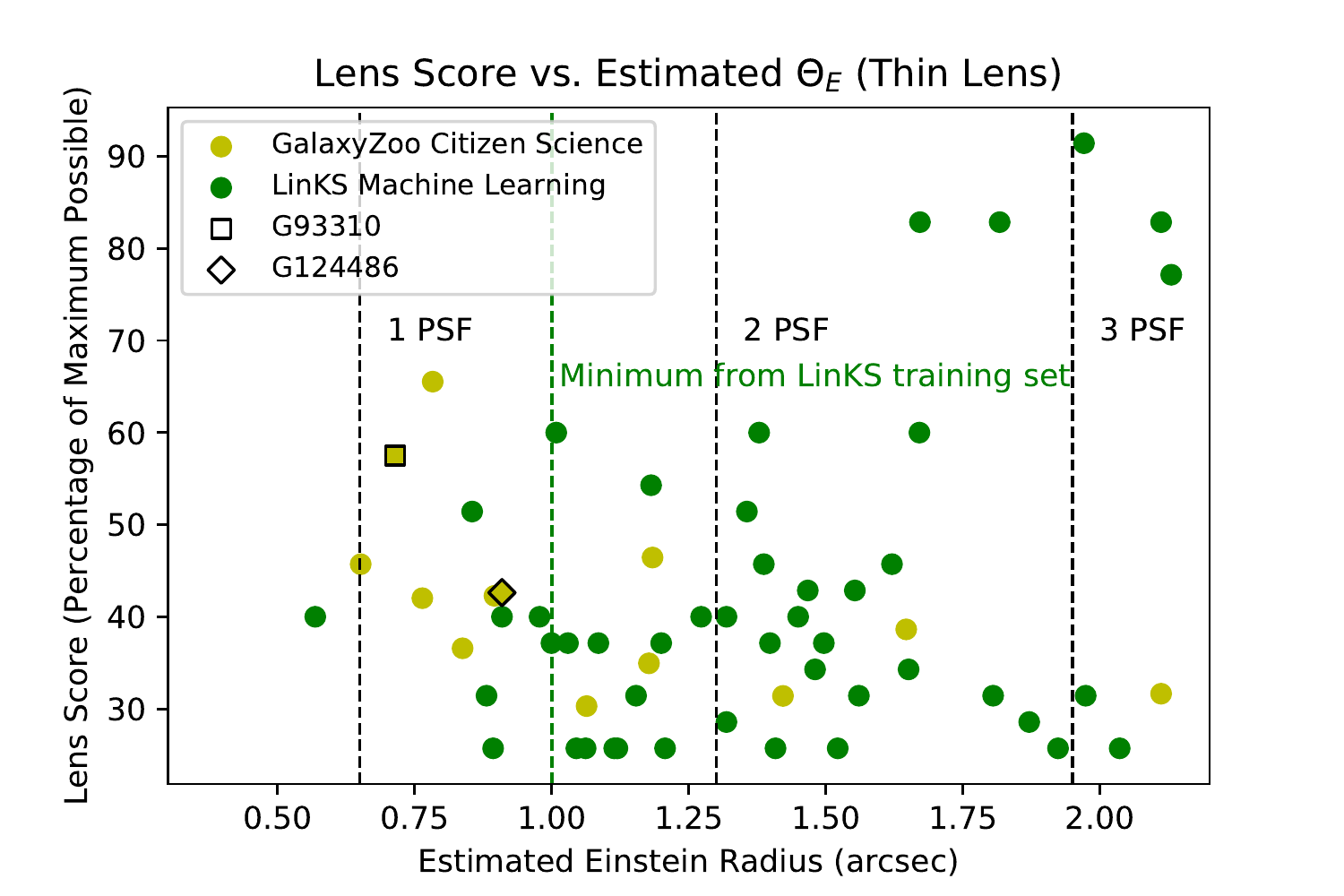}
    \includegraphics[width=0.45\textwidth]{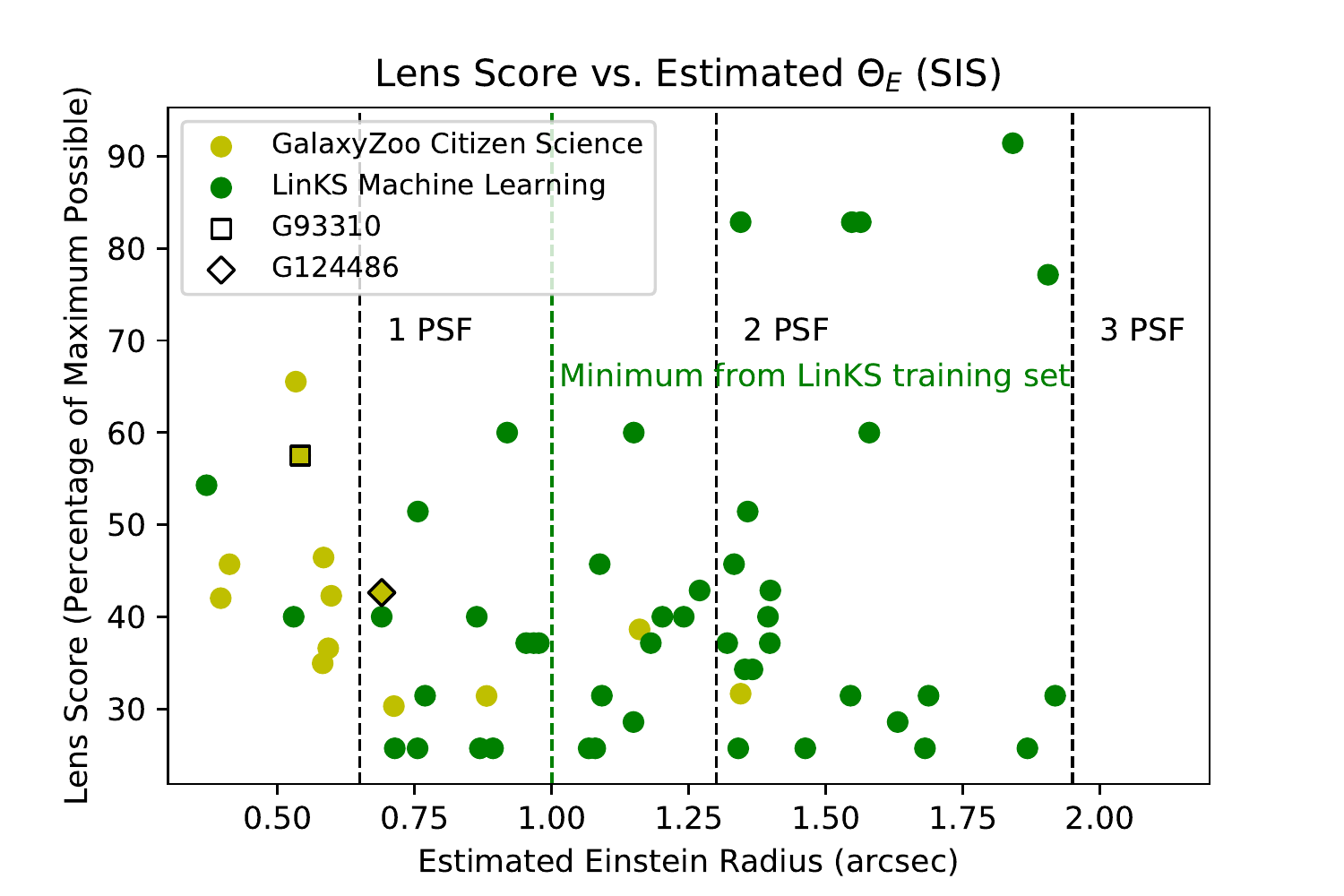}    
    \caption{High scoring LinKS machine learning candidates show estimated Einstein radii above the minimum ($1^{\prime\prime}$) utilized for the training set. The highest scoring LinKS candidates appear to have the largest Einstein radii, due to the maximal separation of lensing features from the image of the foreground galaxy. GalaxyZoo candidates do not appear to show this same trend, though the size of the candidate sample makes generalization difficult.}
    \label{fig:mac_zoo_theta_e}
\end{figure}

Machine learning is an effective method for identifying image features similar to the training set utilized. Its effectiveness is then constrained inherently by the scope of its training set. The catalog analyzed here from \cite{Petrillo18} used simulated images resembling SLACS lenses as their training set, and the candidates it identified are characteristically similar in terms of estimated Einstein radius, stellar mass, and redshift to those identified by SLACS spectroscopy. With a training sample volumetrically skewed towards massive elliptical galaxies (the early-type galaxies sample in KiDS) with large Einstein radii, the resulting identified candidates are equally biased toward higher masses and Einstein radii, as shown in Figures \ref{fig:theta_e} and \ref{fig:mac_zoo_theta_e}. This bias can be amplified through the problem of transfer learning. 

\subsection{Fiber Radius Mass and GAMA Spectroscopic Identification}\label{section_maxmass}

\begin{figure}
    \centering
    \includegraphics[width=0.45\textwidth]{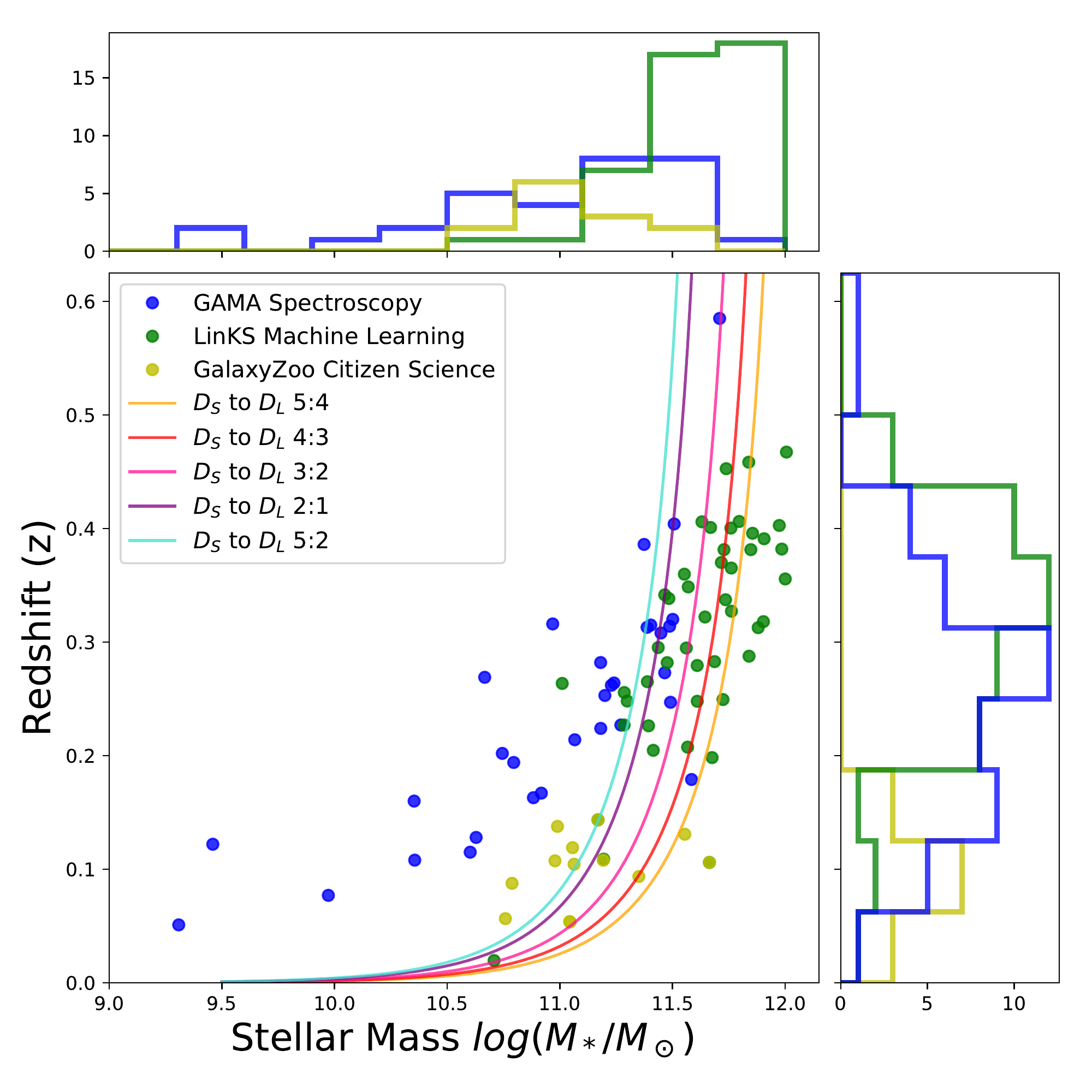}
    \includegraphics[width=0.45\textwidth]{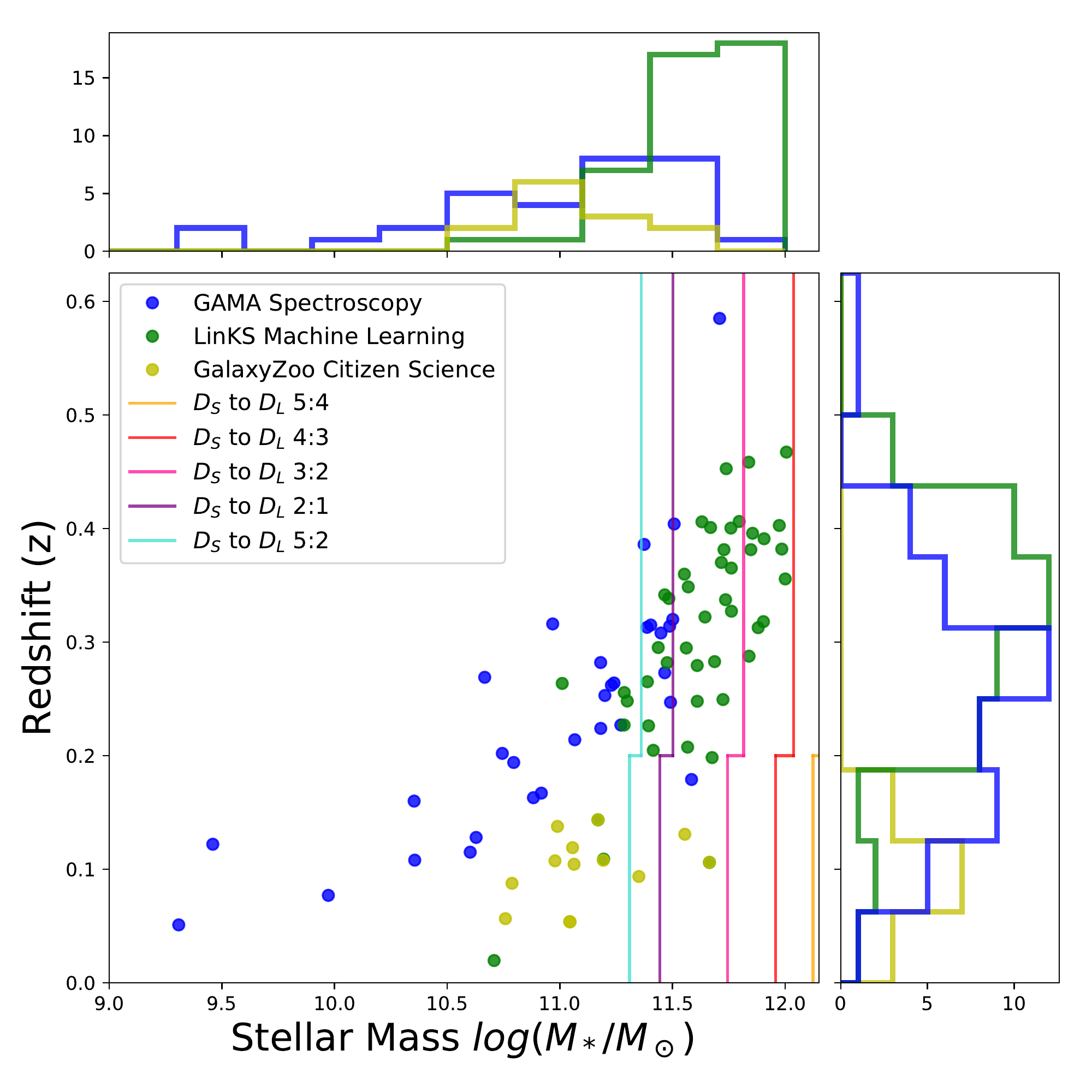}    
    \caption{The data presented here is the same as in Figure \ref{fig:big_plot} with additional elements. These elements are introduced to represent total stellar mass equal to the fiber radius mass ($M_* = M_{fr}$), which results in an Einstein radius of $\theta_E = 1$ arcsecond (the aperture-size of GAMA spectroscopic fibers). (Left) Curves parameterized by the ratio of source distance to lens distance show the total stellar mass and lens candidate redshift where total stellar mass is equal to the fiber radius mass. (Right) Vertical lines reflect the general independence of SIS velocity dispersion Einstein radius estimates (Equation \ref{theta_e_sis_eqn}) on lens redshift, with the exception of the $z \sim 0.2$ break in the fit of stellar velocity dispersions to stellar mass (Equations \ref{sigma_lowz_eqn} and \ref{sigma_intz_eqn}) that results in the horizontal break at this redshift in the plot. For both plots, datapoints that fall along a given curve or line will have a $1^{\prime\prime}$ Einstein radius for the given ratio of source distance to lens distance. Note that unknown source redshifts for LinKS and GalaxyZoo candidates could also affect the Einstein radius estimate, were they available. For this reason, we have shown the fiber radius mass curves and lines for a variety of source to lens distance ratios. Multiple candidates identified by the other two methods have stellar masses that exceed the GAMA spectroscopy fiber radius mass at the given lens candidate redshift.}
    \label{fig:max_mass}
\end{figure}

Lens candidate identification through any spectroscopic method requires sufficient flux from the background galaxy in order to obtain a second spectral match. For lensing systems whose Einstein radius exceeds the radius of the instrument's aperture, the probability of detection goes down significantly and rapidly \citep[][and reference therein]{Sonnenfeld15}. For GAMA spectroscopy we therefore introduce a characteristic mass that we call the \textit{fiber radius mass} and define it as follows: $M_{fr}$ is the precise total stellar mass of a lensing system (of which the lens and source redshift are known) for which the estimated Einstein radius equals the radius of the spectroscopic fiber, assuming the fiber has been positioned precisely at the center of the lensing object, i.e. $\theta_E(M_*=M_{fr}) = 1$ arcsec. For GAMA spectroscopy, this represents a soft upper constraint on the total stellar mass that corresponds to a total enclosed mass $M_E$ or stellar velocity disperion $\sigma_*$ that will contribute to an Einstein radius that will fit within the instrument's aperture. 

Taking into account GAMA spectroscopy's $1^{\prime\prime}$-radius aperture, we derive expressions for stellar mass that correspond to $\theta_E = 1$ arcsecond using the equations of Section \ref{section_theta_e_estimates}. The left plot of Figure \ref{fig:max_mass} shows colored curves parameterized by the ratio of source distance to lens distance that show the stellar mass and lens candidate redshift which result in thin-lens Einstein radius estimates equal to the fiber radius of $1^{\prime\prime}$. This relation of lens distance to stellar mass is taken from Equations (\ref{theta_e_pm_eqn} and \ref{enclosed_mass_eqn}) and is shown for various ratios of $A= \frac{D_S}{D_L}$ represented by the colored curves.
\begin{equation}
    \frac{D_L}{Mpc} = \left(\frac{A-1}{A}\right)\left(0.011\times 10^{-8.09}\right)\left(\frac{M_*}{M\odot}\right)^{1.25}
\end{equation}
Redshift is easily converted from angular diameter distance.

For SIS models that estimate $\theta_E$ from velocity dispersion, the Einstein radius is not inherently proportional to the lens candidate distance, but instead to the ratio of the distances of the lens and source. We use Equations (\ref{theta_e_sis_eqn}, \ref{sigma_lowz_eqn}, and \ref{sigma_intz_eqn}) to derive the total stellar mass that corresponds to $\theta_E=1$ arcsec for these models.

\begin{align}
M_* &= \left(\frac{A-1}{A}\right)^{\frac{-1}{0.586}}(8.52\times10^{10}) & z &< 0.2 \\
M_* &= \left(\frac{A-1}{A}\right)^{\frac{-1}{0.562}}(9.26\times10^{10}) & z &\geq 0.2 
\end{align}

This results in lines of constant stellar mass at which, for the given redshift range and ratio $A= \frac{D_S}{D_L}$, $M_*=M_{fr}$ and the Einstein radius is 1 arcsecond.

For a given ratio of source distance to lens distance, $\theta_E$ estimates of candidates that fall to the right of the associated curve will exceed the radius of the GAMA spectroscopy fiber (1"). Multiple candidates from the other two methods have stellar masses exceeding the curves for given ratios $A$. This indicates that GAMA spectroscopy may have little chance of observing enough flux from the background source galaxy to acquire the second redshift match required for identification.
It is important to note that this is not a hard limit on the \textit{possibility} of lens candidate identification. The geometry and surface brightness of the features rarely resemble the perfect Einstein ring whose features lie exactly along the Einstein radius. These features extend, stretch, and become asymmetric across the center of the galaxy and can result in sufficient flux for background source detection. In addition, the placement of the aperture is not always precisely centered on the object. These possibilities in addition to chance alignment can and do allow the detection of lensing objects with Einstein radii larger than the fiber. However, as we have previously stated, the probability of detection diminishes quickly when the Einstein radius extends beyond the fiber radius.

\subsection{Low-Mass Lensing Galaxies}\label{section_lowmass}

GalaxyZoo citizen science and GAMA spectroscopy found an encouraging number of candidates at lower masses than those identified through the LinKS machine learning method considered here, as well as many previous lens studies that focused on galaxies fitting the description of SDSS LRGs \citep{Eisenstein01}. These candidates could potentially be analyzed in order to study the structure of galaxies and their dark matter content in this lower range of galaxy mass. However, each arc from GalaxyZoo will need to be confirmed spectroscopically before modeling these as lensing systems, and more samples are needed in order to study this parameter space with meaningful results.

\subsection{Candidates and LRG Color-Magnitude Selection}\label{section_candidateLRG}

The color-magnitude selection criteria for SDSS-LRGs detailed in Section \ref{section_selection} were applied (with the empirical factor set to 1 to represent the most lenient selection) to GAMA Lambdar AB photometry of the final candidate samples of GAMA spectroscopy, LinKS machine learning, and GalaxyZoo citizen science methods.

Only three of the 47 GAMA spectroscopy candidates pass the LRG cuts. All but one candidate has a magnitude in the r-band $< 20$. 43 candidates fail the $c_{par}$ color parameter criterion, which represents a sliding luminosity threshold as a function of redshift, by a mean of 1.3 and median of 1.2 mag. This could possibly indicate that there are more strong lenses to be found by broadening the search beyond the assumption of LRG dominance, a result also found in \cite{Li20c}.

39 of the 47 LinKS machine learning candidates pass the LRG cuts. 3 are at $z > 0.4$, and therefore do not follow the criteria. The remaining 5 fail the criterion relating to the $c_{par}$ color parameter. These candidates all lie in the redshift range $z \sim 0.25-0.3$ and fail the color selection criterion by a mean of 0.47 and median of 0.26 mag. \cite{Eisenstein01} points out that the LRG color-magnitude selection becomes less appropriate as object redshift decreases well below $z < 0.4$. One of these candidates is G124486, one of the candidates that overlaps with the GalaxyZoo sample at a redshift of $z = 0.144$. Note that these 47 candidates were all originally selected using the same color-magnitude criteria applied to a different set of photometric measurements.

8 of 13 GalaxyZoo citizen science candidates pass the LRG cuts. The 5 candidates fail the $c_{par}$ color parameter criterion by a mean of 0.21 and median of 0.11 mag. As mentioned above, these LRG cuts are not well-suited to the redshift range of GalaxyZoo candidates. This could possibly indicate that they are less luminous or less red, further evidence of viable strong lens candidates outside the realm of the LRG assumption. Alternatively, these candidates may pass LRG cuts that are adjusted to be more appropriate to the redshift range. \cite{Petrillo18} makes no mention of any adjustment to color-magnitude selection criteria for galaxies at $ z < 0.15$.

\subsection{Other Lens Searches and Future Efforts}

\begin{figure*}
    \centering
        \includegraphics[width=0.49\textwidth]{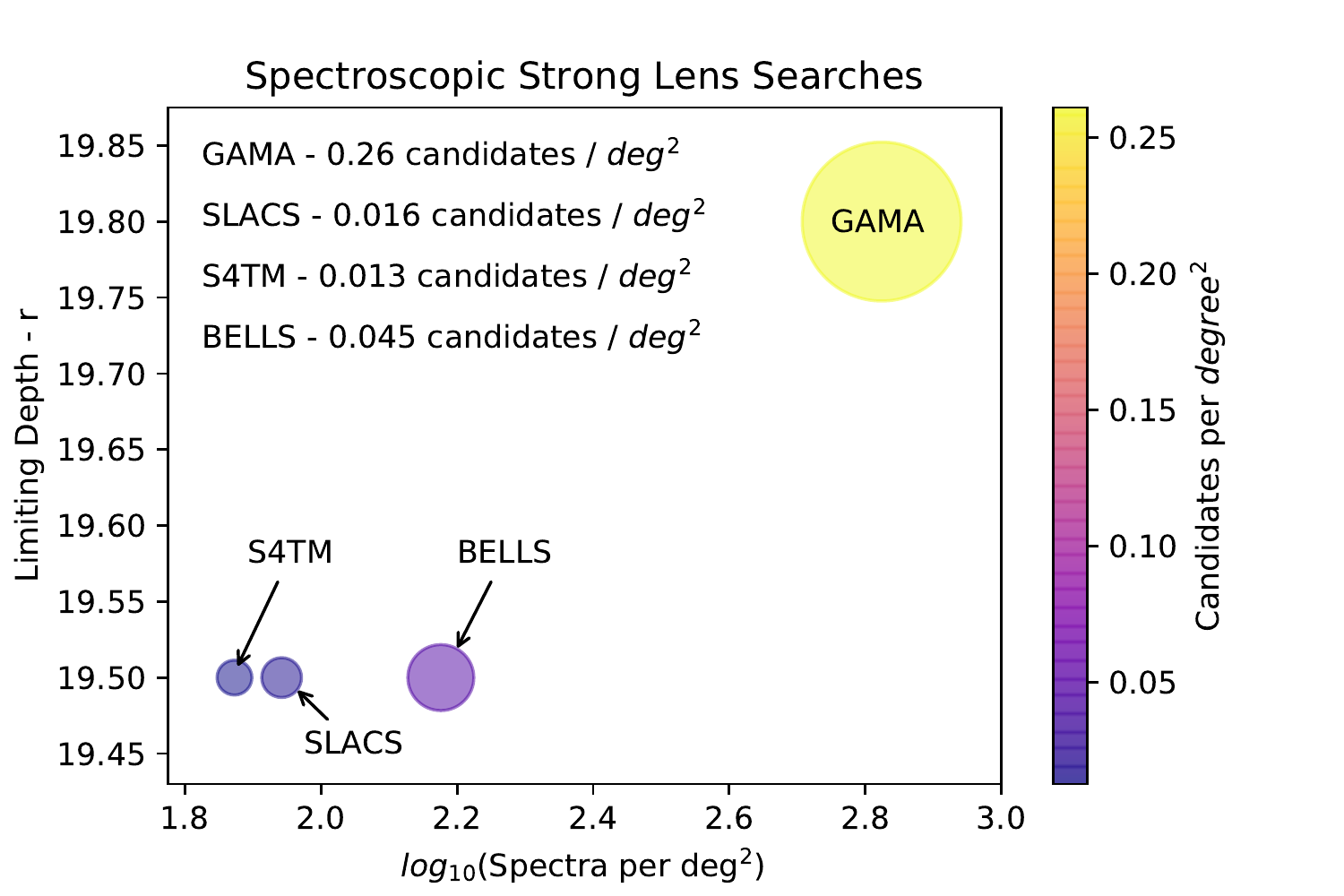}
        \includegraphics[width=0.49\textwidth]{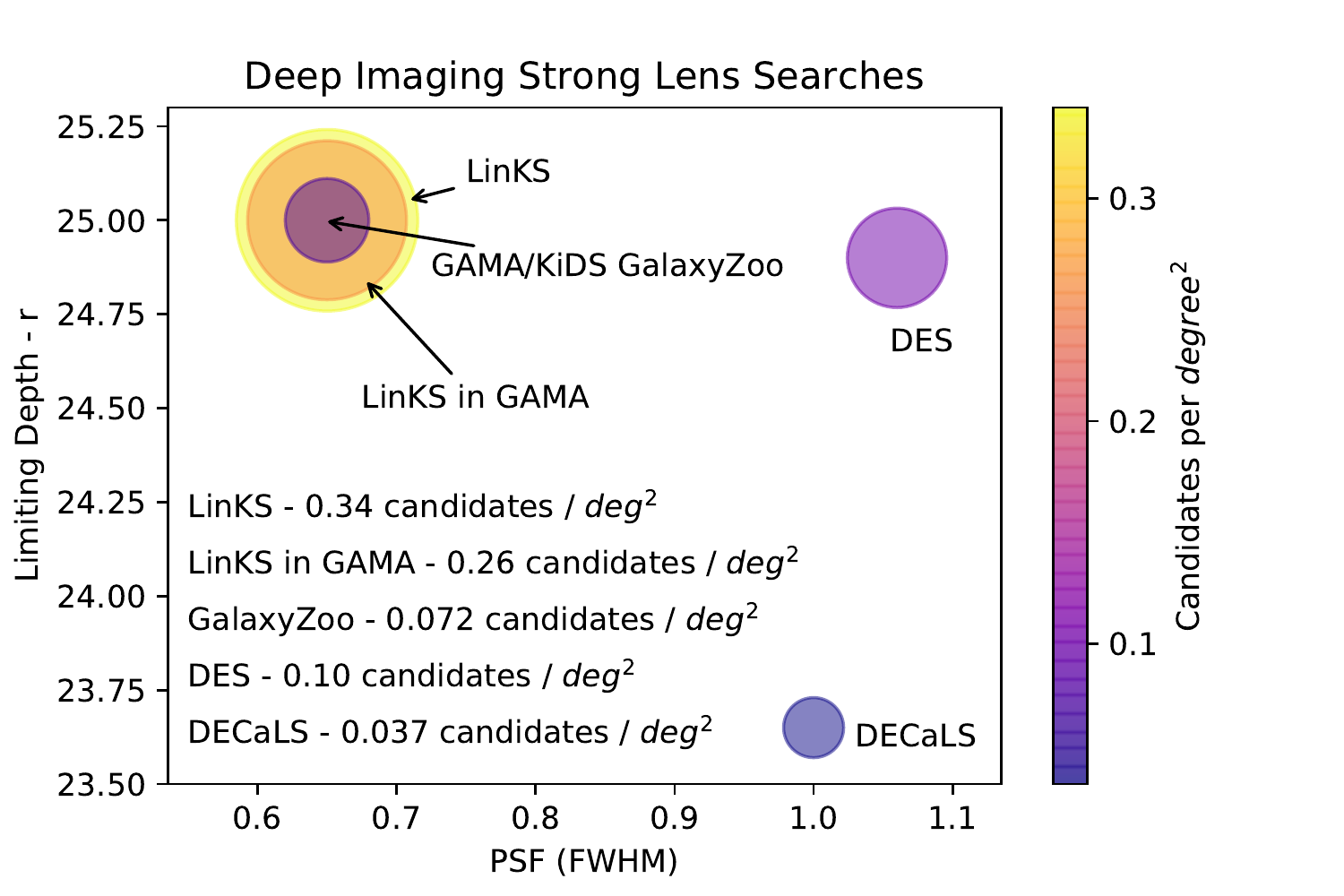}
    \caption{A comparison between the different identifications of strong galaxy-galaxy lens candidates in different surveys, using blended spectroscopy (left) and deep imaging (right). The left figure shows SLACS \protect\citep{slacs1}, S4TM \protect\citep{Shu17} from SDSS-DR7 \citep{SDSS-DR7} and BELLS \protect\citep{Brownstein12} spectroscopic searches compared to the GAMA spectroscopy candidates selected in this work from \protect\citep{Holwerda15}. The right figure shows the LinKS survey from \protect\cite{Petrillo18} and the candidates presented in this paper from GAMA/KiDS GalaxyZoo citizen science, compared with deep imaging searches in DES \protect\citep{Jacobs19} and DECaLS \citep{Huang20b}. We note that SDSS spectroscopic searches aimed for clean rather than complete samples (requiring high signal-to-noise) and GAMA emphasizes completeness throughout the survey. S4TM and SLACS report the number of observed candidates for an HST Snapshot program which implies a longer candidate list. The candidate number per unit area is therefore a lower limit.   }
    \label{fig:survey_success}
\end{figure*}

Similar efforts, mostly machine learning, have found strong lens candidates in deep imaging surveys \citep[e.g.,][in HSC, DECaLS and DES respectively]{Speagle19,Huang20b,Jacobs19}. Figure \ref{fig:big_plot} shows the GAMA equatorial lens candidates of this work and includes two candidates previously identified by DECaLS, SLACS \citep{Bolton08} in the GAMA equatorial regions that had a match in RA/DEC to the GAMA catalog. Both of these candidates were identified by GalaxyZoo citizen science, and one was also identified by LinKS machine learning. This lends confidence to GalaxyZoo's ability to return high quality lens candidates. Figure \ref{fig:survey_success} shows the numbers of identified lens candidates per unit of sky against the survey characteristics. The majority of theses studies aimed for a clean (reliable but incomplete) rather than complete sample of galaxy-galaxy lenses. Our results show that the combination of methods implies a much higher sky density of lenses in a given survey. Key for spectroscopic survey identification is completeness, a key driver of GAMA survey \citep{Driver09,Robotham10,Baldry12}, and an automated redshift finder \citep{Baldry14}. For imaging surveys, the key drivers are spatial resolution (subarcsecond seeing) and survey depth. The on-sky density of strong lenses is critical if one wants to estimate, for example, the rates of strongly lensed events such as supernovae (Holwerda et al. {\em submitted}).

\subsubsection{\cite{Li20c} Machine Learning ``Bright Galaxy" Candidates}

\begin{figure}
    \centering
    \includegraphics[width=0.45\textwidth]{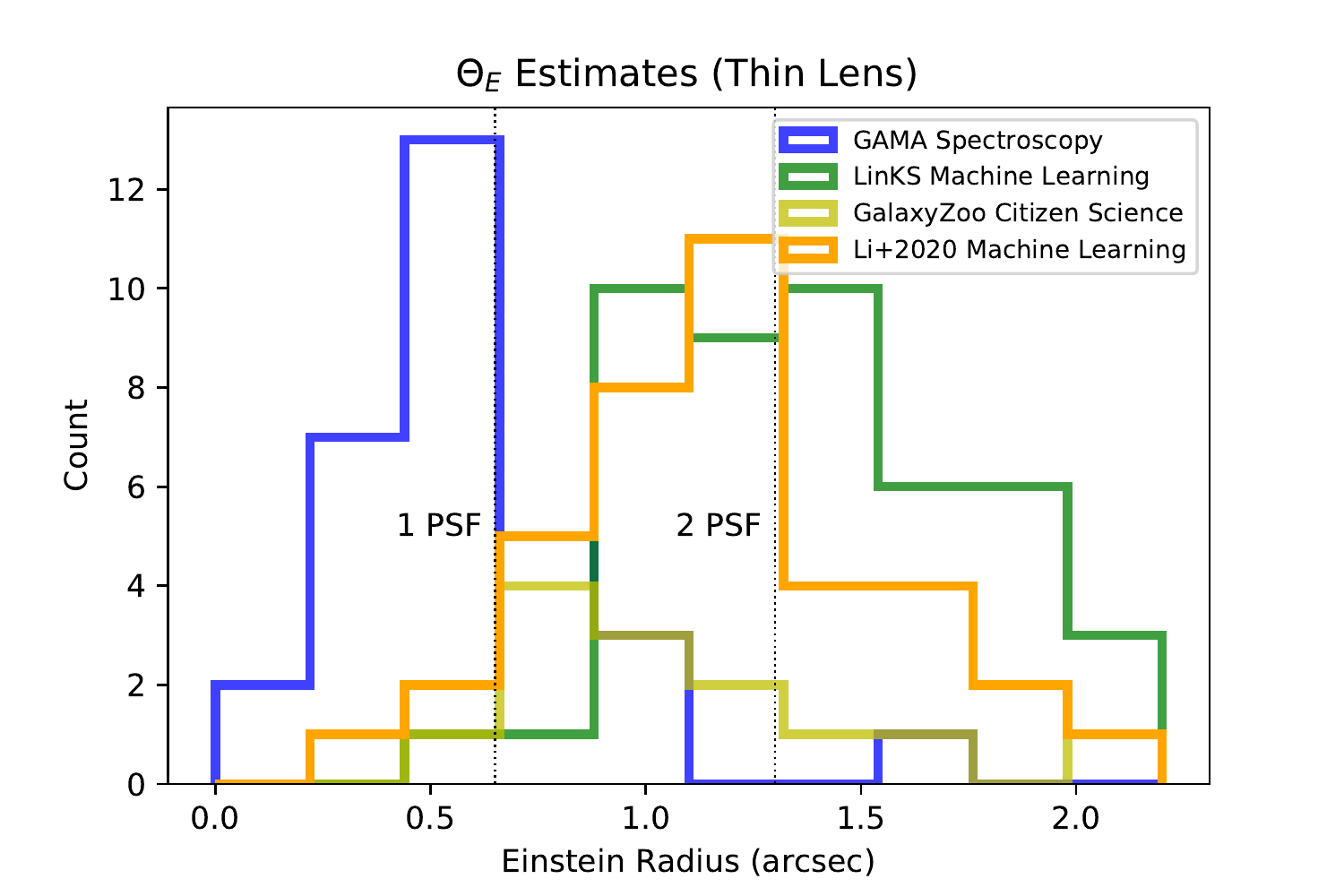}
    \includegraphics[width=0.45\textwidth]{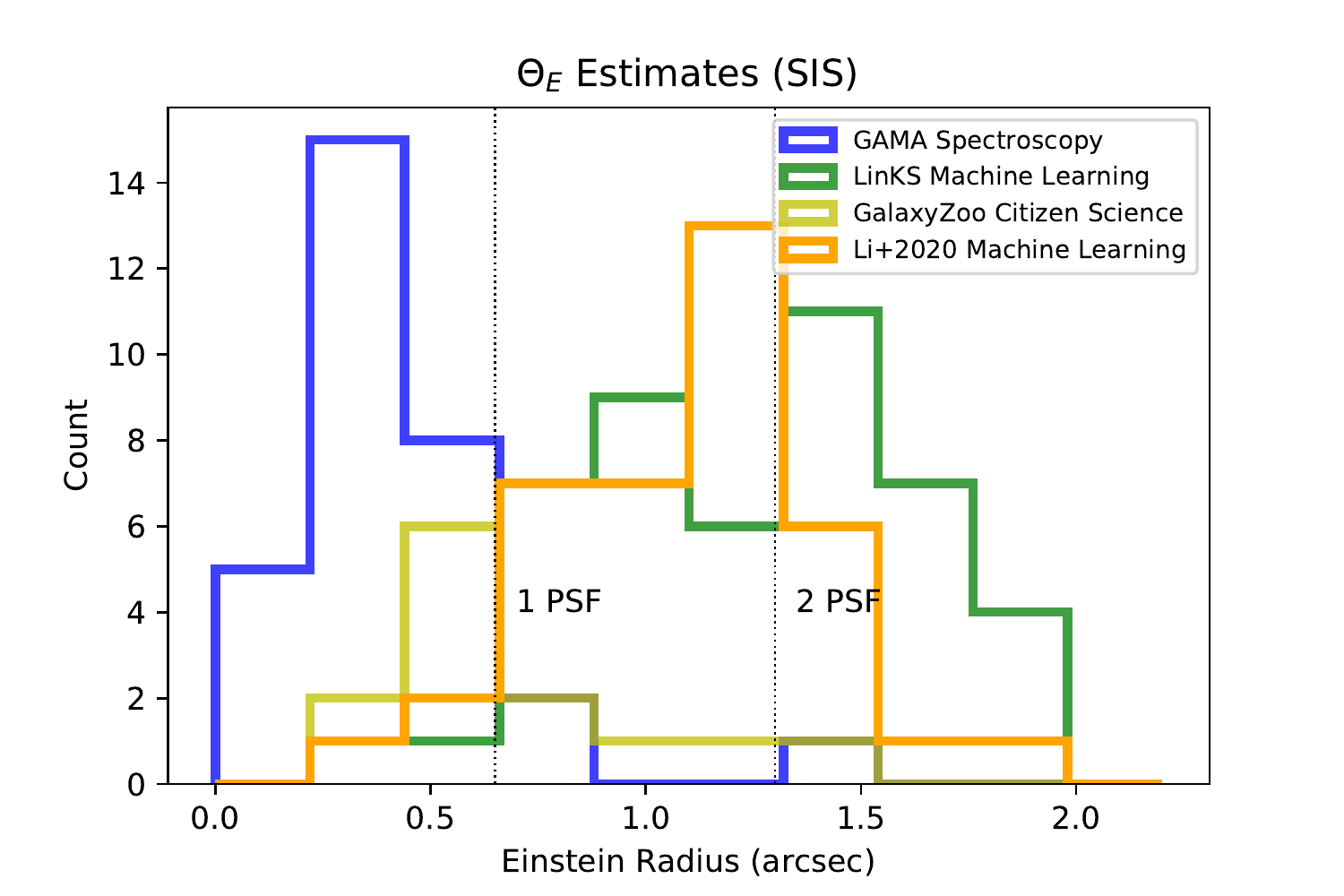}
    \includegraphics[width=0.6\textwidth]{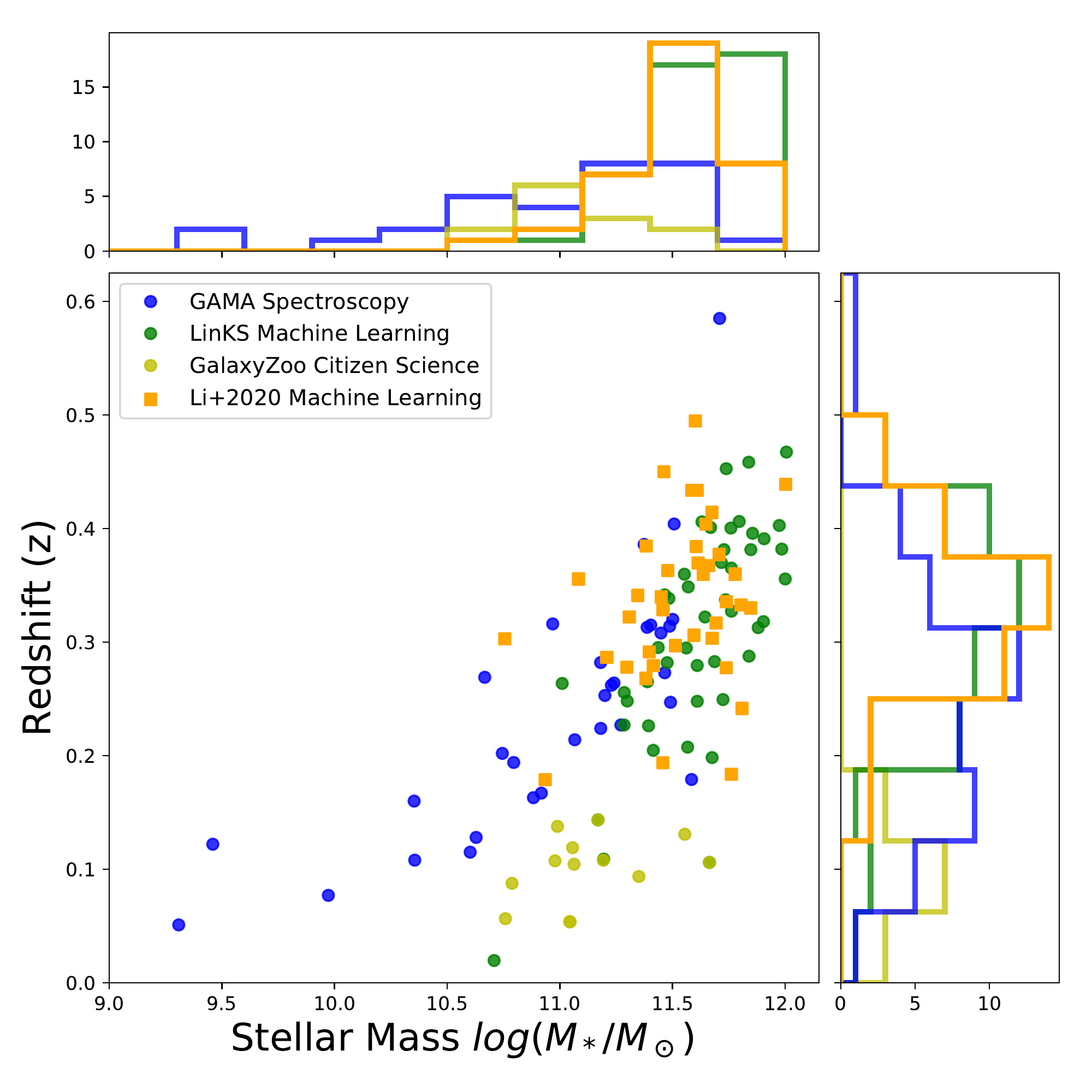}
    \caption{(Top) Thin-lens and SIS Einstein radius estimates for ``Bright Galaxy" candidates from \cite{Li20c} in comparison with the three methods examined in this study and (Bottom) stellar mass--redshift space for the same candidates.}
    \label{fig:li}
\end{figure}

During the preparation of this paper, \cite{Li20c} published the results of their CNN's selection of strong lens candidates in the KiDS DR4 data using a technique similar, though slightly modified from the LinKS machine learning method. They offer qualitative comparison to \cite{Petrillo18} and \cite{Petrillo19}, noting that their application of the CNN to an expanded sample of ``bright galaxies" (BGs) without LRG cuts doubled the number of quality lens candidates identified by their algorithm.  Their result supports this work's assertion that the expansion of machine learning's scope beyond the standard assumption of LRGs will improve the completeness of such automated lens-finding methods. \cite{Li20c} further note their intention to follow with more quantitative comparison with the LinKS candidates in a future paper, as well as commenting on the value of comparison with other lens-finding methods. 48 of their BG candidates fall within the 180 $\textnormal{deg}^2$ considered in this study and have a match by RA/DEC to an object in GAMA. 39 of these have a reliable stellar mass estimate from GAMA {\sc LAMBDAR} photometry, and we apply the same procedure for Einstein radius estimatation outlined in Section \ref{section_theta_e_estimates}. The BG candidates of \cite{Li20c} are shown in Figure \ref{fig:li} in comparison to the candidates considered in this study. They tend to lie within a similar region of parameter space as the LinKS machine learning candidates we have considered in this paper with $z \sim 0.2 - 0.5$ (mean and median 0.33) $\log(M_*/M_{\odot})\sim 11-12$ (mean 11.55, median 11.61). These candidates do not overlap with any candidates identified here. The follow-up study of that data in comparison with the data from this study in more detail is worth conducting.

\subsubsection{Hubble and Future Surveys}

\begin{figure*}
    \centering
    \includegraphics[width=0.45\textwidth]{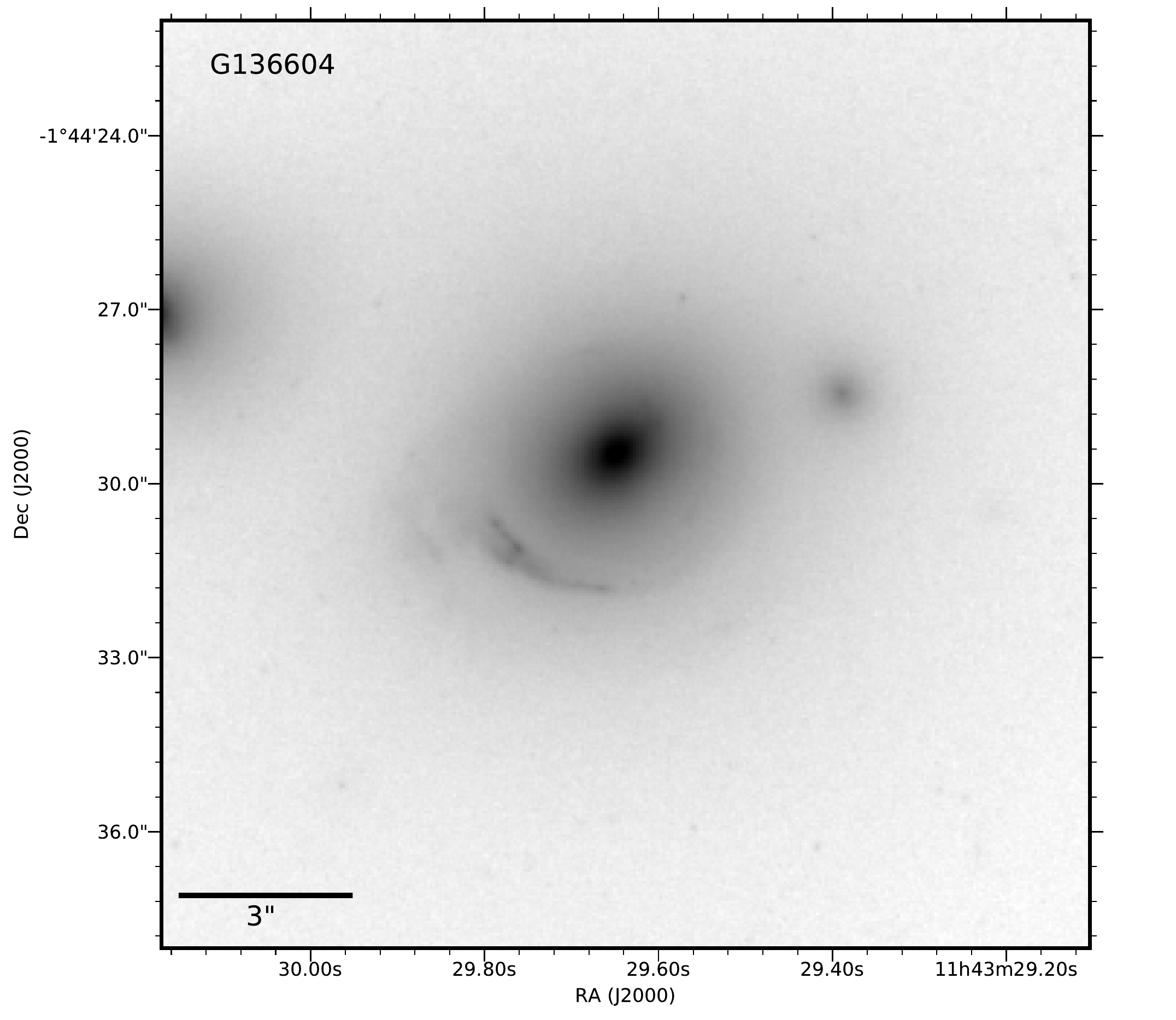}
    \includegraphics[width=0.45\textwidth]{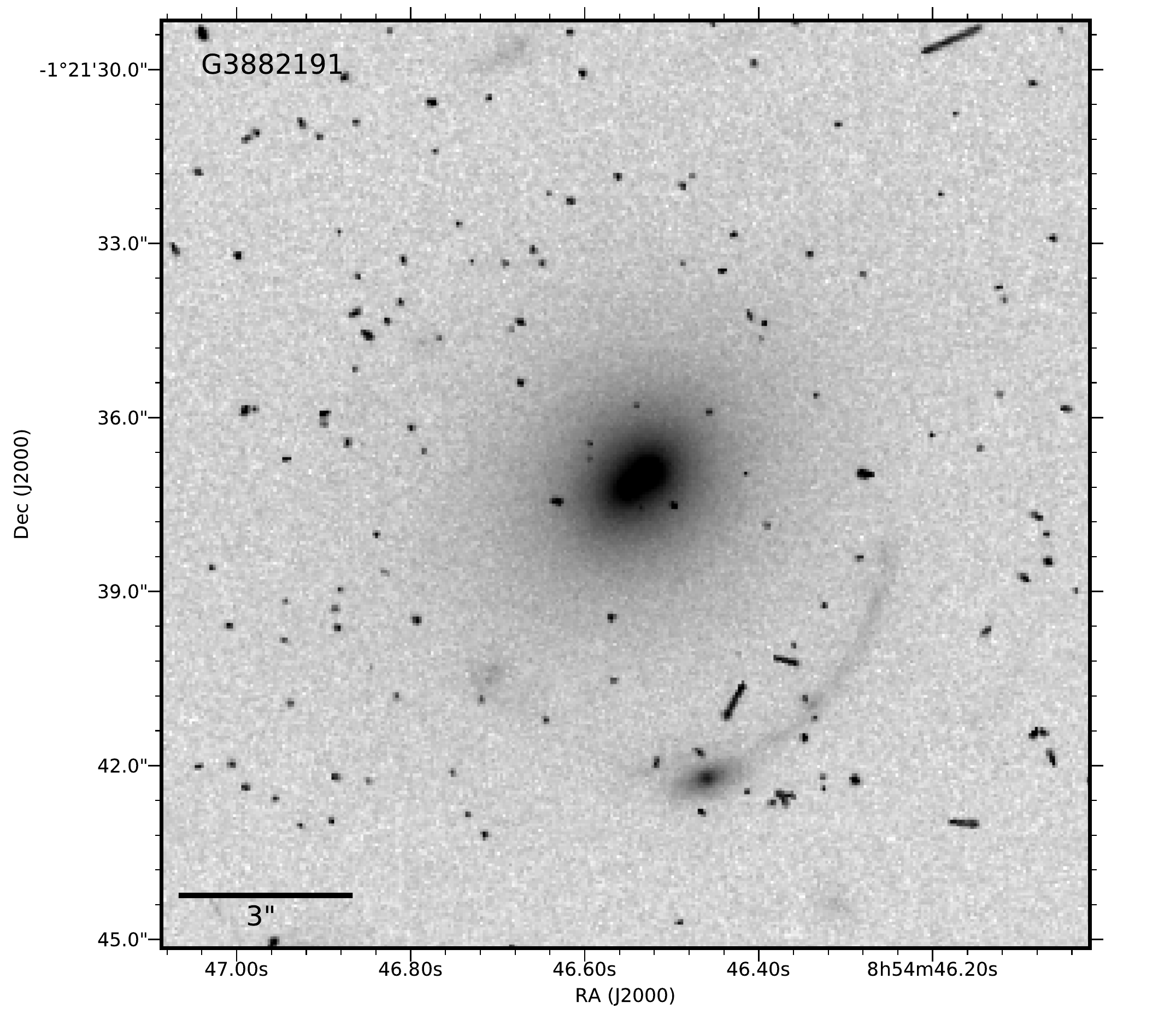}
    \includegraphics[width=0.45\textwidth]{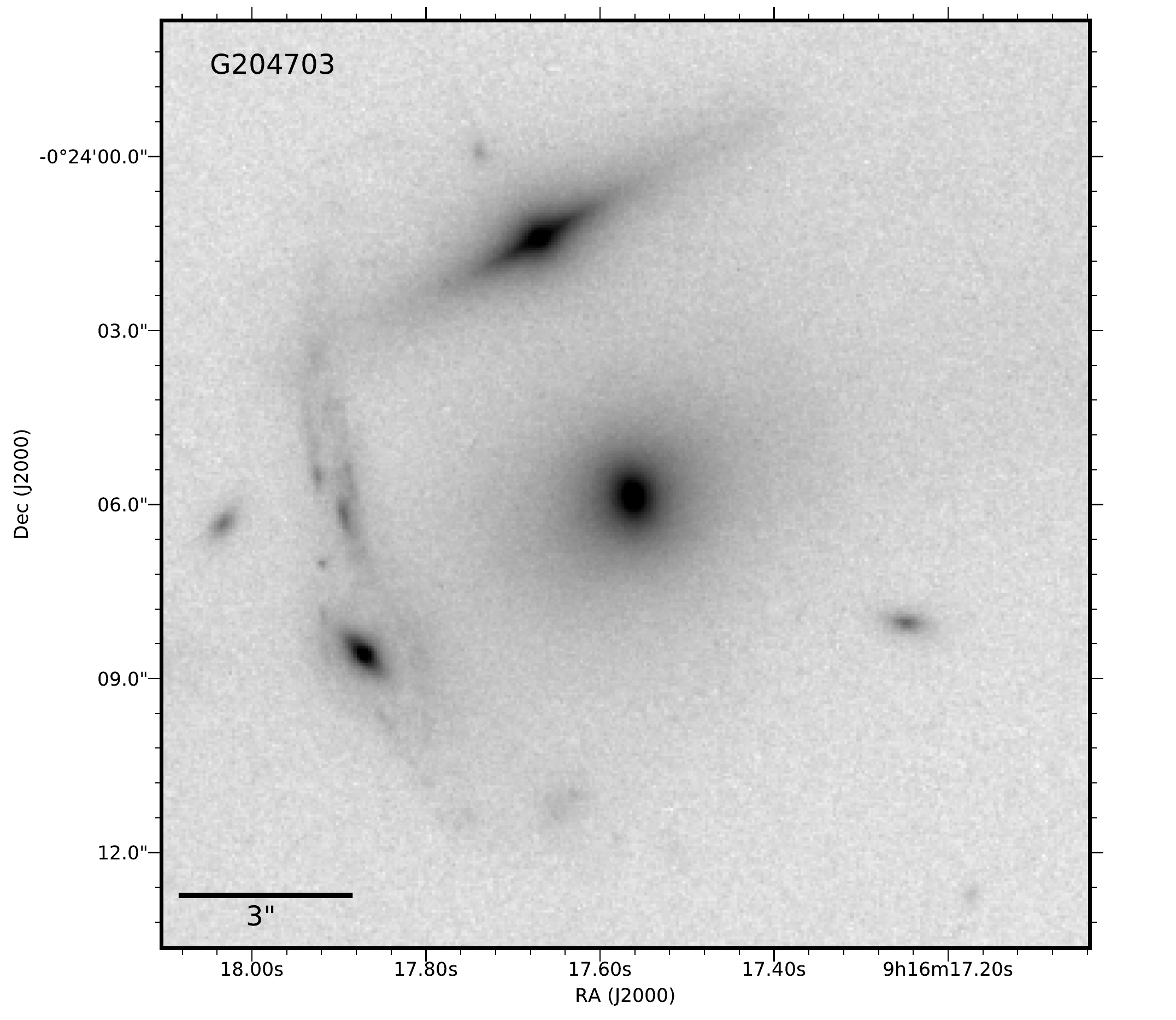}
    \includegraphics[width=0.45\textwidth]{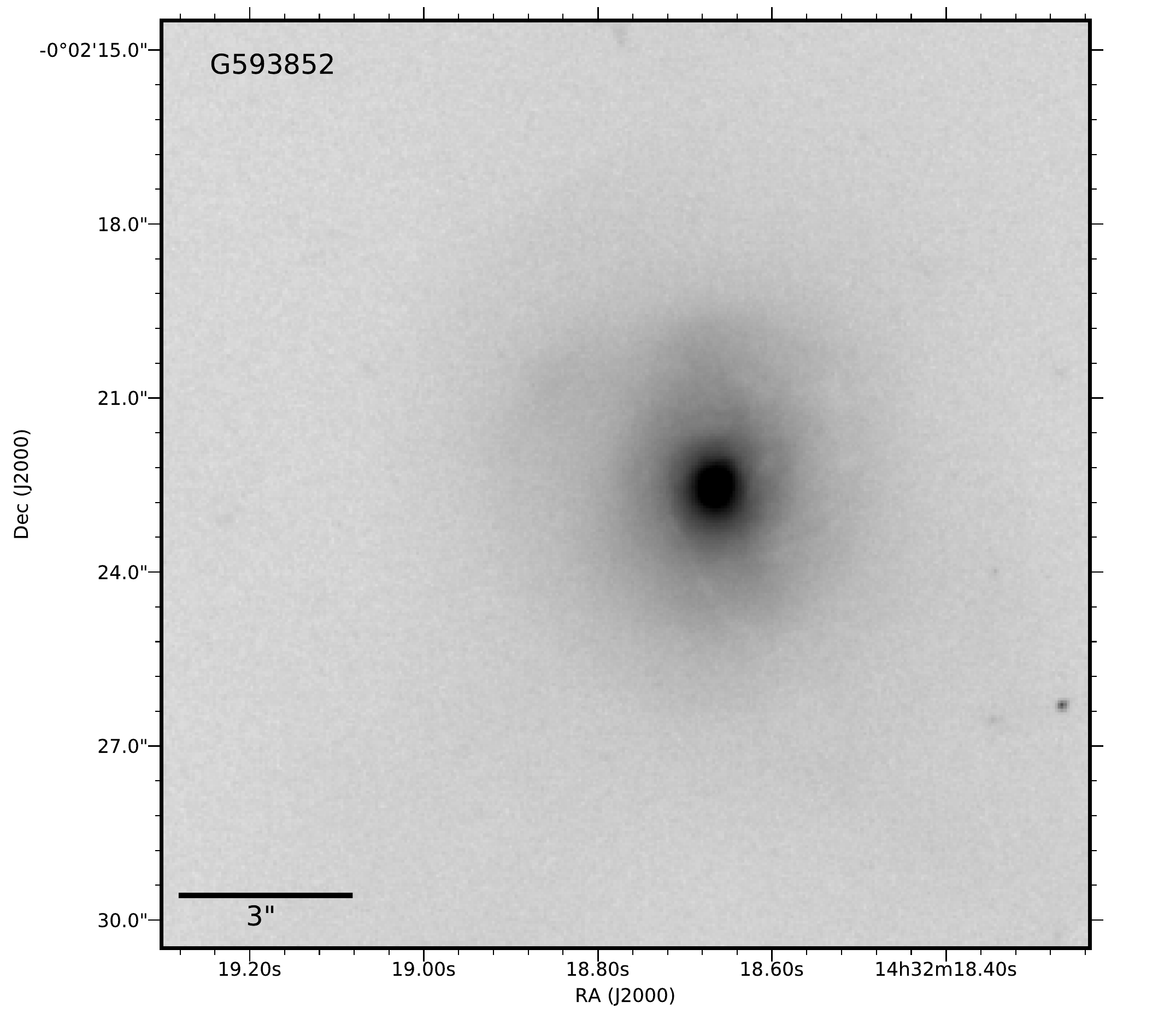}
    \includegraphics[width=0.45\textwidth]{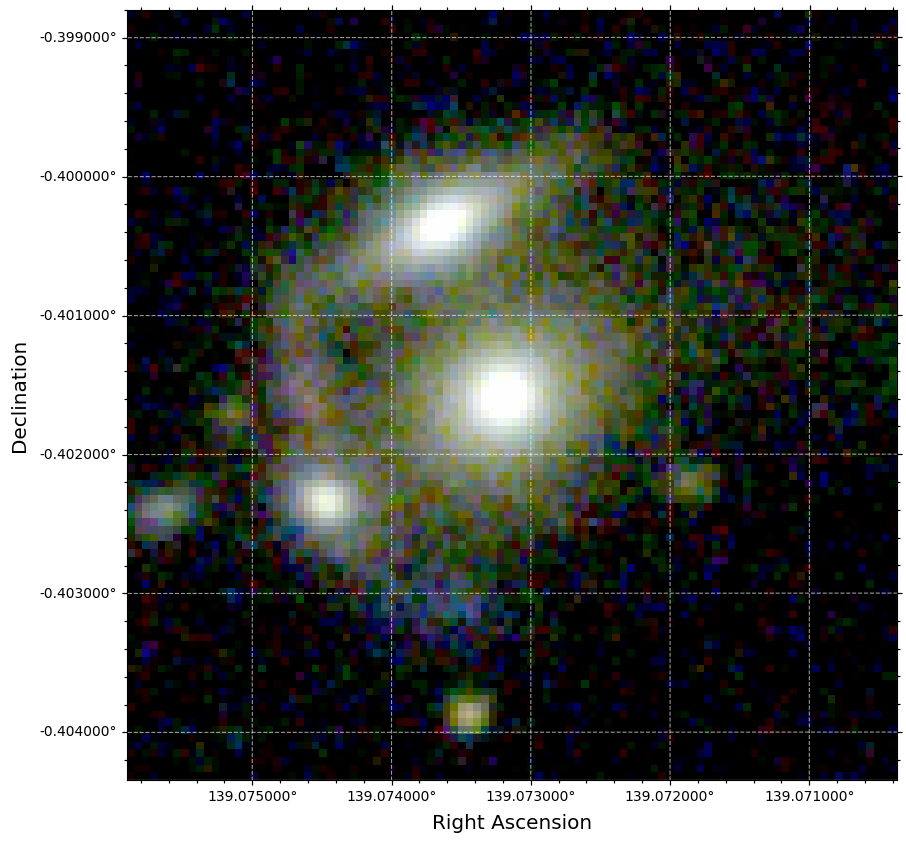}
    \includegraphics[width=0.45\textwidth]{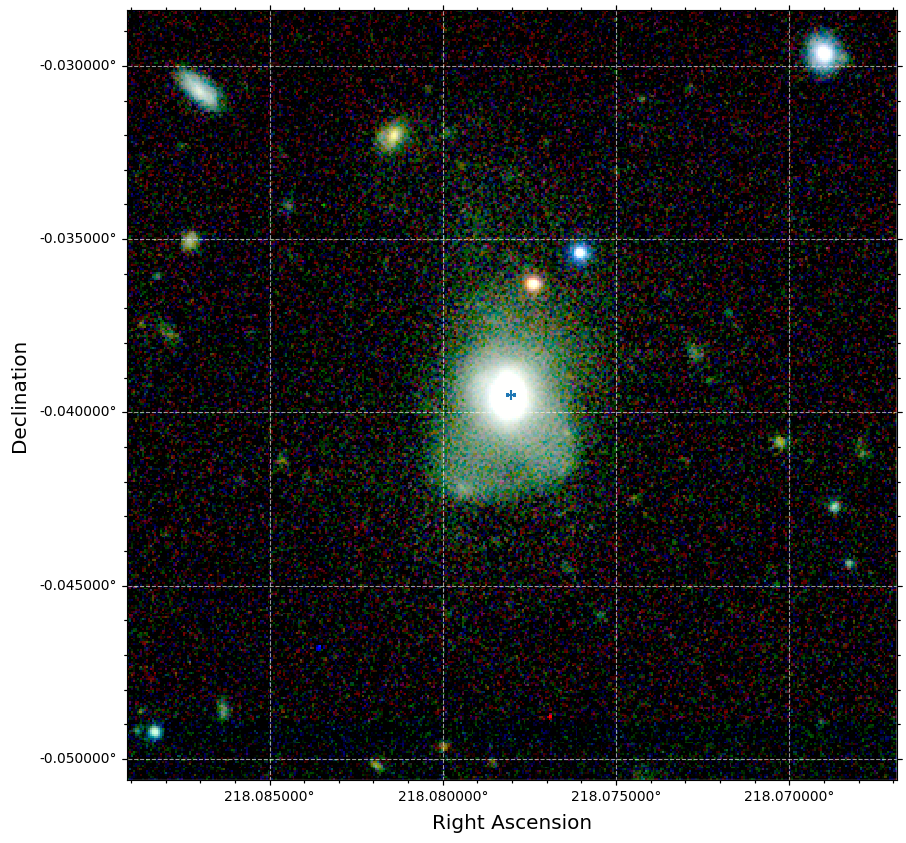}
    \caption{Examples of lensing candidates found in the Hubble Legacy Archive (HLA). Grayscale image in F814W, except G593852, which is in F625W. F625W is significantly narrower than F814W, which results in the lower contrast of the image of G593852. G593852 and G204703 are presented next to their corresponding KiDS cutout images. G136604 was identified by SLACS, LinKS machine learning, and GalaxyZoo citizen science. The arc shown next to G204703 may be due to cluster lensing. G3882191 was previously identified as a strong lensing galaxy group in SL2S \citep{Limousin09,Limousin10} and is recovered by LinKS machine learning.}
    \label{fig:hla}
\end{figure*}

The Hubble Legacy Archive (HLA) \citep{Whitmore16} contains imaging information on four of the lensing candidates presented here in Table \ref{t:hla} and Figure \ref{fig:hla}. G136604 was identified in both the LinKS machine learning and GalaxyZoo citizen science selections (KiDS cutout shown also in Figure \ref{fig:mac_zoo_overlap}, left) and was a known strong lens from the SLACS survey \citep{slacs7,slacs10,Barnabe11,Bolton12,Czoske12,Shu15}. The second example, G3882191, was identified as a strong lensing galaxy group in SL2S \cite{Limousin09,Limousin10} and has been recovered by LinKS machine learning. Because G3882191 is located at the center of a massive dark matter halo associated with a galaxy group, it has a particularly large measured Einstein radius as a result ($5.48 ^{\prime\prime}$). This is much greater than the estimates we compute from GAMA data ($\sim 2^{\prime\prime}$) using the adopted empirical fits to the lensing parameters as described in Section \ref{section_theta_e_estimates}. This perhaps exposes a limitation on the analysis we have applied to these candidates if they are located at the center of a much larger lensing galaxy group. On the other hand, the LinKS method has correctly identified this as a lensing system, showing that the selection based on Einstein radius is still valid for this candidate. Follow-up could examine more closely the role of environment in strong lensing systems, but this consideration is outside the scope of this paper. G204703 is part of a cluster, and the arc is partially the result of cluster strong lensing \citep{Zitrin16,Zitrin17,Dessauges-Zavadsky17a,Newman18b}. One of the GalaxyZoo identifications, G593852, is in the HLA, but the arc is faint in the Hubble single filter image (Figure \ref{fig:hla}). These images illustrate that these two selection methods find valid strong lens candidates and show that Hubble images on all lens candidates could validate their nature and make detailed lensing models possible. 

\begin{table}
\caption{Strong lenses with archival Hubble data available from the Hubble Legacy Archive.  }
\begin{center}
\begin{tabular}{l l l l}
GAMAID & RA & DEC & Method \\
\hline
G136604 & 175.87349 & -1.74167 & ML \& GZ \\
G204703 & 139.07321 & -0.40157 & ML \\
G3882191 & 133.69397 & -1.36032  & ML \\
G593852 & 218.0782 & -0.03958 & GZ \\
\hline
\end{tabular}
\end{center}
\label{t:hla}
\end{table}%

With improved spatial resolution for wide-field imaging (WFIRST and EUCLID), the machine learning and citizen science identification will overlap more in mass and redshift with the spectroscopic identifications of strong lens candidates. Wide-field integral field spectroscopy (e.g. SAMI, MUSE, and 4MOST WAVES wide surveys) could potentially identify strong lenses from a mix of spectroscopic and imaging information (curved arc in the data-cube). Future large telescopes (GMT/ELT/TMT) could follow-up the higher redshift or lower mass lens candidates, and the WEAVE and 4MOST instruments are set to select or confirm many candidates spectroscopically.

\begin{figure}
    \centering
    \includegraphics[width=0.45\columnwidth]{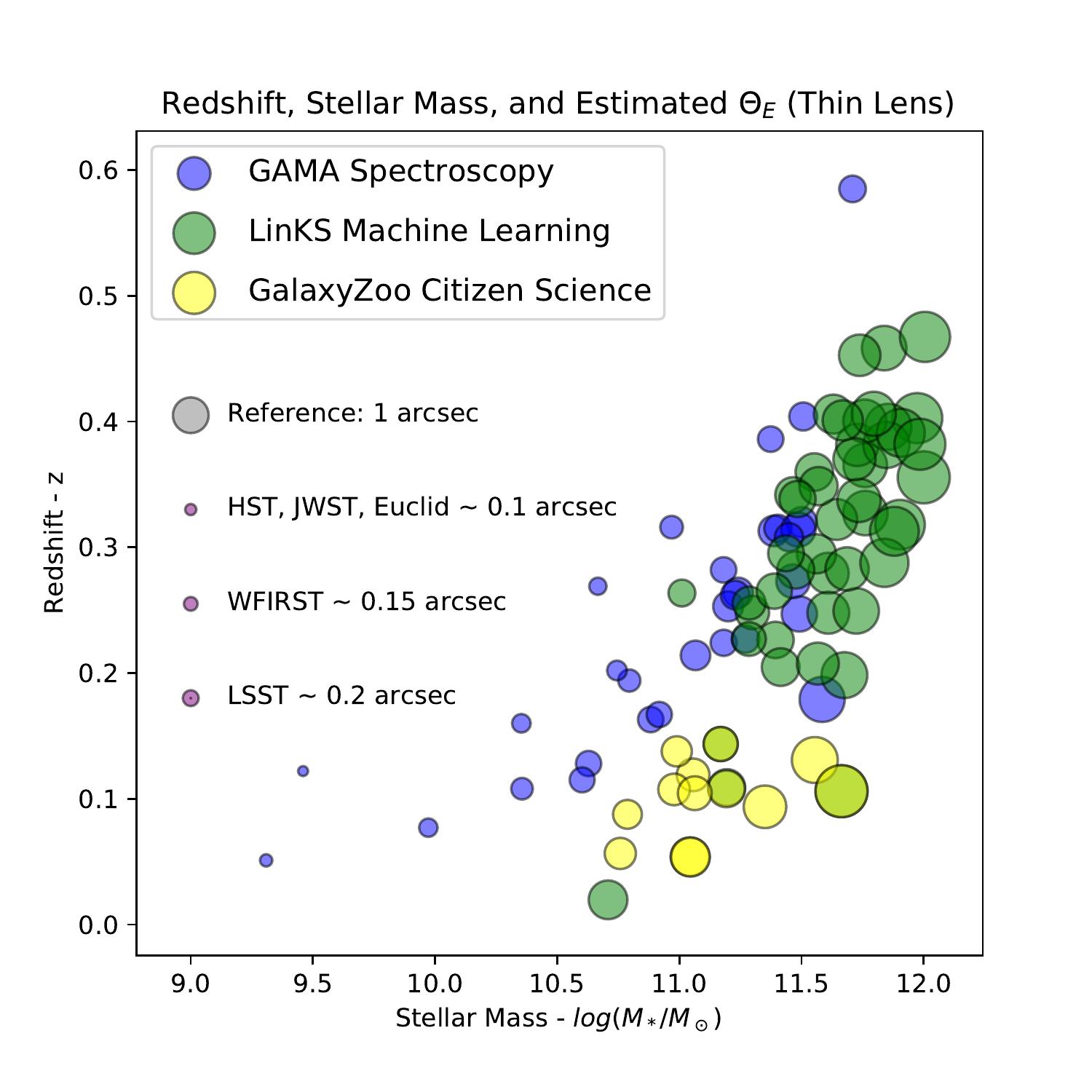}
    \includegraphics[width=0.45\columnwidth]{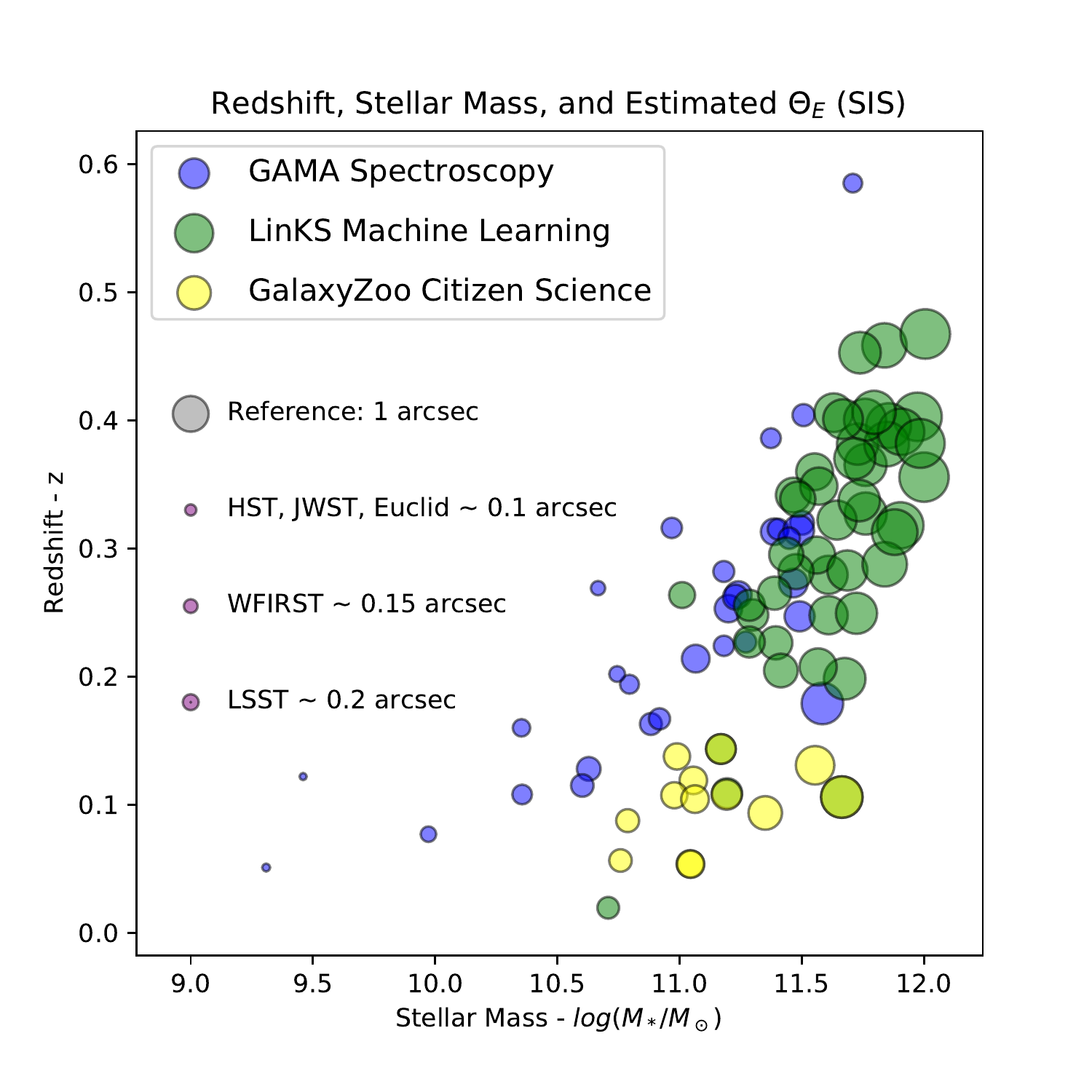}
    \caption{The data presented here is the same as in Figure \ref{fig:big_plot} with marker-size corresponding to estimated Einstein radius. The largest of estimated Einstein radii tend to be at higher mass and lower lens redshift. The gray reference marker indicates the benchmark of $\theta_E = 1$ arcsecond (corresponding to the aperture-size of GAMA spectroscopic fibers). Next generation surveys (JWST, Euclid, WFIRST, LSST) will have optical resolutions comparable to HST, on the order of 0.1-0.2 arcseconds. In theory, under the right conditions, these surveys should be able to resolve features from even some of the smallest estimated Einstein radii considered in this survey.}
    \label{fig:theta_e_mass_redshift}
\end{figure}

For future identification of galaxy-galaxy lens candidates (e.g. LSST or WFIRST imaging), a hybrid approach between citizen science and machine learning \citep[see][]{Beck18} could result in a more complete and reliable yield of strong lensing candidates. This is in our view a better approach than to try to validate galaxy-galaxy lens candidates using different methods. 
A worthwhile next step will be to design and implement a CNN with a more complete training set using our findings and the work of \cite{Petrillo19} and \cite{Li20c} to experiment with its application to KiDS imaging and HST archival data and to prepare for future wide-field imaging surveys. Figure \ref{fig:theta_e_mass_redshift} shows that several of the candidates unidentified by LinKS machine learning could be recovered in these deeper surveys that will be able to resolve the features of lenses with even the smallest of estimated Einstein radii from this study. We believe the deliberate expansion of the training set to include lenses with color features beyond the LRG assumption will improve the completeness a step further than that achieved by \cite{Li20c}. Detection and modeling of these smaller lenses may reveal insights into dark matter fractions in the lower mass regime. Alternatively, an entirely new selection technique could result in a selection of similar completeness as the three presented here.

\section{Conclusions}

Our analysis of the data generated by the three methods leads us to the following conclusions regarding the biases and advantages of the methods as well as the relation of each to the others:
\begin{itemize}
    {\item The three specific methods analysed here are ineffective means of vetting candidate samples obtained by either of the other two methods due to the lack of overlap between candidate properties, which we illustrate in Figures \ref{fig:venn_hardcut} and \ref{fig:big_plot}.
    \item The differences in candidate properties arise primarily from selection functions inherent to the parent sample selections and procedures of each method; specifically:
    \begin{itemize}
        \item Blended spectra identifies only those candidate galaxies whose lens features contribute significant flux within the radius of the fiber aperture collecting the spectrum. GAMA's depth and completeness extend candidate selection to a lower mass range than those identified by previous strong lens surveys, including those conducted within SDSS. A wider aperture or integral field spectroscopy (e.g. SAMI or MUSE) would allow the possibility of identifying lower redshift or more massive galaxies than the GAMA spectroscopy candidates examined here.
        \item Machine learning finds candidates with similar features to its training sets, i.e. well-separated lens and arc. This limitation can be improved upon with higher-resolution images and a wider variation in training sets, including lens galaxies that do not conform to LRG characteristics.
        \item GalaxyZoo's upper threshold on redshift limits its range of applicability, and the wide focus of its classification stage introduces significant challenges to candidate quality assessment. Higher-resolution images and a higher redshift cutoff together would allow for the inclusion of more distant galaxies, e.g. for WFIRST, LSST, or Euclid citizen science efforts.
    \end{itemize}
    \item Machine learning has the promise to be the most efficient automated lens identification technique, and substantial effort must be made to improve these algorithms ahead of the next generation of galaxy surveys.
    \item In the meantime, other methods -- including blended spectra and citizen science -- remain useful for extending and diversifying the catalog of lens candidates available for study and for their application in the training of machine learning algorithms}.
\end{itemize}

Strong gravitational lenses offer a unique laboratory to be used in cosmology measures and the next level of accuracy in estimates of dark matter distribution and substructure. The much greater samples needed (orders of magnitude higher than the current number of identified lenses) require automated identification. Using a combined systematic approach to improve the scope of machine learning's applicability, a more complete census of these rare objects can be achieved for the next generation of imaging surveys (e.g. WFIRST, Euclid, and LSST). 

\section*{Acknowledgements}


S. Knabel acknowledges the support of the Summer Research Opportunity Program (SROP) and the Undergraduate Research Grant (URG) by the University of Louisville's Office of the Executive Vice President for Research and Innovation (EVPRI). 
The material is based upon work supported by NASA Kentucky under NASA award No: NNX15AR69H (R. Steele). 
This work was supported by a NASA Keck PI Data Award, administered by the NASA Exoplanet Science Institute. 
M. Bilicki is supported by the Polish Ministry of Science and Higher
Education through grant DIR/WK/2018/12, and by the Polish National
Science Center through grants no. 2018/30/E/ST9/00698 and
2018/31/G/ST9/03388.

\section{Appendix}
%
%
%

\begin{table}[htp]
\caption{GalaxyZoo Citizen Science Lens Candidates}
\begin{center}
\begin{tabular}{llllllll}
\hline
GAMA ID   & Lens Score & $\log(M_*/M_\odot)$ & z & $\theta_{E,TL}$ (arcsec) & $\theta_{E,SIS}$ (arcsec) & RA (deg)  & DEC (deg)  \\ 
\hline
        485873 & 34.95\% & 11.04 & 0.054 & 1.177 & 0.582 & 217.75 & -1.80 \\ 
        84050 & 36.58\% & 11.06 & 0.119 & 0.837 & 0.592 & 175.80 & 0.48 \\ 
        55245 & 31.41\% & 11.35 & 0.094 & 1.422 & 0.881 & 181.08 & -0.32 \\ 
        70282 & 65.54\% & 10.98 & 0.107 & 0.783 & 0.533 & 179.40 & 0.13 \\ 
        185451 & 30.30\% & 11.19 & 0.108 & 1.064 & 0.712 & 180.28 & -1.61 \\ 
        124486 & 42.62\% & 11.17 & 0.144 & 0.909 & 0.690 & 179.73 & -2.52 \\ 
        593852 & 38.63\% & 11.55 & 0.131 & 1.647 & 1.160 & 218.08 & -0.04 \\ 
        93310 & 57.51\% & 10.99 & 0.138 & 0.714 & 0.541 & 219.92 & 0.51 \\ 
        136604 & 31.65\% & 11.66 & 0.106 & 2.112 & 1.345 & 175.87 & -1.74 \\ 
        600421 & 42.02\% & 10.76 & 0.057 & 0.764 & 0.396 & 135.49 & 0.28 \\ 
        574423 & 46.43\% & 11.05 & 0.054 & 1.184 & 0.584 & 135.76 & -0.20 \\ 
        324764 & 42.28\% & 11.06 & 0.104 & 0.895 & 0.598 & 137.20 & 1.73 \\ 
        342699 & 45.71\% & 10.79 & 0.088 & 0.651 & 0.412 & 216.90 & 2.13 \\ 
        
\hline

\end{tabular}
\end{center}
\label{GZ_table}
\end{table}%

\subsection{Alternate Cuts}
{\label{section_alternate_cuts}}

\begin{figure*}
    \centering
    \includegraphics[width=\textwidth]{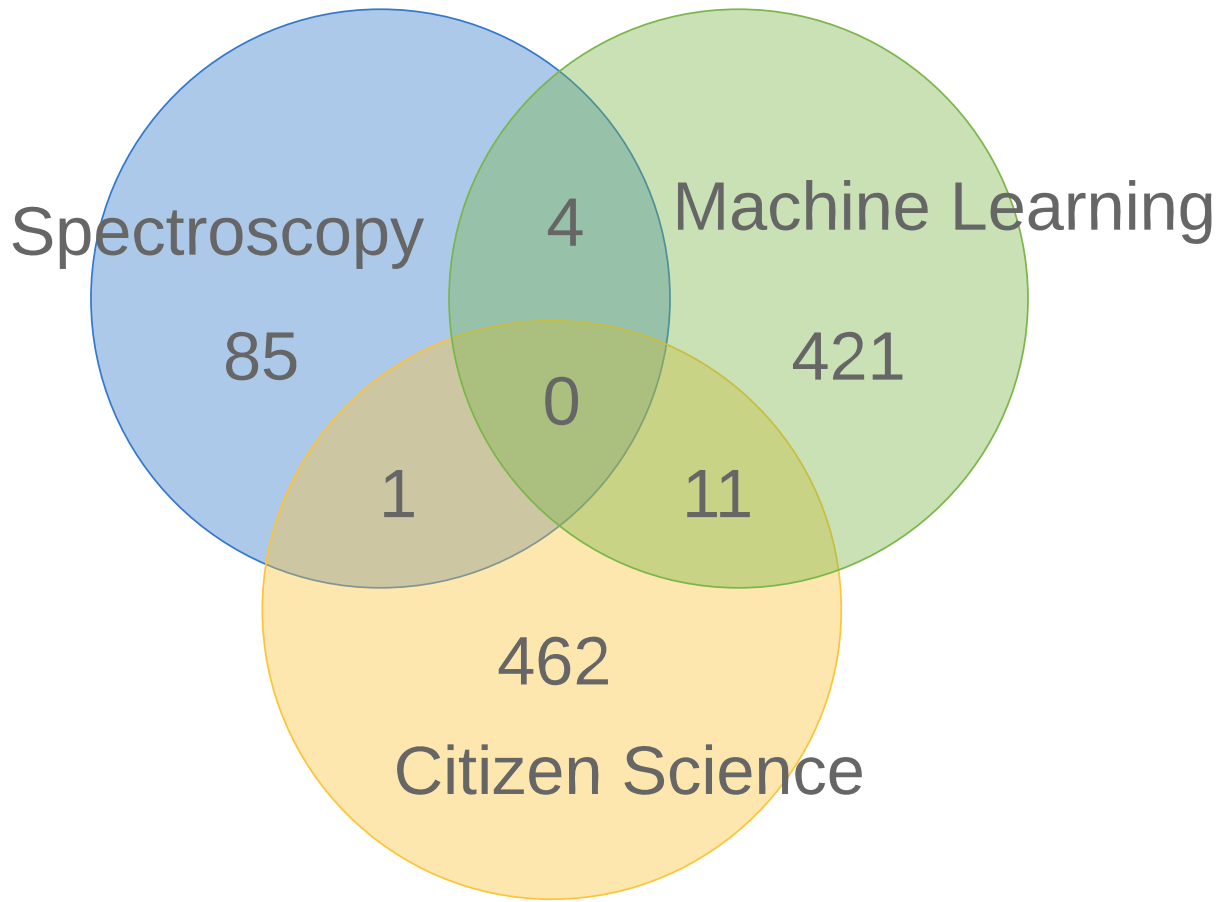}
    \caption{Venn diagram showing the number of lens candidates identified by each of the three methods with more lenient cuts to both GalaxyZoo citizen science and LinKS machine learning. Candidates with ``Lens or arc" scores above 20\% and LinKS machine learning candidates with lens scores greater than 0 are included in this selection. Overlapping regions indicate the number of lens candidates identified by both (or all three) candidates.}
    \label{fig:venn_softcut}
\end{figure*}

The final cuts were intended to ensure a comparable and reliable catalog of lens candidates for each method analyzed. The numbers reflected in our final catalogs will vary if considered under different cuts, so we also examined the overlap considering a more relaxed cut that included all GAMA spectroscopy candidates, all LinKS machine learning candidates with scores greater than 0, and all GalaxyZoo candidates with lens scores greater than 20\%. Figure \ref{fig:venn_softcut} shows that this cut improves overlap to a small extent. However, it introduces a significant increase in false-positives and unreliable candidates, which we considered to outweigh the benefits of the improvement to overlap for the purposes of this study. The overlaps that we see in the more relaxed cuts do appear in the expected parameter space, with masses and redshifts comparable to the two overlap candidates that appear in our final catalogs.

%

\begin{thebibliography}{}
\expandafter\ifx\csname natexlab\endcsname\relax\def\natexlab#1{#1}\fi

\bibitem[{{Abazajian} {et~al.}(2009){Abazajian}, {Adelman-McCarthy},
  {Ag{\"u}eros}, {Allam}, {Allende Prieto}, {An}, {Anderson}, {Anderson},
  {Annis}, {Bahcall}, {Bailer-Jones}, {Barentine}, {Bassett}, {Becker},
  {Beers}, {Bell}, {Belokurov}, {Berlind}, {Berman}, {Bernardi}, {Bickerton},
  {Bizyaev}, {Blakeslee}, {Blanton}, {Bochanski}, {Boroski}, {Brewington},
  {Brinchmann}, {Brinkmann}, {Brunner}, {Budav{\'a}ri}, {Carey}, {Carliles},
  {Carr}, {Castander}, {Cinabro}, {Connolly}, {Csabai}, {Cunha}, {Czarapata},
  {Davenport}, {de Haas}, {Dilday}, {Doi}, {Eisenstein}, {Evans}, {Evans},
  {Fan}, {Friedman}, {Frieman}, {Fukugita}, {G{\"a}nsicke}, {Gates},
  {Gillespie}, {Gilmore}, {Gonzalez}, {Gonzalez}, {Grebel}, {Gunn},
  {Gy{\"o}ry}, {Hall}, {Harding}, {Harris}, {Harvanek}, {Hawley}, {Hayes},
  {Heckman}, {Hendry}, {Hennessy}, {Hindsley}, {Hoblitt}, {Hogan}, {Hogg},
  {Holtzman}, {Hyde}, {Ichikawa}, {Ichikawa}, {Im}, {Ivezi{\'c}}, {Jester},
  {Jiang}, {Johnson}, {Jorgensen}, {Juri{\'c}}, {Kent}, {Kessler}, {Kleinman},
  {Knapp}, {Konishi}, {Kron}, {Krzesinski}, {Kuropatkin}, {Lampeitl},
  {Lebedeva}, {Lee}, {Lee}, {Leger}, {L{\'e}pine}, {Li}, {Lima}, {Lin}, {Long},
  {Loomis}, {Loveday}, {Lupton}, {Magnier}, {Malanushenko}, {Malanushenko},
  {Mandelbaum}, {Margon}, {Marriner}, {Mart{\'{\i}}nez-Delgado}, {Matsubara},
  {McGehee}, {McKay}, {Meiksin}, {Morrison}, {Mullally}, {Munn}, {Murphy},
  {Nash}, {Nebot}, {Neilsen}, {Newberg}, {Newman}, {Nichol}, {Nicinski},
  {Nieto-Santisteban}, {Nitta}, {Okamura}, {Oravetz}, {Ostriker}, {Owen},
  {Padmanabhan}, {Pan}, {Park}, {Pauls}, {Peoples}, {Percival}, {Pier}, {Pope},
  {Pourbaix}, {Price}, {Purger}, {Quinn}, {Raddick}, {Fiorentin}, {Richards},
  {Richmond}, {Riess}, {Rix}, {Rockosi}, {Sako}, {Schlegel}, {Schneider},
  {Scholz}, {Schreiber}, {Schwope}, {Seljak}, {Sesar}, {Sheldon}, {Shimasaku},
  {Sibley}, {Simmons}, {Sivarani}, {Smith}, {Smith}, {Smol{\v c}i{\'c}},
  {Snedden}, {Stebbins}, {Steinmetz}, {Stoughton}, {Strauss}, {Subba Rao},
  {Suto}, {Szalay}, {Szapudi}, {Szkody}, {Tanaka}, {Tegmark}, {Teodoro},
  {Thakar}, {Tremonti}, {Tucker}, {Uomoto}, {Vanden Berk}, {Vandenberg},
  {Vidrih}, {Vogeley}, {Voges}, {Vogt}, {Wadadekar}, {Watters}, {Weinberg},
  {West}, {White}, {Wilhite}, {Wonders}, {Yanny}, {Yocum}, {York}, {Zehavi},
  {Zibetti}, \& {Zucker}}]{SDSS-DR7}
{Abazajian}, K.~N., {Adelman-McCarthy}, J.~K., {Ag{\"u}eros}, M.~A., {et~al.}
  2009, \apjs, 182, 543

\bibitem[{{Alam} {et~al.}(2015){Alam}, {Albareti}, {Allende Prieto}, {Anders},
  {Anderson}, {Anderton}, {Andrews}, {Armengaud}, {Aubourg}, {Bailey}, \&
  et~al.}]{Alam15}
{Alam}, S., {Albareti}, F.~D., {Allende Prieto}, C., {et~al.} 2015, \apjs, 219,
  12

\bibitem[{{Arneson} {et~al.}(2012){Arneson}, {Brownstein}, \&
  {Bolton}}]{Arneson12}
{Arneson}, R.~A., {Brownstein}, J.~R., \& {Bolton}, A.~S. 2012, \apj, 753, 4

\bibitem[{{Auger} {et~al.}(2009{\natexlab{a}}){Auger}, {Treu}, {Bolton},
  {Gavazzi}, {Koopmans}, {Marshall}, {Bundy}, \& {Moustakas}}]{slacs9}
{Auger}, M.~W., {Treu}, T., {Bolton}, A.~S., {et~al.} 2009{\natexlab{a}}, \apj,
  705, 1099

\bibitem[{{Auger} {et~al.}(2009{\natexlab{b}}){Auger}, {Treu}, {Bolton},
  {Gavazzi}, {Koopmans}, {Marshall}, {Bundy}, \& {Moustakas}}]{Auger09}
---. 2009{\natexlab{b}}, \apj, 705, 1099

\bibitem[{{Auger} {et~al.}(2010{\natexlab{a}}){Auger}, {Treu}, {Bolton},
  {Gavazzi}, {Koopmans}, {Marshall}, {Moustakas}, \& {Burles}}]{Auger10}
---. 2010{\natexlab{a}}, \apj, 724, 511

\bibitem[{{Auger} {et~al.}(2010{\natexlab{b}}){Auger}, {Treu}, {Bolton},
  {Gavazzi}, {Koopmans}, {Marshall}, {Moustakas}, \& {Burles}}]{slacs10}
---. 2010{\natexlab{b}}, \apj, 724, 511

\bibitem[{{Baldry} {et~al.}(2012){Baldry}, {Driver}, {Loveday}, {Taylor},
  {Kelvin}, {Liske}, {Norberg}, {Robotham}, {Brough}, {Hopkins}, {Bamford},
  {Peacock}, {Bland-Hawthorn}, {Conselice}, {Croom}, {Jones}, {Parkinson},
  {Popescu}, {Prescott}, {Sharp}, \& {Tuffs}}]{Baldry12}
{Baldry}, I.~K., {Driver}, S.~P., {Loveday}, J., {et~al.} 2012, \mnras, 421,
  621

\bibitem[{{Baldry} {et~al.}(2014){Baldry}, {Alpaslan}, {Bauer},
  {Bland-Hawthorn}, {Brough}, {Cluver}, {Croom}, {Davies}, {Driver},
  {Gunawardhana}, {Holwerda}, {Hopkins}, {Kelvin}, {Liske},
  {L{\'o}pez-S{\'a}nchez}, {Loveday}, {Norberg}, {Peacock}, {Robotham}, \&
  {Taylor}}]{Baldry14}
{Baldry}, I.~K., {Alpaslan}, M., {Bauer}, A.~E., {et~al.} 2014, \mnras, 441,
  2440

\bibitem[{{Baldry} {et~al.}(2018){Baldry}, {Liske}, {Brown}, {Robotham},
  {Driver}, {Dunne}, {Alpaslan}, {Brough}, {Cluver}, {Eardley}, {Farrow},
  {Heymans}, {Hildebrandt}, {Hopkins}, {Kelvin}, {Loveday}, {Moffett},
  {Norberg}, {Owers}, {Taylor}, {Wright}, {Bamford}, {Bland-Hawthorn},
  {Bourne}, {Bremer}, {Colless}, {Conselice}, {Croom}, {Davies}, {Foster},
  {Grootes}, {Holwerda}, {Jones}, {Kafle}, {Kuijken}, {Lara-Lopez},
  {L{\'o}pez-S{\'a}nchez}, {Meyer}, {Phillipps}, {Sutherland}, {van Kampen}, \&
  {Wilkins}}]{Baldry18}
{Baldry}, I.~K., {Liske}, J., {Brown}, M.~J.~I., {et~al.} 2018, \mnras, 474,
  3875

\bibitem[{{Barnab{\`e}} {et~al.}(2011){Barnab{\`e}}, {Czoske}, {Koopmans},
  {Treu}, \& {Bolton}}]{Barnabe11}
{Barnab{\`e}}, M., {Czoske}, O., {Koopmans}, L.~V.~E., {Treu}, T., \& {Bolton},
  A.~S. 2011, \mnras, 415, 2215

\bibitem[{{Beck} {et~al.}(2018){Beck}, {Scarlata}, {Fortson}, {Lintott},
  {Simmons}, {Galloway}, {Willett}, {Dickinson}, {Masters}, {Marshall}, \&
  {Wright}}]{Beck18}
{Beck}, M.~R., {Scarlata}, C., {Fortson}, L.~F., {et~al.} 2018, \mnras, 476,
  5516

\bibitem[{{Bolton} {et~al.}(2008{\natexlab{a}}){Bolton}, {Burles}, {Koopmans},
  {Treu}, {Gavazzi}, {Moustakas}, {Wayth}, \& {Schlegel}}]{slacs5}
{Bolton}, A.~S., {Burles}, S., {Koopmans}, L. V.~E., {et~al.}
  2008{\natexlab{a}}, \apj, 682, 964

\bibitem[{{Bolton} {et~al.}(2008{\natexlab{b}}){Bolton}, {Burles}, {Koopmans},
  {Treu}, {Gavazzi}, {Moustakas}, {Wayth}, \& {Schlegel}}]{Bolton08}
---. 2008{\natexlab{b}}, \apj, 682, 964

\bibitem[{{Bolton} {et~al.}(2006){Bolton}, {Burles}, {Koopmans}, {Treu}, \&
  {Moustakas}}]{slacs1}
{Bolton}, A.~S., {Burles}, S., {Koopmans}, L.~V.~E., {Treu}, T., \&
  {Moustakas}, L.~A. 2006, \apj, 638, 703

\bibitem[{{Bolton} {et~al.}(2004){Bolton}, {Burles}, {Schlegel}, {Eisenstein},
  \& {Brinkmann}}]{Bolton04}
{Bolton}, A.~S., {Burles}, S., {Schlegel}, D.~J., {Eisenstein}, D.~J., \&
  {Brinkmann}, J. 2004, \aj, 127, 1860

\bibitem[{{Bolton} {et~al.}(2008{\natexlab{c}}){Bolton}, {Treu}, {Koopmans},
  {Gavazzi}, {Moustakas}, {Burles}, {Schlegel}, \& {Wayth}}]{slacs7}
{Bolton}, A.~S., {Treu}, T., {Koopmans}, L.~V.~E., {et~al.} 2008{\natexlab{c}},
  \apj, 684, 248

\bibitem[{{Bolton} {et~al.}(2012){Bolton}, {Schlegel}, {Aubourg}, {Bailey},
  {Bhardwaj}, {Brownstein}, {Burles}, {Chen}, {Dawson}, {Eisenstein}, {Gunn},
  {Knapp}, {Loomis}, {Lupton}, {Maraston}, {Muna}, {Myers}, {Olmstead},
  {Padmanabhan}, {P{\^a}ris}, {Percival}, {Petitjean}, {Rockosi}, {Ross},
  {Schneider}, {Shu}, {Strauss}, {Thomas}, {Tremonti}, {Wake}, {Weaver}, \&
  {Wood-Vasey}}]{Bolton12}
{Bolton}, A.~S., {Schlegel}, D.~J., {Aubourg}, {\'E}., {et~al.} 2012, \aj, 144,
  144

\bibitem[{{Brownstein} {et~al.}(2012){Brownstein}, {Bolton}, {Schlegel},
  {Eisenstein}, {Kochanek}, {Connolly}, {Maraston}, {Pandey}, {Seitz}, {Wake},
  {Wood-Vasey}, {Brinkmann}, {Schneider}, \& {Weaver}}]{Brownstein12}
{Brownstein}, J.~R., {Bolton}, A.~S., {Schlegel}, D.~J., {et~al.} 2012, \apj,
  744, 41

\bibitem[{{Bruzual} \& {Charlot}(2003)}]{Bruzual03}
{Bruzual}, G., \& {Charlot}, S. 2003, \mnras, 344, 1000

\bibitem[{{Calzetti} \& {Heckman}(1999)}]{Calzetti99}
{Calzetti}, D., \& {Heckman}, T.~M. 1999, \apj, 519, 27

\bibitem[{{Chan} {et~al.}(2016){Chan}, {Suyu}, {More}, {Oguri}, {Chiueh},
  {Coupon}, {Hsieh}, {Komiyama}, {Miyazaki}, {Murayama}, {Nishizawa}, {Price},
  {Tait}, {Terai}, {Utsumi}, \& {Wang}}]{Chan16}
{Chan}, J.~H.~H., {Suyu}, S.~H., {More}, A., {et~al.} 2016, \apj, 832, 135

\bibitem[{{Chen} {et~al.}(2019){Chen}, {Fassnacht}, {Suyu}, {Rusu}, {Chan},
  {Wong}, {Auger}, {Hilbert}, {Bonvin}, {Birrer}, {Millon}, {Koopmans},
  {Lagattuta}, {McKean}, {Vegetti}, {Courbin}, {Ding}, {Halkola}, {Jee},
  {Shajib}, {Sluse}, {Sonnenfeld}, \& {Treu}}]{Chen19b}
{Chen}, G.~C.-F., {Fassnacht}, C.~D., {Suyu}, S.~H., {et~al.} 2019, \mnras

\bibitem[{{Cluver} {et~al.}(2014){Cluver}, {Jarrett}, {Hopkins}, {Driver},
  {Liske}, {Gunawardhana}, {Taylor}, {Robotham}, {Alpaslan}, {Baldry}, {Brown},
  {Peacock}, {Popescu}, {Tuffs}, {Bauer}, {Bland-Hawthorn}, {Colless},
  {Holwerda}, {Lara-L{\'o}pez}, {Leschinski}, {L{\'o}pez-S{\'a}nchez},
  {Norberg}, {Owers}, {Wang}, \& {Wilkins}}]{Cluver14}
{Cluver}, M.~E., {Jarrett}, T.~H., {Hopkins}, A.~M., {et~al.} 2014, \apj, 782,
  90

\bibitem[{{Collett} {et~al.}(2018){Collett}, {Oldham}, {Smith}, {Auger},
  {Westfall}, {Bacon}, {Nichol}, {Masters}, {Koyama}, \& {van den
  Bosch}}]{Collett18}
{Collett}, T.~E., {Oldham}, L.~J., {Smith}, R.~J., {et~al.} 2018, Science, 360,
  1342

\bibitem[{{Collier} {et~al.}(2018{\natexlab{a}}){Collier}, {Smith}, \&
  {Lucey}}]{Collier18b}
{Collier}, W.~P., {Smith}, R.~J., \& {Lucey}, J.~R. 2018{\natexlab{a}}, \mnras,
  478, 1595

\bibitem[{{Collier} {et~al.}(2018{\natexlab{b}}){Collier}, {Smith}, \&
  {Lucey}}]{Collier18a}
---. 2018{\natexlab{b}}, \mnras, 473, 1103

\bibitem[{{Cyr-Racine} {et~al.}(2019){Cyr-Racine}, {Keeton}, \&
  {Moustakas}}]{Cyr-Racine19}
{Cyr-Racine}, F.-Y., {Keeton}, C.~R., \& {Moustakas}, L.~A. 2019, \prd, 100,
  023013

\bibitem[{{Czoske} {et~al.}(2012){Czoske}, {Barnab{\`e}}, {Koopmans}, {Treu},
  \& {Bolton}}]{Czoske12}
{Czoske}, O., {Barnab{\`e}}, M., {Koopmans}, L.~V.~E., {Treu}, T., \& {Bolton},
  A.~S. 2012, \mnras, 419, 656

\bibitem[{{de Jong} {et~al.}(2013){de Jong}, {Verdoes Kleijn}, {Kuijken}, \&
  {Valentijn}}]{de-Jong13}
{de Jong}, J.~T.~A., {Verdoes Kleijn}, G.~A., {Kuijken}, K.~H., \& {Valentijn},
  E.~A. 2013, Experimental Astronomy, 35, 25

\bibitem[{{de Jong} {et~al.}(2015){de Jong}, {Verdoes Kleijn}, {Boxhoorn},
  {Buddelmeijer}, {Capaccioli}, {Getman}, {Grado}, {Helmich}, {Huang},
  {Irisarri}, {Kuijken}, {La Barbera}, {McFarland}, {Napolitano}, {Radovich},
  {Sikkema}, {Valentijn}, {Begeman}, {Brescia}, {Cavuoti}, {Choi}, {Cordes},
  {Covone}, {Dall'Ora}, {Hildebrandt}, {Longo}, {Nakajima}, {Paolillo},
  {Puddu}, {Rifatto}, {Tortora}, {van Uitert}, {Buddendiek},
  {Harnois-D{\'e}raps}, {Erben}, {Eriksen}, {Heymans}, {Hoekstra}, {Joachimi},
  {Kitching}, {Klaes}, {Koopmans}, {K{\"o}hlinger}, {Roy}, {Sif{\'o}n},
  {Schneider}, {Sutherland}, {Viola}, \& {Vriend}}]{de-Jong15}
{de Jong}, J.~T.~A., {Verdoes Kleijn}, G.~A., {Boxhoorn}, D.~R., {et~al.} 2015,
  \aap, 582, A62

\bibitem[{{de Jong} {et~al.}(2017){de Jong}, {Kleijn}, {Erben}, {Hildebrandt},
  {Kuijken}, {Sikkema}, {Brescia}, {Bilicki}, {Napolitano}, {Amaro}, {Begeman},
  {Boxhoorn}, {Buddelmeijer}, {Cavuoti}, {Getman}, {Grado}, {Helmich}, {Huang},
  {Irisarri}, {La Barbera}, {Longo}, {McFarland}, {Nakajima}, {Paolillo},
  {Puddu}, {Radovich}, {Rifatto}, {Tortora}, {Valentijn}, {Vellucci}, {Vriend},
  {Amon}, {Blake}, {Choi}, {Conti}, {Gwyn}, {Herbonnet}, {Heymans}, {Hoekstra},
  {Klaes}, {Merten}, {Miller}, {Schneider}, \& {Viola}}]{de-Jong17}
{de Jong}, J.~T.~A., {Kleijn}, G.~A.~V., {Erben}, T., {et~al.} 2017, \aap, 604,
  A134

\bibitem[{{De Lucia} {et~al.}(2006){De Lucia}, {Springel}, {White}, {Croton},
  \& {Kauffmann}}]{De-Lucia06}
{De Lucia}, G., {Springel}, V., {White}, S.~D.~M., {Croton}, D., \&
  {Kauffmann}, G. 2006, \mnras, 366, 499

\bibitem[{{Dessauges-Zavadsky} {et~al.}(2017){Dessauges-Zavadsky}, {Zamojski},
  {Rujopakarn}, {Richard}, {Sklias}, {Schaerer}, {Combes}, {Ebeling}, {Rawle},
  {Egami}, {Boone}, {Cl{\'e}ment}, {Kneib}, {Nyland}, \&
  {Walth}}]{Dessauges-Zavadsky17a}
{Dessauges-Zavadsky}, M., {Zamojski}, M., {Rujopakarn}, W., {et~al.} 2017,
  \aap, 605, A81

\bibitem[{{Driver} {et~al.}(2009){Driver}, {Norberg}, {Baldry}, {Bamford},
  {Hopkins}, {Liske}, {Loveday}, {Peacock}, {Hill}, {Kelvin}, {Robotham},
  {Cross}, {Parkinson}, {Prescott}, {Conselice}, {Dunne}, {Brough}, {Jones},
  {Sharp}, {van Kampen}, {Oliver}, {Roseboom}, {Bland-Hawthorn}, {Croom},
  {Ellis}, {Cameron}, {Cole}, {Frenk}, {Couch}, {Alister}, {Proctor}, {De
  Propris}, {Doyle}, {Edmondson}, {Nichol}, {Thomas}, {Eales}, {Jarvis},
  {Kuijken}, {Lahav}, {Madore}, {Seibert}, {Meyer}, {Staveley-Smith},
  {Phillipps}, {Popescu}, {Sansom}, {Sutherland}, {Tuffs}, \&
  {Warren}}]{Driver09}
{Driver}, S.~P., {Norberg}, P., {Baldry}, I.~K., {et~al.} 2009, Astronomy and
  Geophysics, 50, 050000

\bibitem[{{Driver} {et~al.}(2011){Driver}, {Hill}, {Kelvin}, {Robotham},
  {Liske}, {Norberg}, {Baldry}, {Bamford}, {Hopkins}, {Loveday}, {Peacock},
  {Andrae}, {Bland-Hawthorn}, {Brough}, {Brown}, {Cameron}, {Ching}, {Colless},
  {Conselice}, {Croom}, {Cross}, {de Propris}, {Dye}, {Drinkwater}, {Ellis},
  {Graham}, {Grootes}, {Gunawardhana}, {Jones}, {van Kampen}, {Maraston},
  {Nichol}, {Parkinson}, {Phillipps}, {Pimbblet}, {Popescu}, {Prescott},
  {Roseboom}, {Sadler}, {Sansom}, {Sharp}, {Smith}, {Taylor}, {Thomas},
  {Tuffs}, {Wijesinghe}, {Dunne}, {Frenk}, {Jarvis}, {Madore}, {Meyer},
  {Seibert}, {Staveley-Smith}, {Sutherland}, \& {Warren}}]{Driver11}
{Driver}, S.~P., {Hill}, D.~T., {Kelvin}, L.~S., {et~al.} 2011, \mnras, 413,
  971

\bibitem[{{Eisenstein} {et~al.}(2001){Eisenstein}, {Annis}, {Gunn}, {Szalay},
  {Connolly}, {Nichol}, {Bahcall}, {Bernardi}, {Burles}, {Castander},
  {Fukugita}, {Hogg}, {Ivezi{\'c}}, {Knapp}, {Lupton}, {Narayanan}, {Postman},
  {Reichart}, {Richmond}, {Schneider}, {Schlegel}, {Strauss}, {SubbaRao},
  {Tucker}, {Vanden Berk}, {Vogeley}, {Weinberg}, \& {Yanny}}]{Eisenstein01}
{Eisenstein}, D.~J., {Annis}, J., {Gunn}, J.~E., {et~al.} 2001, \aj, 122, 2267

\bibitem[{{Gavazzi} {et~al.}(2014){Gavazzi}, {Marshall}, {Treu}, \&
  {Sonnenfeld}}]{Gavazzi14}
{Gavazzi}, R., {Marshall}, P.~J., {Treu}, T., \& {Sonnenfeld}, A. 2014, \apj,
  785, 144

\bibitem[{{Gavazzi} {et~al.}(2008){Gavazzi}, {Treu}, {Koopmans}, {Bolton},
  {Moustakas}, {Burles}, \& {Marshall}}]{slacs6}
{Gavazzi}, R., {Treu}, T., {Koopmans}, L.~V.~E., {et~al.} 2008, \apj, 677, 1046

\bibitem[{{Gavazzi} {et~al.}(2007){Gavazzi}, {Treu}, {Rhodes}, {Koopmans},
  {Bolton}, {Burles}, {Massey}, \& {Moustakas}}]{slacs4}
{Gavazzi}, R., {Treu}, T., {Rhodes}, J.~D., {et~al.} 2007, \apj, 667, 176

\bibitem[{{Geller} {et~al.}(2016){Geller}, {Hwang}, {Dell'Antonio}, {Zahid},
  {Kurtz}, \& {Fabricant}}]{Geller16}
{Geller}, M.~J., {Hwang}, H.~S., {Dell'Antonio}, I.~P., {et~al.} 2016, \apjs,
  224, 11

\bibitem[{{Geller} {et~al.}(2014){Geller}, {Hwang}, {Fabricant}, {Kurtz},
  {Dell'Antonio}, \& {Zahid}}]{Geller14}
{Geller}, M.~J., {Hwang}, H.~S., {Fabricant}, D.~G., {et~al.} 2014, \apjs, 213,
  35

\bibitem[{{Geller} {et~al.}(2006){Geller}, {Kenyon}, {Barton}, {Jarrett}, \&
  {Kewley}}]{Geller06}
{Geller}, M.~J., {Kenyon}, S.~J., {Barton}, E.~J., {Jarrett}, T.~H., \&
  {Kewley}, L.~J. 2006, \aj, 132, 2243

\bibitem[{{Hilbert} {et~al.}(2008){Hilbert}, {White}, {Hartlap}, \&
  {Schneider}}]{Hilbert08}
{Hilbert}, S., {White}, S. D.~M., {Hartlap}, J., \& {Schneider}, P. 2008,
  \mnras, 386, 1845

\bibitem[{{Holwerda} {et~al.}(2015){Holwerda}, {Baldry}, {Alpaslan}, {Bauer},
  {Bland-Hawthorn}, {Brough}, {Brown}, {Cluver}, {Conselice}, {Driver},
  {Hopkins}, {Jones}, {L{\'o}pez-S{\'a}nchez}, {Loveday}, {Meyer}, \&
  {Moffett}}]{Holwerda15}
{Holwerda}, B.~W., {Baldry}, I.~K., {Alpaslan}, M., {et~al.} 2015, \mnras, 449,
  4277

\bibitem[{{Holwerda} {et~al.}(2019){Holwerda}, {Kelvin}, {Baldry}, {Lintott},
  {Alpaslan}, {Pimbblet}, {Liske}, {Kitching}, {Bamford}, {de Jong}, {Bilicki},
  {Hopkins}, {Bridge}, {Steele}, {Jacques}, {Goswami}, {Kusmic}, {Roemer},
  {Kruk}, {Popescu}, {Kuijken}, {Wang}, {Wright}, \& {Kitching}}]{Holwerda19}
{Holwerda}, B.~W., {Kelvin}, L., {Baldry}, I., {et~al.} 2019, \aj, 158, 103

\bibitem[{{Hopkins}(2018)}]{Hopkins18}
{Hopkins}, A.~M. 2018, \pasa, 35, 39

\bibitem[{{Huang} {et~al.}(2020{\natexlab{a}}){Huang}, {Storfer}, {Gu}, {Ravi},
  {Pilon}, {Sheu}, {Venguswamy}, {Bankda}, {Dey}, {Landriau}, {Lang},
  {Meisner}, {Moustakas}, {Myers}, {Sajith}, {Schlafly}, \&
  {Schlegel}}]{Huang20a}
{Huang}, X., {Storfer}, C., {Gu}, A., {et~al.} 2020{\natexlab{a}}, arXiv
  e-prints, arXiv:2005.04730

\bibitem[{{Huang} {et~al.}(2020{\natexlab{b}}){Huang}, {Storfer}, {Ravi},
  {Pilon}, {Domingo}, {Schlegel}, {Bailey}, {Dey}, {Gupta}, {Herrera},
  {Juneau}, {Landriau}, {Lang}, {Meisner}, {Moustakas}, {Myers}, {Schlafly},
  {Valdes}, {Weaver}, {Yang}, \& {Y{\`e}che}}]{Huang20b}
{Huang}, X., {Storfer}, C., {Ravi}, V., {et~al.} 2020{\natexlab{b}}, \apj, 894,
  78

\bibitem[{{Jacobs} {et~al.}(2019){Jacobs}, {Collett}, {Glazebrook},
  {Buckley-Geer}, {Diehl}, {Lin}, {McCarthy}, {Qin}, {Odden}, {Caso Escudero},
  {Dial}, {Yung}, {Gaitsch}, {Pellico}, {Lindgren}, {Abbott}, {Annis}, {Avila},
  {Brooks}, {Burke}, {Carnero Rosell}, {Carrasco Kind}, {Carretero}, {da
  Costa}, {De Vicente}, {Fosalba}, {Frieman}, {Garc{\'\i}a-Bellido},
  {Gaztanaga}, {Goldstein}, {Gruen}, {Gruendl}, {Gschwend}, {Hollowood},
  {Honscheid}, {Hoyle}, {James}, {Krause}, {Kuropatkin}, {Lahav}, {Lima},
  {Maia}, {Marshall}, {Miquel}, {Plazas}, {Roodman}, {Sanchez}, {Scarpine},
  {Serrano}, {Sevilla-Noarbe}, {Smith}, {Sobreira}, {Suchyta}, {Swanson},
  {Tarle}, {Vikram}, {Walker}, {Zhang}, \& {DES Collaboration}}]{Jacobs19}
{Jacobs}, C., {Collett}, T., {Glazebrook}, K., {et~al.} 2019, \apjs, 243, 17

\bibitem[{{Kelvin} {et~al.}(2012){Kelvin}, {Driver}, {Robotham}, {Hill},
  {Alpaslan}, {Baldry}, {Bamford}, {Bland-Hawthorn}, {Brough}, {Graham},
  {H{\"a}ussler}, {Hopkins}, {Liske}, {Loveday}, {Norberg}, {Phillipps},
  {Popescu}, {Prescott}, {Taylor}, \& {Tuffs}}]{Kelvin12}
{Kelvin}, L.~S., {Driver}, S.~P., {Robotham}, A.~S.~G., {et~al.} 2012, \mnras,
  421, 1007

\bibitem[{{Kettlety} {et~al.}(2018){Kettlety}, {Hesling}, {Phillipps},
  {Bremer}, {Cluver}, {Taylor}, {Bland -Hawthorn}, {Brough}, {De Propris},
  {Driver}, {Holwerda}, {Kelvin}, {Sutherland }, \& {Wright}}]{Kettlety18}
{Kettlety}, T., {Hesling}, J., {Phillipps}, S., {et~al.} 2018, \mnras, 473, 776

\bibitem[{{Koopmans} {et~al.}(2006){Koopmans}, {Treu}, {Bolton}, {Burles}, \&
  {Moustakas}}]{slacs3}
{Koopmans}, L.~V.~E., {Treu}, T., {Bolton}, A.~S., {Burles}, S., \&
  {Moustakas}, L.~A. 2006, \apj, 649, 599

\bibitem[{{Kuijken} {et~al.}(2019){Kuijken}, {Heymans}, {Dvornik},
  {Hildebrandt}, {de Jong}, {Wright}, {Erben}, {Bilicki}, {Giblin}, {Shan},
  {Getman}, {Grado}, {Hoekstra}, {Miller}, {Napolitano}, {Paolilo}, {Radovich},
  {Schneider}, {Sutherland }, {Tewes}, {Tortora}, {Valentijn}, \& {Verdoes
  Kleijn}}]{Kuijken19}
{Kuijken}, K., {Heymans}, C., {Dvornik}, A., {et~al.} 2019, \aap, 625, A2

\bibitem[{{Li} {et~al.}(2020){Li}, {Napolitano}, {Tortora}, {Spiniello},
  {Koopmans}, {Huang}, {Roy}, {Vernardos}, {Chatterjee}, {Giblin}, {Getman},
  {Radovich}, {Covone}, \& {Kuijken}}]{Li20c}
{Li}, R., {Napolitano}, N.~R., {Tortora}, C., {et~al.} 2020, \apj, 899, 30

\bibitem[{{Limousin} {et~al.}(2009){Limousin}, {Cabanac}, {Gavazzi}, {Kneib},
  {Motta}, {Richard}, {Thanjavur}, {Foex}, {Pello}, {Crampton}, {Faure},
  {Fort}, {Jullo}, {Marshall}, {Mellier}, {More}, {Soucail}, {Suyu},
  {Swinbank}, {Sygnet}, {Tu}, {Valls-Gabaud}, {Verdugo}, \&
  {Willis}}]{Limousin09}
{Limousin}, M., {Cabanac}, R., {Gavazzi}, R., {et~al.} 2009, \aap, 502, 445

\bibitem[{{Limousin} {et~al.}(2010){Limousin}, {Jullo}, {Richard}, {Cabanac},
  {Suyu}, {Halkola}, {Kneib}, {Gavazzi}, \& {Soucail}}]{Limousin10}
{Limousin}, M., {Jullo}, E., {Richard}, J., {et~al.} 2010, \aap, 524, A95

\bibitem[{{Lintott} {et~al.}(2008){Lintott}, {Schawinski}, {Slosar}, {Land},
  {Bamford}, {Thomas}, {Raddick}, {Nichol}, {Szalay}, {Andreescu}, {Murray}, \&
  {Vandenberg}}]{Lintott08}
{Lintott}, C.~J., {Schawinski}, K., {Slosar}, A., {et~al.} 2008, \mnras, 389,
  1179

\bibitem[{{Liske} {et~al.}(2015){Liske}, {Baldry}, {Driver}, {Tuffs},
  {Alpaslan}, {Andrae}, {Brough}, {Cluver}, {Grootes}, {Gunawardhana},
  {Kelvin}, {Loveday}, {Robotham}, {Taylor}, {Bamford}, {Bland-Hawthorn},
  {Brown}, {Drinkwater}, {Hopkins}, {Meyer}, {Norberg}, {Peacock}, {Agius},
  {Andrews}, {Bauer}, {Ching}, {Colless}, {Conselice}, {Croom}, {Davies}, {De
  Propris}, {Dunne}, {Eardley}, {Ellis}, {Foster}, {Frenk}, {H{\"a}u{\ss}ler},
  {Holwerda}, {Howlett}, {Ibarra}, {Jarvis}, {Jones}, {Kafle}, {Lacey},
  {Lange}, {Lara-L{\'o}pez}, {L{\'o}pez-S{\'a}nchez}, {Maddox}, {Madore},
  {McNaught-Roberts}, {Moffett}, {Nichol}, {Owers}, {Palamara}, {Penny},
  {Phillipps}, {Pimbblet}, {Popescu}, {Prescott}, {Proctor}, {Sadler},
  {Sansom}, {Seibert}, {Sharp}, {Sutherland}, {V{\'a}zquez-Mata}, {van Kampen},
  {Wilkins}, {Williams}, \& {Wright}}]{Liske15}
{Liske}, J., {Baldry}, I.~K., {Driver}, S.~P., {et~al.} 2015, \mnras, 452, 2087

\bibitem[{{Marshall} {et~al.}(2016){Marshall}, {Verma}, {More}, {Davis},
  {More}, {Kapadia}, {Parrish}, {Snyder}, {Wilcox}, {Baeten}, {Macmillan},
  {Cornen}, {Baumer}, {Simpson}, {Lintott}, {Miller}, {Paget}, {Simpson},
  {Smith}, {K{\"u}ng}, {Saha}, \& {Collett}}]{Marshall16}
{Marshall}, P.~J., {Verma}, A., {More}, A., {et~al.} 2016, \mnras, 455, 1171

\bibitem[{{Newman} {et~al.}(2018){Newman}, {Belli}, {Ellis}, \&
  {Patel}}]{Newman18b}
{Newman}, A.~B., {Belli}, S., {Ellis}, R.~S., \& {Patel}, S.~G. 2018, \apj,
  862, 125

\bibitem[{{Petrillo} {et~al.}(2017){Petrillo}, {Tortora}, {Chatterjee},
  {Vernardos}, {Koopmans}, {Verdoes Kleijn}, {Napolitano}, {Covone},
  {Schneider}, {Grado}, \& {McFarland}}]{Petrillo17}
{Petrillo}, C.~E., {Tortora}, C., {Chatterjee}, S., {et~al.} 2017, \mnras, 472,
  1129

\bibitem[{{Petrillo} {et~al.}(2018){Petrillo}, {Tortora}, {Chatterjee},
  {Vernardos}, {Koopmans}, {Kleijn}, {Napolitano}, {Covone}, {Kelvin}, \&
  {Hopkins}}]{Petrillo18}
---. 2018

\bibitem[{{Petrillo} {et~al.}(2019){Petrillo}, {Tortora}, {Vernardos},
  {Koopmans}, {Verdoes Kleijn}, {Bilicki}, {Napolitano}, {Chatterjee},
  {Covone}, {Dvornik}, {Erben}, {Getman}, {Giblin}, {Heymans}, {de Jong},
  {Kuijken}, {Schneider}, {Shan}, {Spiniello}, \& {Wright}}]{Petrillo19}
{Petrillo}, C.~E., {Tortora}, C., {Vernardos}, G., {et~al.} 2019, \mnras, 484,
  3879

\bibitem[{{Riess} {et~al.}(2011){Riess}, {Macri}, {Casertano}, {Lampeitl},
  {Ferguson}, {Filippenko}, {Jha}, {Li}, \& {Chornock}}]{Riess11}
{Riess}, A.~G., {Macri}, L., {Casertano}, S., {et~al.} 2011, \apj, 730, 119

\bibitem[{{Robotham} {et~al.}(2010){Robotham}, {Driver}, {Norberg}, {Baldry},
  {Bamford}, {Hopkins}, {Liske}, {Loveday}, {Peacock}, {Cameron}, {Croom},
  {Doyle}, {Frenk}, {Hill}, {Jones}, {van Kampen}, {Kelvin}, {Kuijken},
  {Nichol}, {Parkinson}, {Popescu}, {Prescott}, {Sharp}, {Sutherland},
  {Thomas}, \& {Tuffs}}]{Robotham10}
{Robotham}, A., {Driver}, S.~P., {Norberg}, P., {et~al.} 2010, \pasa, 27, 76

\bibitem[{{Shu} {et~al.}(2015{\natexlab{a}}){Shu}, {Bolton}, {Brownstein},
  {Montero-Dorta}, {Koopmans}, {Treu}, {Gavazzi}, {Auger}, {Czoske},
  {Marshall}, \& {Moustakas}}]{slacs12}
{Shu}, Y., {Bolton}, A.~S., {Brownstein}, J.~R., {et~al.} 2015{\natexlab{a}},
  \apj, 803, 71

\bibitem[{{Shu} {et~al.}(2015{\natexlab{b}}){Shu}, {Bolton}, {Brownstein},
  {Montero-Dorta}, {Koopmans}, {Treu}, {Gavazzi}, {Auger}, {Czoske},
  {Marshall}, \& {Moustakas}}]{Shu15}
---. 2015{\natexlab{b}}, \apj, 803, 71

\bibitem[{{Shu} {et~al.}(2016){Shu}, {Bolton}, {Kochanek}, {Oguri},
  {Perez-Fournon}, {Zheng}, {Mao}, {Montero-Dorta}, {Brownstein},
  {Marques-Chaves}, \& {Menard}}]{Shu16}
{Shu}, Y., {Bolton}, A.~S., {Kochanek}, C.~S., {et~al.} 2016, ArXiv e-prints

\bibitem[{{Shu} {et~al.}(2017){Shu}, {Brownstein}, {Bolton}, {Koopmans},
  {Treu}, {Montero-Dorta}, {Auger}, {Czoske}, {Gavazzi}, {Marshall}, \&
  {Moustakas}}]{Shu17}
{Shu}, Y., {Brownstein}, J.~R., {Bolton}, A.~S., {et~al.} 2017

\bibitem[{{Sonnenfeld} {et~al.}(2015){Sonnenfeld}, {Treu}, {Marshall}, {Suyu},
  {Gavazzi}, {Auger}, \& {Nipoti}}]{Sonnenfeld15}
{Sonnenfeld}, A., {Treu}, T., {Marshall}, P.~J., {et~al.} 2015, \apj, 800, 94

\bibitem[{{Speagle} {et~al.}(2019){Speagle}, {Leauthaud}, {Huang}, {Bradshaw},
  {Ardila}, {Capak}, {Eisenstein}, {Masters}, {Mand elbaum}, {More}, {Simet},
  \& {Sif{\'o}n}}]{Speagle19}
{Speagle}, J.~S., {Leauthaud}, A., {Huang}, S., {et~al.} 2019, \mnras, 490,
  5658

\bibitem[{{Suyu} {et~al.}(2017){Suyu}, {Bonvin}, {Courbin}, {Fassnacht},
  {Rusu}, {Sluse}, {Treu}, {Wong}, {Auger}, {Ding}, {Hilbert}, {Marshall},
  {Rumbaugh}, {Sonnenfeld}, {Tewes}, {Tihhonova}, {Agnello}, {Blandford},
  {Chen}, {Collett}, {Koopmans}, {Liao}, {Meylan}, \& {Spiniello}}]{Suyu17}
{Suyu}, S.~H., {Bonvin}, V., {Courbin}, F., {et~al.} 2017, \mnras, 468, 2590

\bibitem[{{Taylor} {et~al.}(2011){Taylor}, {Hopkins}, {Baldry}, {Brown},
  {Driver}, {Kelvin}, {Hill}, {Robotham}, {Bland-Hawthorn}, {Jones}, {Sharp},
  {Thomas}, {Liske}, {Loveday}, {Norberg}, {Peacock}, {Bamford}, {Brough},
  {Colless}, {Cameron}, {Conselice}, {Croom}, {Frenk}, {Gunawardhana},
  {Kuijken}, {Nichol}, {Parkinson}, {Phillipps}, {Pimbblet}, {Popescu},
  {Prescott}, {Sutherland}, {Tuffs}, {van Kampen}, \& {Wijesinghe}}]{Taylor11}
{Taylor}, E.~N., {Hopkins}, A.~M., {Baldry}, I.~K., {et~al.} 2011, \mnras, 418,
  1587

\bibitem[{{Taylor} {et~al.}(2020){Taylor}, {Cluver}, {Duffy}, {Gurri},
  {Hoekstra}, {Sonnenfeld}, {Bremer}, {Brouwer}, {Chisari}, {Dvornik}, {Erben},
  {Hildebrandt}, {Hopkins}, {Kelvin}, {Phillipps}, {Robotham}, {Sifon},
  {Vakili}, \& {Wright}}]{Taylor20}
{Taylor}, E.~N., {Cluver}, M.~E., {Duffy}, A., {et~al.} 2020, arXiv e-prints,
  arXiv:2006.10040

\bibitem[{{Tortora} {et~al.}(2018){Tortora}, {Koopmans}, {Napolitano}, \&
  {Valentijn}}]{Tortora18a}
{Tortora}, C., {Koopmans}, L.~V.~E., {Napolitano}, N.~R., \& {Valentijn}, E.~A.
  2018, \mnras, 473, 2324

\bibitem[{{Tortora} {et~al.}(2010){Tortora}, {Napolitano}, {Cardone},
  {Capaccioli}, {Jetzer}, \& {Molinaro}}]{Tortora10}
{Tortora}, C., {Napolitano}, N.~R., {Cardone}, V.~F., {et~al.} 2010, ArXiv
  e-prints

\bibitem[{{Treu} {et~al.}(2009){Treu}, {Gavazzi}, {Gorecki}, {Marshall},
  {Koopmans}, {Bolton}, {Moustakas}, \& {Burles}}]{slacs8}
{Treu}, T., {Gavazzi}, R., {Gorecki}, A., {et~al.} 2009, \apj, 690, 670

\bibitem[{{Treu} {et~al.}(2006){Treu}, {Koopmans}, {Bolton}, {Burles}, \&
  {Moustakas}}]{slacs2}
{Treu}, T., {Koopmans}, L.~V., {Bolton}, A.~S., {Burles}, S., \& {Moustakas},
  L.~A. 2006, \apj, 640, 662

\bibitem[{{Vegetti} {et~al.}(2012){Vegetti}, {Lagattuta}, {McKean}, {Auger},
  {Fassnacht}, \& {Koopmans}}]{Vegetti12}
{Vegetti}, S., {Lagattuta}, D.~J., {McKean}, J.~P., {et~al.} 2012, \nat, 481,
  341

\bibitem[{{Verlinde}(2017)}]{Verlinde17}
{Verlinde}, E. 2017, SciPost Physics, 2, 016

\bibitem[{{Whitmore} {et~al.}(2016){Whitmore}, {Allam}, {Budav{\'a}ri},
  {Casertano}, {Downes}, {Donaldson}, {Fall}, {Lubow}, {Quick}, {Strolger},
  {Wallace}, \& {White}}]{Whitmore16}
{Whitmore}, B.~C., {Allam}, S.~S., {Budav{\'a}ri}, T., {et~al.} 2016, \aj, 151,
  134

\bibitem[{{Wright} {et~al.}(2018){Wright}, {Driver}, \& {Robotham}}]{Wright18}
{Wright}, A.~H., {Driver}, S.~P., \& {Robotham}, A.~S.~G. 2018, \mnras

\bibitem[{{Wright} {et~al.}(2016){Wright}, {Robotham}, {Bourne}, {Driver},
  {Dunne}, {Maddox}, {Alpaslan}, {Andrews}, {Bauer}, {Bland-Hawthorn},
  {Brough}, {Brown}, {Clarke}, {Cluver}, {Davies}, {Grootes}, {Holwerda},
  {Hopkins}, {Jarrett}, {Kafle}, {Lange}, {Liske}, {Loveday}, {Moffett},
  {Norberg}, {Popescu}, {Smith}, {Taylor}, {Tuffs}, {Wang}, \&
  {Wilkins}}]{Wright16}
{Wright}, A.~H., {Robotham}, A.~S.~G., {Bourne}, N., {et~al.} 2016, \mnras,
  460, 765

\bibitem[{{Zahid} {et~al.}(2016){Zahid}, {Geller}, {Fabricant}, \&
  {Hwang}}]{Zahid16}
{Zahid}, H.~J., {Geller}, M.~J., {Fabricant}, D.~G., \& {Hwang}, H.~S. 2016,
  \apj, 832, 203

\bibitem[{{Zitrin}(2017)}]{Zitrin17}
{Zitrin}, A. 2017, \apj, 834, 45

\bibitem[{{Zitrin} \& {Broadhurst}(2016)}]{Zitrin16}
{Zitrin}, A., \& {Broadhurst}, T. 2016, \apj, 833, 25

\end{thebibliography}



\end{document}